\documentclass[twocolumn,showpacs,aps,prd,floatfix]{revtex4}
\usepackage{graphicx}
\newcommand{\beq}{\begin{equation}}
\newcommand{\eeq}{\end{equation}}
\newcommand{\beqn}{\begin{eqnarray}}
\newcommand{\eeqn}{\end{eqnarray}}

\newcommand{\ve}[1]{\mbox{\boldmath $#1$}}

\newcommand{\usch}{\hat{u}}
\newcommand{\rsch}{\hat{r}}
\newcommand{\uiso}{u}
\newcommand{\riso}{r}
\newcommand{\viso}{v}
\newcommand{\radj}{r_0}
\newcommand{\Rss}{\hat{R}_s}
\newcommand{\Rs}{R_s}
\newcommand{\Racc}{R_a}
\newcommand{\RHL}{R_{BHL}}
\newcommand{\vcloud}{V_{\infty}}
\newcommand{\sscloud}{a_{\infty}}
\newcommand{\vkep}{V_k}
\newcommand{\clocal}{c_l}

\usepackage{dcolumn}
\usepackage{bm}
\usepackage{epsf}

\begin{document}
\bibliographystyle{apsrev}
\title{Binary Black Hole Mergers in Gaseous
  Environments: 

``Binary Bondi'' and ``Binary Bondi-Hoyle-Lyttleton'' Accretion}

\author{Brian D. Farris}

\author{Yuk Tung Liu}

\author{Stuart L.\ Shapiro}

\altaffiliation{Also at Department of Astronomy \& NCSA, University of Illinois
at Urbana-Champaign, Urbana, IL 61801}

\affiliation{Department of Physics, University of Illinois at
Urbana-Champaign, Urbana, IL~61801}

\begin{abstract}
Merging supermassive black hole-black hole (BHBH) binaries produced in galaxy mergers are promising sources of detectable gravitational waves.  If such a merger takes place in a gaseous environment, there is a possibility of a simultaneous detection of electromagnetic and gravitational radiation, as the stirring, shock heating and accretion of the gas may produce variability and enhancements in the electromagnetic flux.  Such a simultaneous detection can provide a wealth of opportunities to study gravitational physics, accretion physics, and cosmology.  We investigate this scenario by performing fully general relativistic, hydrodynamic simulations of merging, equal-mass, nonspinning BHBH binaries embedded in gas clouds.  We evolve the metric using the Baumgarte-Shapiro-Shibata-Nakamura (BSSN) formulation with standard moving puncture gauge conditions and handle the hydrodynamics via a high-resolution shock-capturing (HRSC) scheme.  We consider both ``binary Bondi accretion'' in which the binary is at rest relative to the ambient gas cloud, as well as ``binary Bondi-Hoyle-Lyttleton accretion'' in which the binary moves relative to the gas cloud. The gas cloud is assumed to be homogeneous far from the binary and governed by a $\Gamma$-law equation of state. We vary $\Gamma$ between $4/3$ and $5/3$.  For each simulation, we compute the gas flow and accretion rate and estimate the electromagnetic luminosity due to bremsstrahlung and synchrotron emission.  We find evidence for significant enhancements in both the accretion rate and luminosity over values for a single black hole of the same mass as the binary.  We estimate that this luminosity enhancement should be detectable by LSST for a $10^6 M_{\odot}$ binary in a hot gas cloud of density $n \sim 10 \mbox{ cm}^{-3}$ and temperature $T \sim 10^6 \mbox{ K}$ at $z=1$, reaching a maximum of $L\sim 3 \times 10^{43} \mbox{erg } \mbox{s}^{-1}$, with the emission peaking in the visible band.  
\end{abstract}

\maketitle

\section{Introduction}
All bulge galaxies (including the Milky Way) are believed to contain
a central supermassive black hole (SMBH) with a mass $M$ between $10^4 M_{\odot}$ and  $10^9
M_{\odot}$ \cite{richstone98,peterson00,ferrarese05}.  It is also believed that galaxy mergers commonly lead to the formation of a
massive black hole-black hole (BHBH) binary system in the merged remnants \cite{begelman80,roos81}.  In
the standard picture, the BHBH binary separation decreases first through dynamical
friction due to distant stellar encounters, then through gravitational slingshot interactions in which
nearby stars are ejected at velocities comparable to the binary's orbital velocity, and
finally through the emission of gravitational radiation, leading to
coalescence of the binary \cite{merritt05}.
These low-frequency gravitational waves will be detectable by LISA
and will contain a wealth of information about the inspiral.  However,
it has been argued that the gaseous accretion flow which forms around
the binary can be a source of electromagnetic radiation as well
\cite{milosavljevic04}.  This raises the exciting possibility of a
simultaneous detection of electromagnetic and gravitational waves from BHBH mergers.  

This picture is supported by a number of observed AGNs that may be harboring BHBH binaries.  VLBI observations of the elliptical galaxy 0402+379 have discovered two radio sources at a projected separation of only 7 pc.  The existence of jets, as well as variability associated with BH activity, indicate that the system may be a BHBH binary \cite{maness04,rodriguez06,rodriguez09}.  Another candidate is OJ 287, a BL Lac object whose light curve shows variability with a period of $\sim 12$~yr.  It is believed that this may be a massive BHBH binary in which the smaller BH orbits with a period of 12~yr, penetrating the disk of the primary and giving rise to the observed variability \cite{lehto96,valtonen06,valtonen08}.  The quasar SDSS 092712.65+294344 is not believed to be a binary system, but rather a recoiling BH which is the product of a binary merger.  This suggestion is supported by a systematic shift of $2650~\mbox{km s}^{-1}$ in its emission lines \cite{adelman07,komossa08}.  Another candidate is quasar SDSS J153636.22+044127.0, in which two broad line emission systems are observed, separated in velocity by $3500~\mbox{km s}^{-1}$.  This observation has been interpreted as a BHBH binary system in which each object has its own emission system \cite{boroson09}.  

Information from a simultaneous detection of electromagnetic and gravitational waves may be useful for studying fundamental aspects of gravitational physics.  For example, in some modified gravity scenarios, the propagation velocity for gravitons may differ from that of photons \cite{kocsis08,deffayet07}.  Additionally,
the measurement of the luminosity distance from the gravitational wave signal at an accuracy of $1-10\%$, coupled with the redshift information from the electromagnetic detection, could serve as a cosmological ``standard siren'' of unprecedented accuracy (better than $\sim 1 \%$) \cite{holz05}.  Such detections may also combine accurate measurements of BH spins and masses obtained from gravitational wave signals with electromagnetic observations to probe BH accretion physics in great detail \cite{kocsis06}.

Several mechanisms for electromagnetic emission from accretion disks
around merging BHBH binaries have been proposed.  In one
scenario, the inner edge of the accretion disk is identified as the radius
at which the viscous torque on the gas balances the gravitational torque from
binary.  This radius is between 1.5 and 2 times the orbital separation \cite{artymowicz94,gunther02,escala05,macfadyen08} and encompasses a hollowed region in the disk.  Late in the inspiral the BHBH binary decouples from the disk and coalesces.  For a binary of mass $M \approx 10^6 M_{\odot}$ accreting at $10\%$ of the Eddington rate, the subsequent evolution of this disk, assumed optically thick, gives rise to a source which initially peaks in the {\it UV} band and hardens to {\it EUV} and soft {\it X}-ray emission at late times \cite{milosavljevic04,shapiro09}.  
Additionally, there is a sudden
change in the mass of the binary during the final stage of the merger,
as gravitational waves carry away a few percent of the mass.  The
abrupt change in the potential creates waves in the disk which can
grow into shocks and increase the luminosity of the disk in a
characteristic way \cite{corrales09,oneill09,megevand09}, giving rise to a potentially detectable prompt {\it X}-ray signal.  Another possibility is that the
merged BH remnant experiences a recoil velocity which may, in principle, be as high as several thousand km
$\mathrm{s}^{-1}$ \cite{campanelli07}, although it is likely to be much lower ($< 200 \ \mbox{km}/\mbox{s}$) in most galaxy mergers \cite{bogdanovic07}.  This recoiling BH may ``shake'' or penetrate
the disk, creating shocks which could give rise to a transient
signal. 

Various methods have been used to model plausible sources of electromagnetic emission from BH mergers.  In one approach, the dynamics of the inspiral is ignored, focusing on the effect of BH kicks and/or BH mass loss on the hydrodynamical flow \cite{anderson09,megevand09,schnittman08,shields08,lippai08,rossi09,oneill09,corrales09}.  In another approach, the behavior of the gas is modeled by following the motion of collisionless ``particle tracers'' on geodesics \cite{vanmeter09}.  Only recently has a fully relativistic, dynamical simulation of a BHBH binary merger in a hydrodynamic setting been performed \cite{bode09}.

In this paper we study BHBH binary mergers in the presence of ambient gas.  Modeling such systems requires fully general relativistic dynamical simulations, including relativistic hydrodynamics.  The development of stable algorithms to integrate Einstein's field equations of general relativity numerically in $3+1$ dimensions, such as the BSSN formalism~\cite{bssn_shibata,bssn_stu} and the generalized harmonic approach~\cite{pretorius05a}, together with the identification of suitable gauge conditions, has enabled several pioneering simulations that demonstrate how to track the late inspiral, plunge and merger of a BHBH binary in vacuum~\cite{pretorius05b,campanelli06,baker06}.  More refined follow-up simulations of these strong-field, late phases, joined onto analytic, post-Newtonian calculations of the early inspiral epoch~\cite{blanchet08}, are now capable of producing accurate gravitational waveforms for merging BHBH binaries with companions spanning a range of mass ratios and spins.

With the problem of gravitational wave emission from a vacuum BHBH binary inspiral well in hand, it is now important to turn to the problem of electromagnetic emission from BHBH binary coalescence in an astrophysically realistic gaseous environment.  

When gas accretes onto a SMBH in the center of a galaxy the specific angular momentum of the gas $\tilde{L}$ may, in many cases, be much larger than that of a circular orbit near the horizon, $\tilde{L} \sim 2M c$ \cite{novikov73} leading to a disk-like accretion flow.  For this reason, simulations to date have focused on disk-like accretion.  However, it has been argued that for {\it hot} flows, in which the gas
is near the galaxy virial temperature, the gas is supported by pressure against free infall and the flow may well be
described by the spherical ``Bondi'' accretion model \cite{merritt05}.  In this so-called ``cooling-flow model of quasar fueling'', intergalactic gas during the early stages of galaxy formation is accelerated toward the center of dark matter halos and is shock-heated to the virial temperature.  This gas accretes then onto the growing SMBH at nearly the Eddington rate.  For a $10^6 M_{\odot}$ BH, the gas is expected to have a density $n \sim 10 \mbox{ cm}^{-3}$ and temperature $T \sim 10^6~\mbox{K} - 10^8~\mbox{K}$ \cite{merritt05,nulsen00}.

Classical ``Bondi accretion'' refers to spherically symmetric, steady-state accretion of an adiabatic gas onto a stationary star.  The gas is assumed to be homogeneous and at rest far from the star and flow adiabatically with a $\Gamma$-law equation of state (EOS).  The problem was first studied in Newtonian gravitation for a point mass by Bondi \cite{bondi52}, and later extended to accretion onto a BH in full general-relativity \cite{michel72,shapiro73,shapiro_book_83}.  Accretion onto a star moving with constant velocity through a cloud which is asymptotically uniform and at rest was first studied qualitatively in Newtonian gravity by Hoyle and Lyttleton \cite{hoyle39} and later extended by Bondi and Hoyle \cite{bondi44}.  The general-relativistic version for accretion onto a single BH has been studied via numerical simulations \cite{petrich89,hunt71,shima85,font98}.  We refer to this scenario as the ``Bondi-Hoyle-Lyttleton accretion''  (BHL) problem.  

This paper is the first in a series of papers in which we will explore hydrodynamic gas flows around merging BHBH binaries.  Here we focus on {\it hot} flows with zero net angular momentum and restrict our attention to simulations in which the binary is placed in a gas cloud which has constant density and temperature at infinity.  We treat two cases: one in which the gas is asymptotically at rest (binary Bondi accretion), the other where the gas is moving with constant velocity with respect to the BHBH center of mass (binary BHL accretion).  Bondi and BHL accretion onto single BHs represent classic problems which are well understood.  We seek to use this understanding as a foundation for tackling the problem of accretion onto a merging BHBH binary.  The initial conditions of the gas during the merger of two supermassive BHs in realistic astrophysical environments is still an open question and subject to uncertainties in cosmological structure and galaxy formation scenarios, and in our understanding of the formation history and role of massive BHs in this process.  The binary Bondi and BHL problems provide excellent settings in which to begin a rigorous probe of the binary accretion problem.

The structure of the paper is as follows.  In Sec.~\ref{sec:computational_challenge} we discuss the unique computational challenge posed by the vast dynamical range characterizing our problem and we introduce our technique to tackle it. In Secs.~\ref{sec:basic_eqns} and \ref{sec:numerical}, we briefly outline the basic gravitational field and matter evolution equations and their
specific implementation in our relativistic, hydrodynamic numerical
scheme. Here we also provide an overview of our initial data, 
gauge conditions, and diagnostics.  
In Sec.~\ref{sec:code_tests}, we review code tests which were performed to validate our numerical scheme.  In Sec.~\ref{sec:stat_bh} we review the analytic Bondi solution for a single, stationary BH and demonstrate our ability to reproduce those results.  We also compare with previous relativistic simulations of BHL accretion onto a single BH.  In Sec.~\ref{sec:binary_bh}, we summarize the results of our binary BHBH merger
simulations.  In Sec.~\ref{sec:discussion} we summarize our findings
and comment on future directions.  We adopt geometrized units and set $G=c=1$.

\section{Computational Challenge}
\label{sec:computational_challenge}
Simulating realistic accretion flows is extremely challenging due to
the enormous dynamic range in the relevant length scales defining the problem.
One important length scale is set by the ADM mass $M$ of the central gravitating source,
$R=M$.  This is the length scale at which relativistic effects become
significant.  Another important length scale is the transonic radius, $\Rs \sim M / \sscloud^2$, where $\sscloud$ is the asymptotic sound speed in the gas cloud.  This corresponds to the radius inside of which the accretion flow onto a stationary central mass $M$ becomes supersonic.  When the central gravitating object is moving supersonically with velocity $\vcloud$ relative to the gas cloud, we follow~\cite{hoyle39} and define the characteristic length scale,
\begin{equation}
\RHL = \frac{M}{\vcloud^2} \ .
\end{equation}
Gas approaching the central object with impact parameter less than $\sim \RHL$ will be captured.  The two length scales $\Rs$ and $\RHL$ can be combined into a single characteristic radius~\cite{bondi52},
\begin{equation}
  \Racc \equiv \frac{M}{(\sscloud^2 + \vcloud^2)} \ ,
\end{equation}
within which gas is bound to $M$ and will be accreted.  Another important length scale for the BHBH problem is the binaray separation, $d$. 
When $d \ll \Racc$, the accretion at radius $r \gg d$ is basically 
BHL flow onto a central gravitating object of 
mass equal to the total ADM mass of the binary. When $d \gg \Racc$, the 
accretion in regions near each BH is again a BHL 
flow but for a source of mass equal to the mass 
of a single BH.

The typical densities and temperatures for interstellar gas in a {\it hot} Bondi-like accretion flow are $n_{\infty} \sim 10~\mbox{cm}^{-3}$ and $T\sim10^6K$ respectively \cite{merritt05}.  Assuming that $\vcloud \lesssim \sscloud$, we find that $\Racc \sim 10^6 M$.  With
current computational resources, it is impossible to perform a relativistic simulation that follows the complete binary inspiral from separation $d \sim \Racc$ to $d \sim M$, assuming a realistic asymptotic gas temperature.
We address this issue by first performing ``prototype'' simulations in which we artificially increase the asymptotic temperatures in order to make the accretion radius $\Racc$ much smaller.  We then use these results and scaling to perform ``realistic'' simulations in which we treat realistic temperatures, but restrict our grid to domains much smaller than $\Racc$.  The details of these approaches are given in Section~\ref{sec:binary_bh}.

\section{Basic Equations}
\label{sec:basic_eqns}
Throughout this paper, we use Latin indices to denote spatial components (1-3) and Greek indices to denote spacetime components (0-3). 

The basic gravitational field and relativistic hydrodynamics equations are discussed in \cite{bssn_stu,mhd_code_paper} where their numerical implementation is described and code tests are summarized.  Here, we briefly sketch these equations and their implementation.
\subsection{Evolution of gravitational fields}
We write the spacetime metric in the standard $3+1$ form:
\beq
ds^2 = - \alpha^2 dt^2 + \gamma_{ij}(dx^i + \beta^i dt)(dx^j + \beta^j dt) .
\eeq
where $\alpha$, $\beta^i$, and $\gamma_{ij}$ are the lapse, shift, and
spatial metric, respectively.  The extrinsic curvature $K_{ij}$ is
given by
\begin{equation}
\label{Kij}
(\partial_t - {\mathcal{L}}_{\beta})\gamma_{ij} = -2\alpha K_{ij},
\end{equation}
where ${\mathcal{L}}_{\beta}$ is the Lie derivative with respect to
$\beta^i$.  We evolve $\gamma_{ij}$ and $K_{ij}$
using the BSSN formulation~\cite{bssn_shibata,bssn_stu}.  The fundamental variables for
BSSN evolution are
\begin{eqnarray}
  \label{phidef}
  \phi &\equiv& \frac{1}{12}\ln[\det(\gamma_{ij})]\ , \\
  \tilde\gamma_{ij} &\equiv& e^{-4\phi}\gamma_{ij}\ , \\
  K &\equiv& \gamma^{ij}K_{ij}\ , \\
  \tilde A_{ij} &\equiv& e^{-4\phi}(K_{ij} - \frac{1}{3}\gamma_{ij}K)\ , \\
  \tilde\Gamma^i &\equiv& -\tilde\gamma^{ij}{}_{,j}\ .
\end{eqnarray}
The evolution and constraint equations for
these fields are summarized in~\cite{bssn_shibata,bssn_stu}.  We assume in this paper that the mass of the gas is negligible compared to the mass of the BHs, thus we do not include matter source terms in our metric evolution equations.

Adding Kreiss-Oliger dissipation to the BSSN evolution equations outside the BH can reduce high-frequency numerical noise associated with AMR refinement interfaces \cite{baker06}.  We use this technique here and have found it useful in reducing Hamiltonian and momentum constraint violations.

We adopt the standard puncture gauge conditions: an advective ``1+log'' slicing condition for lapse and a ``Gamma-freezing'' condition for shift \cite{van_meter_06}.  Thus we have
\beqn
\partial_0 \alpha &=& - 2 \alpha K ,\\
\partial_0 \beta^i &=& (3/4) B^i , \\
\partial_0 B^i &=& \partial_0 \tilde{\Gamma}^i - \eta B^i \ , 
\eeqn
where $\partial_0 = \partial_t - \beta^j \partial_j$.  The $\eta$ parameter is set to $0.5/M$ in all simulations.

\subsection{Evolution of hydrodynamic fields}

The fundamental matter variables are the rest-mass density $\rho_0 \equiv n m_B$, specific internal energy $\epsilon$, pressure $P$, and four-velocity $u^{\mu}$.  The stress-energy tensor for an ideal gas is given by
\[
T_{\mu \nu} = \rho_0 h u_{\mu} u_{\nu} + P g_{\mu \nu} \ ,
\]
where $h=1+\epsilon + P/\rho_0$ is the specific enthalpy.  We evolve the ``conservative'' variables $\rho_*$, $\tilde{S}_i$, and $\tilde{\tau}$.  They are defined as
\beqn
\rho_* &=& - \sqrt{\gamma} \rho_0 n_{\mu} u^{\mu}\\
\tilde{S}_i &=& \sqrt{\gamma} T_{\mu \nu} n^{\mu} \gamma^{\nu}{}_i\\
\tilde{\tau} &=& \sqrt{\gamma} T_{\mu \nu} n^{\mu} n^{\nu} - \rho_* \ .
\eeqn
Here $n^{\alpha} = (\alpha^{-1},-\alpha^{-1}\beta^i)$ is the timelike unit vector normal to the $t=\mbox{constant}$ time slices.
Evolution equations are given by Eqs. (34), (36), and (38) of \cite{mhd_code_paper}.
\beqn
\partial_t \rho_* + \partial_j ( \rho_* v^j) &=& 0, \ \ \ \ \\\
\partial_t \tilde{S}_i + \partial_j (\alpha \sqrt{\gamma} T^j{}_i) &=& \frac{1}{2} \alpha \sqrt{\gamma} T^{\alpha \beta} \partial_i g_{\alpha \beta},\\
\partial_t \tilde{\tau} + \partial_i ( \alpha^2 \sqrt{\gamma} T^{0i} - \rho_* v^i) &=& s \ ,
\eeqn
where $\gamma \equiv \mbox{det}(\gamma_{ij}) = e^{12 \phi}$ and $v^i \equiv u^i/u^0$ is the fluid's 3-velocity.  The energy source term $s$ is given by 
\beqn
s &=& -\alpha \sqrt{\gamma} T^{\mu \nu} \nabla_{\nu} n_{\mu} \nonumber\\
&=& \alpha \sqrt{\gamma} [ ( T^{00} \beta^i \beta^j + 2 T^{0i} \beta^j + T^{ij}) K_{ij} \nonumber \\
&& -(T^{00} \beta^i + T^{0i}) \partial_i \alpha] \ .
\eeqn

\subsection{Equation of State}
To complete the system of equations, we must specify an equation of state (EOS).  While our code can handle any EOS of the form $P=P(\rho_0,\epsilon)$, we adopt a standard $\Gamma$-law EOS,
\begin{equation}
  \label{EOS}
  P=(\Gamma-1)\rho_0 \epsilon .
\end{equation}
We perform simulations with $\Gamma=4/3$, $5/3$, and $13/9$.  The choice of $\Gamma = 4/3$ is appropriate in the high-temperature limit in which both the electrons and baryons become relativistic ($kT \gtrsim m_B c^2 \sim10^{13} K$).  The choice of $\Gamma=5/3$ is appropriate in the opposite limit in which both electrons and baryons remain nonrelativistic ($kT \lesssim m_e c^2 \sim 10^{10} K$).  The choice of $\Gamma=13/9$ is appropriate when the electrons are relativistic while the baryons remain nonrelativistic ($m_e c^2 \sim 10^{10} K \lesssim kT \lesssim m_B c^2 \sim10^{13} K$)\cite{shapiro73}.  While it is not expected that temperatures in a realistic accretion flow will be high enough to make the matter behave as a $\Gamma=4/3$ gas, we include it in our study as a limiting case (it may also be useful to model a radiation-dominated thermal gas as a $\Gamma=4/3$ EOS).  In general, $\Gamma$ will also depend on the chemical composition of the gas.  However, we show here that it is a good approximation to simply assume a pure gas of ionized hydrogen in assigning $\Gamma$ for the matter-dominated regime of interest here.  

Take, for example, a fully ionized mixture of hydrogen and helium in which the electrons are relativistic while the nucleons remain nonrelativistic, and let $X$ be the fractional abundance by number of hydrogen ions.  The thermal energy density can be expressed as
\begin{equation}
\label{abundance_therm}
\rho_0 \epsilon = X(3 + 3/2) n k T + (1-X)(6+3/2) n k T \ .
\end{equation}
Here, each relativistic electron contributes a factor of $3nkT$, and each nonrelativistic nucleus contributes a factor of $(3/2) n k T$, where $n \equiv \rho_0 / m_B$ is the baryon number density of the fluid.  The pressure can be expressed as
\begin{equation}
\label{abundance_press}
P = X 2nkT + (1-X)3 nkT \ .
\end{equation}
Combining Eq.~(\ref{EOS}),  Eq.~(\ref{abundance_therm}) and Eq.~(\ref{abundance_press}) we find that


\begin{equation}
\Gamma = \frac{(13/9)X+ (7/3)(1-X)}{X+(5/3)(1-X)} \ .
\end{equation}

Adopting cosmological abundances~\cite{fields06}, we set $X=0.92$, which
gives $\Gamma = 1.439$.  Because this is close to $\Gamma=13/9=1.444$, the value for $X=1$, we henceforth set $\Gamma=13/9$ for simplicity.  In some cases, we allow for a temperature dependent transition from $\Gamma=5/3$ to $\Gamma=13/9$ as the electrons become relativistic as they approach the BH horizons (see Sec.~\ref{effective_index} for details).  Throughout this paper, we define temperature by
\begin{equation}
P=2 n k T \ ,
\end{equation}
appropriate for pure ionized hydrogen.

\section{Numerical Methods}
\label{sec:numerical}
\subsection{Initial Data}
\label{initialdata}
For each simulation discussed in this paper, we consider the gas to be adiabatic (apart from shocks) with uniform density and pressure at infinity.  We take the gas either be asymptotically at rest (binary Bondi accretion), or moving uniformly (binary BHL accretion).  We use the {\tt TWOPUNCTURES} code \cite{ansorg04} to construct the initial data for the binary BHBH metric.  We choose the bare mass and momentum of the punctures according to \cite{tichy04} in order to ensure that the BHBH binary orbit is initially close to quasicircular. 

We restrict our analysis to equal-mass, nonspinning BHs in this paper.  However, we note that upon merger, the remnant settles down to a spinning black hole with ADM mass $M_f = 0.95 M$.  Here $M_f$ denotes the final ADM mass, while $M$ denotes the initial ADM mass of the binary.   We also measure a final spin parameter $a/M_f=0.69$, in good agreement with the estimate of \cite{tichy08}.  For this case, there is no recoil velocity.

For all of our runs we use the analytic relativistic Bondi solution for accretion onto a stationary Schwarzschild BH as initial data.  This solution is reviewed in Sec.~\ref{sec:stat_bh}.  To implement this data, we treat the binary as a single gravitating object of mass $M$, where $M$ is the ADM mass of the binary, and apply the analytic Bondi solution everywhere outside a radius $\radj$.  We follow the method of \cite{faber07} and adjust the fluid parameters within the radius $\radj$ according to
\beq
\rho_0(r) = \rho_0(r_0) + \left.\frac{d \rho_0}{d r} \right|_{\radj}\frac{r^2-\radj^2}{2 \radj} \ .
\label{rho_inside_r0}
\eeq
This recipe ensures that the density and its first derivative are continuous across $\radj$.  Eq.~(\ref{rho_inside_r0}) is, of course, not the correct initial data for a quasistationary flow inside $r_0$, but we find that the system settles into a quasistationary flow within several $\delta t \sim \radj / a(r=\radj)$, where $a$ is the local sound speed.  For all our BHBH runs, we set $\radj / M = 5.5$, though we find that our results are not sensitive to this parameter.  The fluid 4-velocity inside $\radj$ is set to be radially inward, with the magnitude set according to Eq.~(\ref{continuity}).  The fluid pressure is set according to a polytropic EOS, $P=K \rho_0^{\Gamma}$, with $K=P_\infty/\rho_{0,\infty}^\Gamma$, where $\rho_{0,\infty}$ and $P_\infty$ are the rest-mass density and pressure at infinity, respectively.

For our code tests in which we evolve only a single, stationary BH, our hydrodynamic initial data is constructed in isotropic coordinates, which become singular on the horizon at $r=M/2$.  For these cases, we set $\radj / M=0.6$, and verify the finding of \cite{faber07} that the fluid evolution quickly relaxes to the equilibrium Bondi solution (see Sec.~\ref{single_punc_tests}). 

\subsection{Evolution of metric and matter}
We evolve the BSSN field equations
with fourth-order accurate, centered, finite-difference stencils,
except on shift advection terms, where we use fourth-order accurate
upwind stencils.  We apply Sommerfeld outgoing wave boundary
conditions to all BSSN fields.  Our code is embedded in
the Cactus parallelization framework~\cite{Cactus}, and our
fourth-order Runge-Kutta timestepping is managed by the {\tt MoL}
(Method of Lines) thorn, with a Courant-Friedrichs-Lewy (CFL) factor
set to 0.5 in all BHBH simulations.  We use the
Carpet~\cite{schnetter04} infrastructure to implement the moving-box
adaptive mesh refinement. In all AMR simulations presented here, we
use second-order temporal prolongation, coupled with fifth-order
spatial prolongation. The apparent horizon (AH) of the
BH is computed with the {\tt AHFinderDirect} Cactus
thorn~\cite{thornburg04}.

We write the general relativistic hydrodynamics equations in
conservative form. They are evolved by a high-resolution
shock-capturing (HRSC) technique~\cite{mhd_code_paper} that employs the
monotonized central (MC) reconstruction scheme~\cite{vanleer77} coupled to
the Harten, Lax, and van Leer (HLL) approximate Riemann solver~\cite{HLL}. 
The adopted hydrodynamic scheme is second-order accurate for smooth 
flows, and first-order accurate when discontinuities (e.g.\ shocks) 
arise.  Throughout the
evolution, we impose limits on the pressure to prevent
spurious heating and negative values of the internal energy
$\epsilon$. Specifically, we require $P_{\rm min}\leq P \leq P_{\rm max}$, 
where $P_{\rm max}=10^3 K \rho_0^\Gamma$ and $P_{\rm min}=K 
\rho_0^\Gamma/2$. 
Whenever $P$ exceeds $P_{\rm max}$ or drops below $P_{\rm min}$, we 
reset $P$ to $P_{\rm max}$ or $P_{\rm min}$, respectively.  We check that this fix is applied only inside the apparent horizon, which is causally disconnected from the rest of the grid.

At each timestep, 
we need to recover the ``primitive variables'' 
$\rho_0$, $P$, and $v^i$ from the ``conservative'' variables 
$\rho_*$, $\tilde{\tau}$, and $\tilde{S}_i$. We perform the 
inversion as specified in Eqs.~(57)--(62) of~\cite{mhd_code_paper}, but with a
slightly modified analytic quartic solver from the GNU Scientific
Library that outputs only the real roots.  We use the same technique as
in \cite{etienne08} to ensure that the values of $\tilde{S}_i$ and
$\tilde{\tau}$ yield physically valid primitive variables,
except we reset $\tilde{\tau}$ to
$10^{-10}\tilde{\tau}_{0,{\rm max}}$ (where
$\tilde{\tau}_{0,{\rm max}}$ is the maximum value of $\tilde{\tau}$
initially) when either $\tilde{S}_i$ or $\tilde{\tau}$ is 
unphysical [i.e., violate one of the inequalities~(34)~or~(35)
in \cite{etienne08}]. The restrictions usually apply only to the region near
the puncture inside the horizon.

For all of our ``prototype'' calculations, we set our outer boundary at $328 M$ and use 9 AMR refinement levels.  The maximum resolution near each puncture is $\delta x/M = 0.032$.  For our ``realistic'' calculations, we place our outer boundary at $164 M$ and use 8 AMR refinement levels.  For these cases, the highest resolution near each puncture  is also $\delta x/M = 0.032$.

We model the emission of electromagnetic radiation by treating this radiation loss as a {\it perturbation}, and neglect its influence on the hydrodynamic flow as well as any deviation from adiabaticity.  We also assume that the self-gravity of the gas can be ignored, and we do not include matter source terms in the evolution equations for the gravitational fields.

\subsection{Diagnostics}
\subsubsection{Rest-mass accretion rate}
An important diagnostic is the rest-mass flux through the horizon of each BH.  In order to compute this quantity we begin by defining a function $f(t,x,y,z)$ such that $f(t,x,y,z)=0$ on the 3d hypersurface which corresponds to the world tube of a BH horizon.  Thus, we can write
\beqn
f &=& \sqrt{(x-x_h(t))^2+(y-y_h(t))^2+(z-z_h(t))^2} \nonumber\\
&&- R(t,\theta,\phi) \ .
\eeqn
Here \{$x_h(t),y_h(t),z_h(t)$\} represents the coordinate position of the BH center at coordinate time $t$ and $R(t,\theta,\phi)$ represents the coordinate distance from the BH center to horizon in the $(\theta,\phi)$ direction.  Here $\theta$ and $\phi$ are spherical polar coordinates with the origin set at the BH center.  The rest-mass flux is then given by,

\beq
\label{Mdot_expression}
\dot{M} = - \int \alpha \sqrt{\gamma} \rho_0 u^{\mu} \partial_{\mu}f J d\theta d \phi \ ,
\eeq

where $J$ is the Jacobian
\beq
J= \left| \frac{\partial (t,f,\theta,\phi)}{\partial (t,x,y,z)} \right|^{-1} 
   = \left| \frac{\partial (f,\theta,\phi)}{\partial (x,y,z)} \right|^{-1} \ ,
\eeq
and where the integral is over each BH (apparent) horizon.

The subtlety in this calculation arises from the fact that the position and shape of each horizon changes in time for a dynamical spacetime, so we cannot neglect the time derivatives of $f$ in Eq.~(\ref{Mdot_expression}).  For a full derivation of Eq.~(\ref{Mdot_expression}), see Appendix~\ref{sec:Mdot_appendix}.

We note that the $\dot{M}$ defined above is inherently gauge dependent. This is true even in a 
spacetime with a timelike Killing vector field. In our adopted gauge the sum of mass fluxes across the BH horizons is equal to the total mass flux through a large sphere measured by a distant observer 
in the cases in which the flow is (quasi)stationary. Quasi-stationary flow is realized during the 
BHBH inspiral phase, as well as after the binary merger once the remnant settles down. See Appendix~\ref{sec:Mdot_appendix} 
for a detail discussion on various issues involving $\dot{M}$ and gauge choices.

\subsubsection{Luminosity}

In order to compute the observed electromagnetic radiation exactly, it would be necessary to employ a fully relativistic, radiative-transfer integrator suitable for a dynamical spacetime in $3+1$ dimensions.  While advancements toward constructing such a scheme have been made in various approximations (e.g. \cite{lindquist66,thorne81_pstf,schinder88,shapiro89,schinder89,mezzacappa89,rezzolla94,zampieri96,shapiro96,rezzolla96,balberg00,liebendorfer04,farris08}), this is an extremely challenging task which has not yet been fully accomplished.  In this paper, we instead adopt a crude method for estimating luminosities from our accreting gas, assuming it is an optically thin medium.  Results should be interpreted as order of magnitude estimates only.  Moreover, since we are primarily interested in enhancements and variations of the luminosity which may correlate with detectable gravitational wave signals, even crude estimates can provide valuable information.  

We assume that the radiation propagates through an optically thin gas and we neglect the roles of radiation pressure and radiative cooling on the hydrodynamic evolution.  Our solution is therefore valid only when the luminosity is below the Eddington limit and the radiative flux is below the thermal gas flux.  To calibrate our crude approximation scheme we compare our results to the more exact calculation of optically thin emission from accretion onto a single, static Schwarzschild BH at rest in a gas cloud \cite{shapiro73}.  We first review this solution below, then describe the approximations we make in this paper to simplify the calculation.

The total luminosity $L_{\infty}$ received at infinity from gas accreting steadily onto a stationary Schwarzschild BH is given by \cite{shapiro73}, and accounts for Doppler and gravitational redshifts as well as photon capture from the BH.  The result is 
\beq
L_{\infty} = \int_0^{\infty} L_{\nu_0} \left(\frac{d \nu_0}{d \nu} \right)d \nu \ .
\eeq
Here, $d \nu_0 / d \nu$ accounts for Doppler and gravitational redshifts, and is given by
\beq
\frac{d \nu_0}{d \nu}  = \frac{(1-v^2)^{1/2}}{1-v \ \mbox{cos} \ \Theta'} (1-2M/r)^{1/2} \ ,
\eeq
where $v$ is the proper velocity of the fluid as measured by a stationary observer, the specific luminosity $L_{\nu_0}$ is given by
\beq
\label{lum_per_freq}
L_{\nu_0} = 8 \pi^2 \int_{2M}^{\infty} r^2 dr \int_{-1}^{\mbox{cos} \Theta^*} j_{\nu} \frac{(1-v^2)}{(1-v \ \mbox{cos} \ \Theta')}^2  d(\mbox{cos}\Theta') \ ,
\eeq
and where 
\beq
 |\mathrm{cos}\Theta^*| = \left[\frac{27}{4}\left(\frac{2M}{r}\right)^2\left(\frac{2M}{r}-1\right)+1\right]^{1/2} \ .
\eeq
Here the emissivity $j_{\nu}$ is the specific emissivity (the energy emitted isotropically per time per volume per frequency interval) in the comoving frame of the fluid.  

For dynamical systems containing inspiraling BHBH binaries, simple analytic expressions for the luminosity similar to Eq.~(\ref{lum_per_freq}) do not exist.  In our simulations we employ a crude approximation for the luminosity.  We simply compute 
\begin{equation}
L(t) \approx \int_\mathcal{V} j \ dV \ ,
\end{equation}
where $dV$ is the proper 3-volume element in the fluid.  We exclude from our integration points less than $\xi r_h$ from each horizon.  Here $r_h$ is the apparent horizon radius, $\xi$ is a constant chosen so that for a single, isolated puncture, $r=\xi r_h$ corresponds to the surface of constant Schwarzschild radius $\rsch = 3 M$.  We have made this choice because within this radius, $50\%$ of the emitted radiation by a stationary source will be captured (see Eq.~(24) in \cite{shapiro73}).  For realistic temperatures (see Sec.~\ref{sec:real_bondi}), we find that the contribution to the luminosity is dominated by emission from near the binary.  
Our measured luminosity is thus insensive to the outer limit of integration and we chose to integrate to the outer boundary of our integration domain for definiteness.  
For our high temperature prototype runs, however the gas has a nonnegligible contribution to the luminosity even far from the binary, thus the measurement is sensitive to the outer limit of the integration.  
We have chosen to integrate to $r_{out}=44.7 M$, which corresponds to $r_{out} \approx 2 \Racc$ for our prototype BHBH Bondi runs.   We have verified that for a single, stationary puncture, our luminosity estimate is within a factor of $\sim 4$ of the exact relativistic luminosity quoted in \cite{shapiro73}, which we have separately verified with our code.  Thus, our somewhat crude method of estimating the luminosity is sufficient as an order of magnitude estimate.

In our estimates of electromagnetic emission, we consider both thermal bremsstrahlung (free-free emission) as well as synchrotron emission.  For bremsstrahlung emission we consider both electron-electron and electron-ion processes.  For synchrotron emission, we assume the presence of a small-scale, turbulent B-field whose magnitude is approximated by setting $P=\beta P_M \equiv \beta B^2/(8 \pi)$.  We thus assume that the magnetic pressure is some fraction $1 / \beta$ of its equipartition value.  Simulations of magnetized accretion flows have demonstrated that the magnetic fields do not typically reach their full equipartition value \cite{mckinney04}.  We have chosen $\beta=10$ to account for this.  We provide further details of the bremsstrahlung and synchrotron emissivities which we use in Appendix~\ref{sec:emissivity_appendix}.  
\subsection{Code Tests}
\label{sec:code_tests}

Our general relativistic hydrodynamic code has been 
thoroughly tested by passing a robust suite of tests. These tests include
maintaining stable rotating stars in stationary equilibrium, reproducing
the exact Oppenheimer-Snyder solution for collapse to a BH,
and reproducing analytic solutions for relativistic shocks and spherical Bondi accretion onto isolated BHs~\cite{mhd_code_paper}.  Our code has also been used to simulate the collapse of very
massive, rotating stars to black holes~\cite{liu07};
merging BHBH binaries~\cite{etienne07}, BHNS binaries \cite{etienne08,etienne09}, and relativistic hydrodynamic matter in the
presence of puncture black holes~\cite{faber07}.  Recently, our code has
been generalized to incorporate (optically thick) radiation transport
and its feedback on fluids in dynamical spacetimes~\cite{farris08}. 

All of the above tests and simulations were performed 
on grids with uniform spacing. In some of the simulations, we 
utilized the multiple-transition fisheye transformation~\cite{campanelli06_fisheye} 
so that a uniform computational grid spacing 
corresponds to physical coordinates with spatially varying
resolution. Recently, we have modified our code 
so that we can use the moving-box AMR infrastructure provided by
Carpet~\cite{schnetter04}. To test our new code, we have performed 
shock-tube tests and 3+1 simulations of linear gravitational waves, single 
stationary and boosted puncture BHs, puncture BHBH binaries,  and rapidly and differentially rotating
relativistic stars.  Our AMR code has also been used to perform simulations of BHNS mergers \cite{etienne09}.
  
All of our 3+1 AMR code tests were performed assuming equatorial
symmetry (i.e., symmetry about the $z=0$ orbital plane),  which we assume
in all evolutions presented in this paper.  We have also checked that our AMR code is able to accurately reproduce the analytic Bondi solution, and we have shown good agreement with the results of \cite{petrich89} for BHL accretion.  Results of the latter two tests are summarized in Sec.~\ref{single_punc_tests}.

\section{Accretion Onto a Single BH}
\label{sec:stat_bh}
\subsection{Analytic Relativistic Bondi Solution}
Steady state, adiabatic, spherically symmetric accretion onto point masses was first considered by Bondi for Newtonian gravitation \cite{bondi52}.  This work was later extended to handle accretion onto BHs in general relativity~\cite{michel72,shapiro73,shapiro73b,shapiro74}.  A thorough discussion of the relativistic solution may be found in Appendix~G of \cite{shapiro_book_83}.  Here we briefly summarize this solution.  We consider spherically symmetric, steady-state accretion onto a Schwarzschild BH of mass $M$.  We assume that the BH is at rest in an infinite gas cloud which has rest-mass density $\rho_{\infty}$, pressure $P_{\infty}$, and fluid 4-velocity $u^i=0$ at infinity.  We assume that the gas is adiabatic with an adiabatic index $\Gamma$.  We can solve the equations of relativistic hydrodynamics to derive an exact solution for the accretion flow.

\subsubsection{Key Equations}
\label{key_eqns_sec}
For simplicity, we will derive the solution in Schwarzschild coordinates, then transform to isotropic coordinates.  We begin by recasting the steady-state relativistic continuity and Euler equations into an integral form,  
\begin{eqnarray}
  \label{continuity}
  4 \pi \rho_0 \usch \rsch^2 &\equiv& \dot{M} =\mbox{const}.\\
  \label{euler}
  h^2 \left( 1- \frac{2M}{\rsch} + \usch^2 \right) &\equiv& h_{\infty}^2 = \mbox{const}.
\end{eqnarray}
Here, $\hat{r}$ and $\hat{u}$ denote the radius and radial fluid 4-velocity in Schwarzschild coordinates.

For adiabatic flow, the $\Gamma$-law EOS implies the polytropic relation $P=K \rho_0^{\Gamma}$, with $K=$const. Thus
\begin{equation}
\label{EOS_polytropic}
P =K \rho_0^{\gamma} = ( \Gamma - 1) \rho_0 \frac{h-1}{\Gamma} \ .
\end{equation}
For an adiabatic gas, the speed of sound is,
\begin{equation}
a \equiv \frac{1}{h^{1/2}} \left( \frac{dP}{d \rho_0}\right)^{1/2} = \left( \frac{\Gamma P}{\rho_0 h}\right)^{1/2} \ .
\end{equation}
This leads to the following relations between the enthalpy, the speed of sound, and the temperature
\begin{eqnarray}
\label{ah_relation}
 a^2 &=& (\Gamma - 1) \frac{h-1}{h}\\
\label{ht_relation}
 kT &\equiv&\frac{P}{2n} = \frac{m_B}{2} \frac{\Gamma-1}{\Gamma} (h-1) \ .
\end{eqnarray}
Any solution satisfying Eqs.~(\ref{continuity}) and (\ref{euler}) that maintains the causality constraint $a^2 < 1$ must contain a transonic radius outside the event horizon \cite{shapiro_book_83}.  At the transonic radius the radial velocity $\usch_s$ and the speed of sound are given by
\begin{eqnarray}
\usch^2_s &=& \frac{M}{2 \Rss} \ ,\\
\label{sound_speed_sonic}
a_s^2 &=&  \frac{M/2\Rss}{1-3 M/2\Rss} \ . 
\end{eqnarray}
The transonic radius sets the length scale at which the fluid parameters begin to deviate from their asymptotic values during inward flow.  We derive the transonic radius by plugging Eq.~(\ref{sound_speed_sonic}) into Eq.~(G.30) from Appendix~G of \cite{shapiro_book_83}.  This leads to a polynomial equation which we solve using a numerical root solver.  In the Newtonian limit, in which $\sscloud \ll 1$ and $M/\Rs \ll 1$, we recover the expressions,

\begin{equation}
  \Rs= \left\{ \begin{array}{l l}
      \displaystyle \frac{5-3\Gamma}{4}\frac{M}{\sscloud^2} & \ \ \ 1 \le \Gamma < 5/3 \\
      \displaystyle \frac{3}{4} \frac{M}{\sscloud} & \ \ \ \Gamma = 5/3 \ .
    \end{array}\right.
\end{equation}

Knowing the transonic radius, the accretion rate is given by
\begin{equation}
\label{Mdot_formula}
\dot{M} = 4 \pi \rho_s \usch_s \Rss^2 = 4 \pi \lambda M^2 \rho_{\infty} \sscloud^{-3} \ ,
\end{equation}
where $\lambda = \lambda(\Gamma,a_{\infty})$ is a parameter of order unity, which in the Newtonian limit is given by
\beq
\lambda = \left(\frac{1}{2} \right)^{(\Gamma+1)/2(\Gamma-1)}\left(\frac{5-3\Gamma}{4}\right)^{-(5-3\Gamma)/2(\Gamma-1)} \ .
\eeq
In the relativistic domain, we must solve a cubic equation for to find $\lambda$ (see \cite{shapiro_book_83} for details).
Given values for $\rho_{0,\infty}$ and $T_{\infty}$, we can determine $\dot{M}$ and $h_{\infty}$ using Eqs.~(\ref{ah_relation}), (\ref{ht_relation}), and (\ref{Mdot_formula}).  We can then use Eqs.~(\ref{continuity}) and (\ref{euler}), along with the EOS, Eq.~(\ref{EOS_polytropic}), to obtain the complete analytic solution.

\subsubsection{Transformation to isotropic coordinates}
In order to provide initial data for our simulations, we transform the solution to isotropic coordinates, the coordinate system adopted for our metric (puncture) initial data.  The relation between Schwarzschild radius $\rsch$ and isotropic radius $\riso$ outside the horizon is given by

\begin{eqnarray}
\rsch &=& \riso \left(1+\frac{M}{2\riso}\right)^2\\
\riso &=& \frac{\rsch - M + \sqrt{\rsch(\rsch - 2 M)}}{2} \ .
\end{eqnarray}
Since the flow is purely radial, we have
\begin{eqnarray}
\uiso &=& -u^{\riso} = -\usch^{\rsch} \frac{dr}{d\rsch} = \usch \left( \frac{d \rsch}{d \riso}\right)^{-1} \nonumber\\
&=& \frac{\usch}{(1-M/2\riso)(1+M/2\riso)} \ ,
\end{eqnarray}
where $u^r$ is the radial component of the 4-velocity.  The time component of the 4-velocity, $\uiso^0$, remains unchanged by the transformation.  We find the 3-velocity $v=v^r=u^r/u^0$ from the normalization condition $u^{\alpha}u_{\alpha}=-1$, or
\begin{equation}
- \alpha^2 (\uiso^0)^2 + e^{4\phi}(\uiso^0)^2 \viso^2 = -1
\end{equation}

The above transformation is needed only when assigning fluid initial data for $r > \radj$.  Inside $\radj$ we set the initial data according to Eq.~(\ref{rho_inside_r0}).  We choose $\radj=0.6M$, which is outside the horizon, to avoid the initial coordinate singularity of $u^0$ at the horizon.  There is no singularity during the evolution as the adopted 1+log slicing condition is horizon penetrating.

\subsubsection{Moving Bondi solutions}
We also wish to consider the problem of adiabatic, steady-state accretion onto a BH which moves with constant velocity $\vcloud$ through a gas which is uniform at infinity.  This problem was first studied in Newtonian gravitation in ~\cite{hoyle39} and \cite{bondi44}, and is often referred to as ``Bondi-Hoyle-Lyttleton (BHL) accretion''.  The problem has also been studied numerically in Newtonian theory by \cite{hunt71,shima85}, and for BHs in full general relativity by \cite{petrich89,font98}.  For cases in which the gas is not at rest asymptotically, we construct initial data following the method described in \cite{faber07}, in which we ``boost'' the stationary Bondi solution by taking the initial density at any coordinate point to be approximately the stationary isotropic Bondi solution, and computing the initial velocity field employing a Lorentz transformation for the velocities.  Denoting the stationary Bondi flow velocity by $v^i$, the moving Bondi flow by $v^{i'}$, and the ``boost'' velocity by $\vcloud$, we set
\begin{eqnarray}
\frac{v^{x'}}{\clocal} &=& \frac{\vcloud+\frac{v^x}{\clocal}}{1+\vcloud \frac{v^x}{\clocal}}\\
\frac{v^{y'}}{\clocal} &=& \frac{\frac{v^y}{\clocal}}{\gamma_b(1+\vcloud \frac{v^x}{\clocal})}\\
\frac{v^{z'}}{\clocal} &=&  \frac{\frac{v^z}{\clocal}}{\gamma_b(1+\vcloud \frac{v^x}{\clocal})} \ ,
\end{eqnarray}
where $\gamma_b \equiv (1-\vcloud^2)^{-1/2}$ and $c_l \equiv \alpha/\psi^2$.  Here $\alpha$ is the lapse, and $\psi \equiv e^{\phi}$, with $\phi$ defined in Eq.~(\ref{phidef}).
For any hydrodynamic quantity $g(x,y,z)$, we compute
\begin{equation}
g(x',y',z') = g(\gamma_b x,y,z) \ ,
\end{equation}
which accounts for the coordinate transformation between the $x^{\alpha}$ and $x^{\alpha'}$ frame on the $t'=0$ timeslice.
The above method of computing initial data is only strictly valid asymptotically. This is because we use the Bowen-York initial data for the moving BH metric, which are not obtained by boosting the stationary BH metric. However, we find that deviations propagate away quickly, leaving a stationary flow.  In Sec.~\ref{single_punc_tests} we describe a simulation of Bondi accretion which we perform in a moving reference frame.  In this case, we use the same technique described above to construct hydrodynamic initial data.

\subsubsection{Effective adiabatic index}
\label{effective_index}

\begin{figure}
\vspace{-4mm}
\begin{center}
  \epsfxsize=3.5in
  \leavevmode
  \epsffile{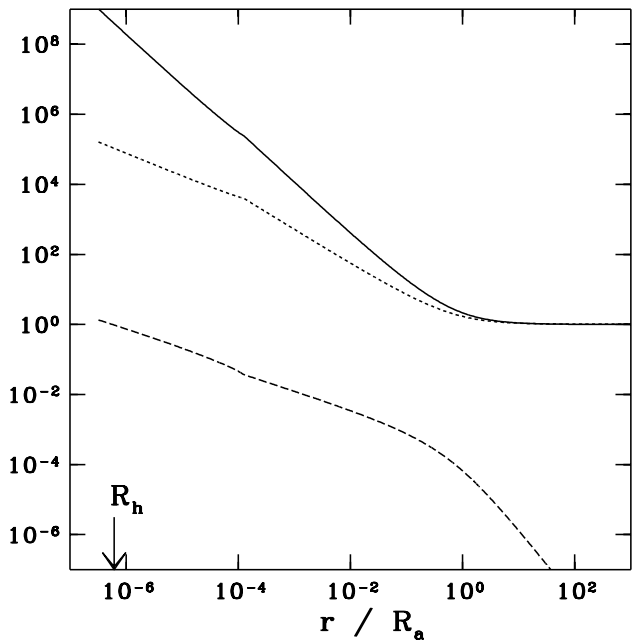}
\end{center}
\caption{Semi-analytic profiles of density, temperature and fluid 4-velocity for spherical, adiabatic Bondi accretion onto a Schwarzschild BH.  The solid line gives $\rho_0 / \rho_{0,\infty}$, the dotted line shows $T/T_{\infty}$, and the dashed line shows the fluid 4-velocity in Schwarzschild coordinates, $\usch$.  The adiabatic index is given by $\Gamma=\Gamma^*$ according to Eq.~(\ref{hybridEOS}).  The asymptotic temperature is $T_{\infty}=10^6 K$.  The horizon is labeled $R_h$.}
\label{fig:semi_analytic_plot}
\end{figure}

In some cases, we consider a gas cloud which the electrons have nonrelativistic temperatures at infinity, but achieve relativistic temperatures near the binary.  In this case we follow \cite{shapiro73} and define an ``effective adiabatic index'' $\Gamma^*$ according to 
\begin{equation}
\label{hybridEOS}
 \Gamma^*= \left\{\begin{array}{l l}
    5/3& \ \ \ k T / m_e c^2 \le 2/3 \\
    13/9 & \ \ \ k T / m_e c^2 > 2/3 \ ,
  \end{array}\right.
\end{equation}
Thereby replacing the actual continuous transition by a simpler discrete transition.
This transition occurs at a Schwarzschild radius~\cite{shapiro_book_83}
\begin{equation}
\frac{r_*}{M} \approx \frac{9}{40} \frac{m_p }{m_e} \approx 400 \ .
\end{equation}
We solve for $r_*$ using Eqs.~(\ref{continuity}), (\ref{euler}), (\ref{ht_relation}), and the fact that $kT(r_*)=(2/3) m_e c^2$.  Using the continuity of $\rho_0$ and $P$ at $r=r_*$, we obtain the full equilibrium flow solution.  In practice, $r_*$ is outside the outer boundary of the computational grid in our simulations, so we still implement a constant $\Gamma$.  However, taking into account this transition alters our initial data significantly, since the outer $\Gamma = 5/3$ region drives $a \lesssim u$ up to $r=r_*$, increasing the gas temperature near the black hole when compared to gas in which $\Gamma=13/9$ everywhere.  Analytic profiles of density, temperature, and fluid velocity for this equation of state are plotted in Fig.~\ref{fig:semi_analytic_plot}.

\subsection{Relativistic Bondi test}
\label{single_punc_tests}
We test our code's ability to accurately simulate hydrodynamic accretion onto a moving puncture.  This is particularly important in an AMR code in which matter crosses moving refinement zone boundaries.  To test our code, we simulated spherical Bondi accretion in a frame in which both the BH and gas cloud are moving with the same velocity.  We consider a BH moving with velocity $V_{BH} \equiv P^x / M = 0.37$, initially situated at the origin.  Here $P^x$ is the momentum of the BH as measured by a stationary coordinate observer at infinity.  We consider a gas with adiabatic index $\Gamma=13/9$ with asymptotic sound speed $\sscloud^2=0.022$ also moving with velocity $v=V_{BH}$ at infinity.  With this choice the transonic radius is given by areal radius $\Racc = 45.5 M$ in the comoving frame.  We evolve for a duration $t = 713.6 M$, which is approximately equal to two free-fall times at the accretion radius.  By this time, the BH has moved to a coordinate location of $x=247 M$.  

\begin{figure}
  \vspace{-4mm}
  \begin{center}
    \epsfxsize=3.5in
    \leavevmode
    \epsffile{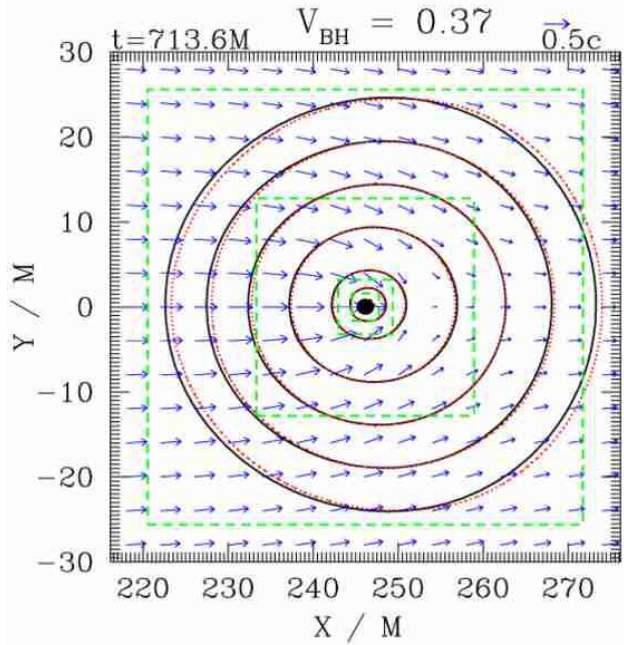}
  \end{center}
  \caption{Snapshots showing spherical Bondi accretion onto a single BH simulated in a frame
    in which the BH and gas move with velocity $V_{BH}= 0.37$.  Contours of
    constant density $\rho_0$ (solid black lines) and constant $d \rho_0 / d \tau$
    (dotted red lines) are shown.  Their overlap indicates agreement with the analytic Bondi
    solution.  The adiabatic index is set to $\Gamma=13/9$ and $a_{\infty}=0.148$.  The green dashed lines show the boundaries of the innermost AMR refinement levels.  Arrows denote velocity vectors.}
  \label{fig:moving_punc_test}
\end{figure}

To assess the agreement with the analytic solution, we compare invariant quantities \cite{faber07}.  Two such invariants are the fluid rest-mass density, $\rho_0$, and the the rate of change of the fluid rest-mass density, as measured by a comoving observer, $d \rho_0 / d \tau$.  We choose a set of 6 areal radii in the comoving frame. At each radius, we compute $\rho_0$ and $d\rho_0/d\tau$ analytically. By construction, contours of $\rho_0$ and $d\rho_0/d\tau$ are spherical and coincide in this frame.  However, because $\rho_0$ and $d \rho_0 / d \tau$ are both coordinate-independent quantities, their contours must continue to coincide in {\it any} coordinate frame.  We use this fact to check that our numerical simulation, performed in a frame in which the BH and gas are moving, matches the analytic solution (see Fig.~\ref{fig:moving_punc_test}).

\subsection{Relativistic Bondi-Hoyle-Lyttleton test}
\begin{figure}
\vspace{-4mm}
\begin{center}
  \epsfxsize=3.5in
  \leavevmode
  \epsffile{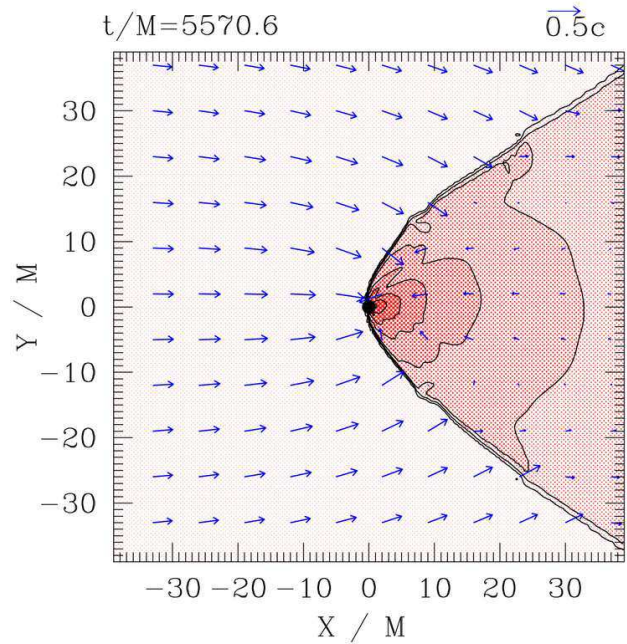}
  \end{center}
\caption{Snapshots of density contours for the BHL accretion code test.  The adiabatic index is $\Gamma=4/3$, the sound speed is  $\sscloud = 0.1$.  Density contours are chosen at $\rho_0=\rho_{0,\infty}10^{0.25 j} \ (j=1,2,....12)$.  Contours of highest density are very near the BH.  Arrows represent velocity vectors.}
\label{fig:petrich_test}
\end{figure}

We will simulate cases of accretion onto binaries in which the BHBH center of mass moves relative to the asymptotic gas cloud.  Here we verify that we can accurately simulate ``BHL'' accretion of a single puncture BH moving at constant velocity through a gas cloud.  While there is no analytic solution for this problem, we compare our findings with results of previous work.  We consider a test case in which a single BH puncture is placed in a cloud with asymptotic sound speed $a_{\infty}=0.1$ and $\Gamma = 4/3$, with the puncture moving supersonically with speed $\vcloud=0.25$ relative to the gas cloud.  We perform our simulation in a frame in which the puncture is at rest and the asymptotic fluid velocity is set to $\vcloud$.  Figure~\ref{fig:petrich_test} shows a snapshot of the stationary-state density and velocity profile from our simulation.  We measure $\dot{M}$ and compare with the results of \cite{petrich89}.  In order to facilitate this comparison, we define a canonical unit of rest-mass accretion flux to be
\beq
\dot{M}_{can} = \frac{4 \pi \lambda M^2\rho_{0,\infty}}{(\vcloud^2+\sscloud^2)^{3/2}} \ ,
\eeq
We find $\dot{M} / \dot{M}_{can} = 2.6$, which is slightly smaller than the value 3.0 reported in~\cite{petrich89}.  The most likely sources of the small discrepancy are the outer boundary condition (we use the stationary spherical Bondi solution with a Lorentz boost, whereas \cite{petrich89} use constant asymptotic values), and the fact that \cite{petrich89} impose an approximate inner boundary condition outside the horizon.  Other differences between these two codes are that we use a 3+1 code with AMR, whereas \cite{petrich89} employ an axisymmetric code.  Also, our outer boundary is placed at $r_{max}/M=820$ for this test, whereas \cite{petrich89} place the outer boundary at $r_{max}/M=140$.

\section{Accretion onto a BHBH binary}
\label{sec:binary_bh}
 
\begin{table*}
\caption{Parameters for BHBH simulations}

\begin{tabular}{lllccccccc}
run \ \ \ &run type \ \ \  &case &$\vcloud$ \ \ \ &$\vcloud / \sscloud$ \ \ &\ \ \ $\Gamma$ \ \ \ \ &$T_{\infty} (K)$ \ \ &$\sscloud$ \ \ & ${}^{\dagger}\Racc/M$ \ \ \ & ${}^{\dagger \dagger}d/M$\\
\hline
\hline
PA1&prototype&Bondi&$0.0$&$0.0$&$13/9$&$8.73 \times 10^{10} $&$0.148$&22.7&$40,20,14,10$\\
PA2&&&&&$4/3$&$9.62  \times 10^{10} $&&&\\
PA3&&&&&$5/3$&$7.44  \times 10^{10} $&&&\\
\hline
PB1&&BHL \ \ \ &$0.1$&0.7&13/9&$8.73  \times 10^{10} $&$0.148$&15.6&$40,10$\\
PB2&&&&&4/3&$9.62  \times 10^{10} $&&&\\
PB3&&&&&5/3&$7.44  \times 10^{10} $&&&\\
\hline
PC1&&BHL \ \ \ &$0.4$&2.7&13/9&$8.73 \times 10^{10}  $&$0.148$&2.74&$40,10$\\
PC2&&&&&4/3&$9.62  \times 10^{10} $&&&\\
PC3&&&&&5/3&$7.44  \times 10^{10} $&&&\\
\hline
\hline
RA1&realistic&Bondi&$0.0$&0.0&$\Gamma^*$&$1 \times 10^6$&$5.53 \times 10^{-4}$&$3.27 \times 10^6$&$10$\\
RA2&&&&&$5/3$&&&&\\
\hline
\hline
\end{tabular}
\vskip 12pt
\begin{minipage}{12cm}
  \raggedright
  ${}^{\dagger} \Racc=(M/2)/(\sscloud^2 + \vcloud^2)^{3/2}$ is the accretion radius for a single BH of mass $M/2$.   
  
  ${}^{\dagger \dagger} d=$ initial binary separation; simulations beginning at $10 M$ proceed to merger.
\end{minipage}    
\label{table:combined}
\end{table*}

\begin{table*}
\caption{Parameters for BHBH simulations}

\begin{tabular}{ccccc|cccc}
\hline
run \ \ &\multicolumn{4}{c}{flow characteristics${}^{\S}$} & \multicolumn{4}{c}{emission characteristics${}^{\S}{}^{\S}$}\\
\hline
&$n_{max}/n_{\infty}$ \ \ &$T_{max}/T_{\infty}$  &
$K_{max}/K_{\infty}$  &${}^{\dagger}\dot{M}_{max}/\dot{M}_a$ \ \ &${}^{\dagger \dagger} L_{ff}^{max}/L_{35} $  & $h \nu_{ff}^{max} $ \ \ & ${}^{\dagger \dagger} L_{syn}^{max}/L_{35} $ \  & $h \nu_{syn}^{max} $  \\
\hline
\hline
PA1 \ \ &631&18.3&1.2&3.7&&&&\\
PA2 \ \ &1743&14.3&1.5&3.8&&&&\\
PA3 \ \ &164&30.1&1.6&2.6&&&&\\
PB1 \ \ &593&18.9&1.2&4.0&&&&\\
PB2 \ \ &1638&15.2&1.7&3.3&\multicolumn{4}{c}{Not Physically Relevant}\\
PB3 \ \ &153&30.8&1.9&2.5&&&&\\
PC1 \ \ &99.2&28.3&6.0&0.7&&&&\\
PC2 \ \ &158&32.2&14.0&0.7&&&&\\
PC3 \ \ &46.6&37.7&6.5&0.8&&&&\\
\hline
\hline
RA1&$1.4\times10^{10}$&$1.8 \times 10^6$&4.7&3.9&$300$&150 \mbox{MeV}&$3 \times 10^8 \beta_1^{-1}$&$80 / (1+z) \ n_1^{1/2} \ T_6^{-3/4} \beta_1^{-1/2}\ \mbox{eV}$\\
RA2&$3.63 \times 10^9$&$2.7 \times 10^6$&13.0&3.3&$400$&230 \mbox{MeV}&$4 \times 10^8 \beta_1^{-1}$&$100 / (1+z) \ n_1^{1/2} \ T_6^{-3/4} \beta_1^{-1/2}\ \mbox{eV}$\\
\hline
\hline
\end{tabular}
\vskip 12pt
\begin{minipage}{14cm}
  \raggedright
  ${}^{\dagger}\dot{M}_a c^2= 4.6 \times 10^{40} \ \lambda_{5/3} \ n_1 \
  T_6^{-3/2} \  M_6^2  \ \mbox{erg s}^{-1}$ is the accretion rate onto a single BH of mass $M/2$ undergoing stationary, spherical Bondi accretion.

   ${}^{\dagger \dagger} L_{35}=10^{35} \ n_1^2 \ T_6^{-3}
   M_6^3 \ \mbox{erg s}^{-1} $
  
   ${}^{\S}$ ``max'' label refers to the characteristic value at the moment of maximum luminosity.

   ${}^{\S \S}$ ``max'' label refers to the maximum values of the luminosities during the merger and the characteristic frequencies at these times.
\end{minipage}    
\label{table:table2}
\end{table*}

As discussed in Sec.~\ref{sec:computational_challenge}, the transonic radius for Bondi accretion with a realistic asymptotic temperature of $10^6 K$ is $\Racc \sim 10^6 M$.  It is beyond the capability of current $3+1$ GR simulations to evolve the binary from initial separation $d > \Racc$ all the way to merger, while resolving a BH horizon of size $\sim M$.  The range of length scales is too large and the total coalescence time too long for such a task.  We approach this issue by performing two types of simulations.  We perform simulations of binaries merging in ``realistic'' gas clouds with asymptotic temperature $T_{\infty}=10^6K$, following only the last phase of the merger in which the binary separation satisfies $d \ll \Racc$.  Our focus here is on identifying observable electromagnetic signals generated by the time-dependent shock heating caused by the binary motion.  We also perform ``prototype'' calculations with artificially high temperatures and sound speeds in order to study how the accretion flow changes as the binary transitions between the following regimes during the inspiral:
\begin{enumerate}
\item ``widely separated regime'', in which $d > \Racc$
\item ``moderately separated regime'', in which $d \approx \Racc$
\item  ``closely separated regime'', in which $d < \Racc$
\end{enumerate}
We set the asymptotic temperature $T_{\infty} \sim 10^{11}K$ in the ``prototype'' calculations to decrease the accretion radius $\Racc$ to a value closer to $d$ so that we can explore all three regimes by combining the results of several numerical simulations.
All of our BHBH simulations are summarized in Table~\ref{table:combined}. Important results are summarized in Table~\ref{table:table2}.  The initial BHs in the binary are all equal-mass and nonspinning. 

\subsection{Scaling}
For a given asymptotic gas temperature $T_{\infty}$ or sound speed $a_{\infty}$, our solution for binary Bondi flow can be scaled to arbitrary density $n_{\infty}\equiv \rho_{0,\infty}/m_B$ (neglecting self-gravity of the gas) and black hole mass $M$.  Hence a single simulation with an arbitrary $n_{\infty}$ and $M$ suffices to determine the solution for any other $n_{\infty}$ or $M$.  Thus, for example, the accretion rate is proportional to $n_{\infty}$ and $M^2$ (see, e.g. Eq.~(\ref{Mdot_formula})), while the 4-velocity $u^{\alpha}$ as a function of coordinates $x^{\alpha}/M$ are independent of $n_{\infty}$ and $M$.  If the asymptotic sound speed is sufficiently low that $a_{\infty} \ll 1$, then the solution can also be scaled to arbitrary $a_{\infty}$ or $T_{\infty}$.  However, once $a_{\infty}$ approaches the speed of light and the transonic radius approaches the horizon, scaling with $a_{\infty}$ or $T_{\infty}$ breaks down.  This behavior is already evident from the relativistic Bondi solution for accretion onto a single BH.

The emergent electromagnetic luminosity and radiation spectrum also exhibit simple scaling.  The form of the scaling relations depend on the temperature and density dependence of the adopted emissivities and will be described later in Sec.~\ref{sec:real_bondi} when we treat realistic asymptotic temperatures and sound speeds.

\subsection{Prototype Cases}
\subsubsection{Binary Bondi Accretion}

\begin{figure}
\begin{center}
\epsfxsize=3.2in
\leavevmode
\vspace{6mm}
\epsffile{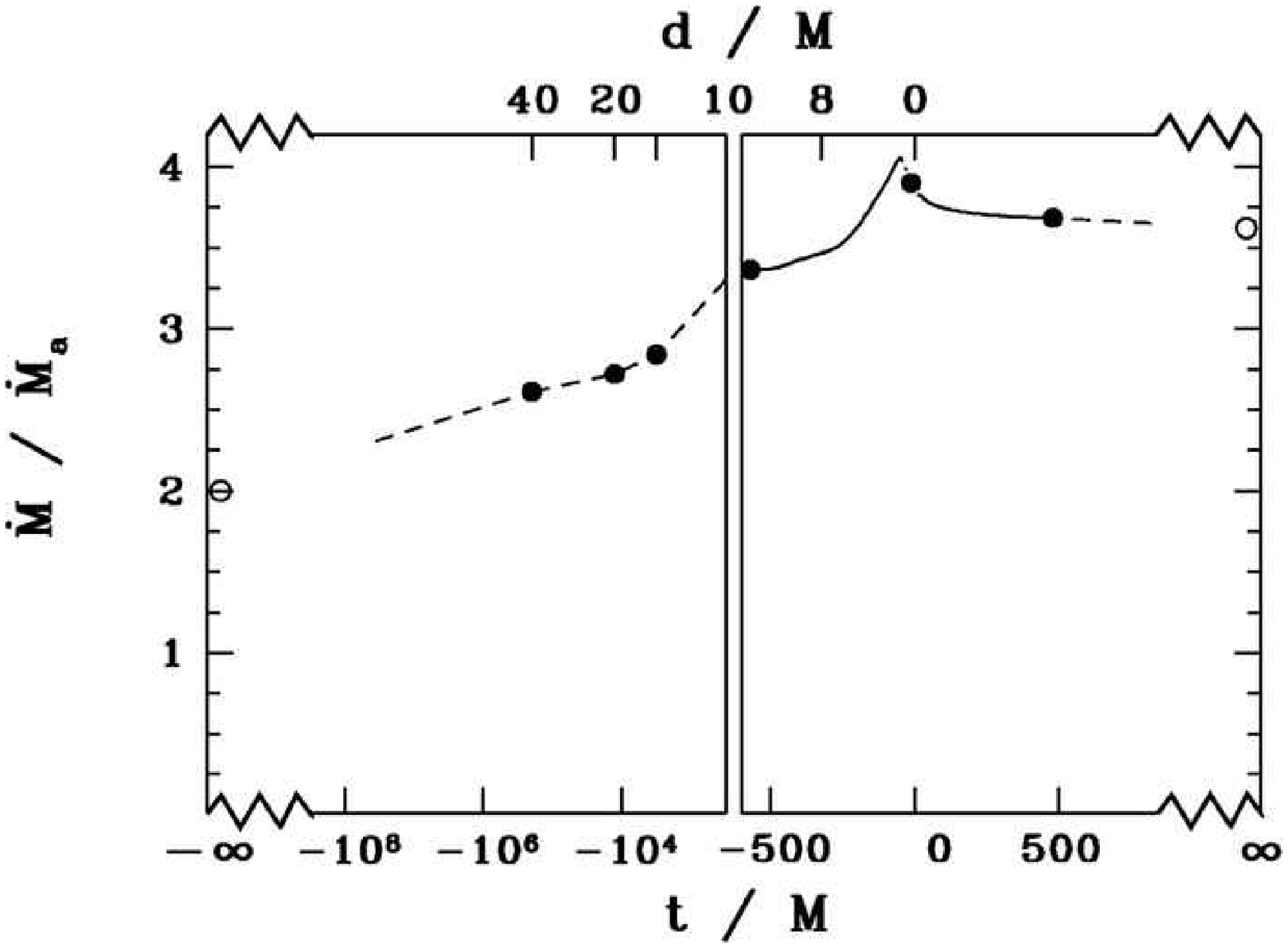}
\epsfxsize=3.2in
\leavevmode
\epsffile{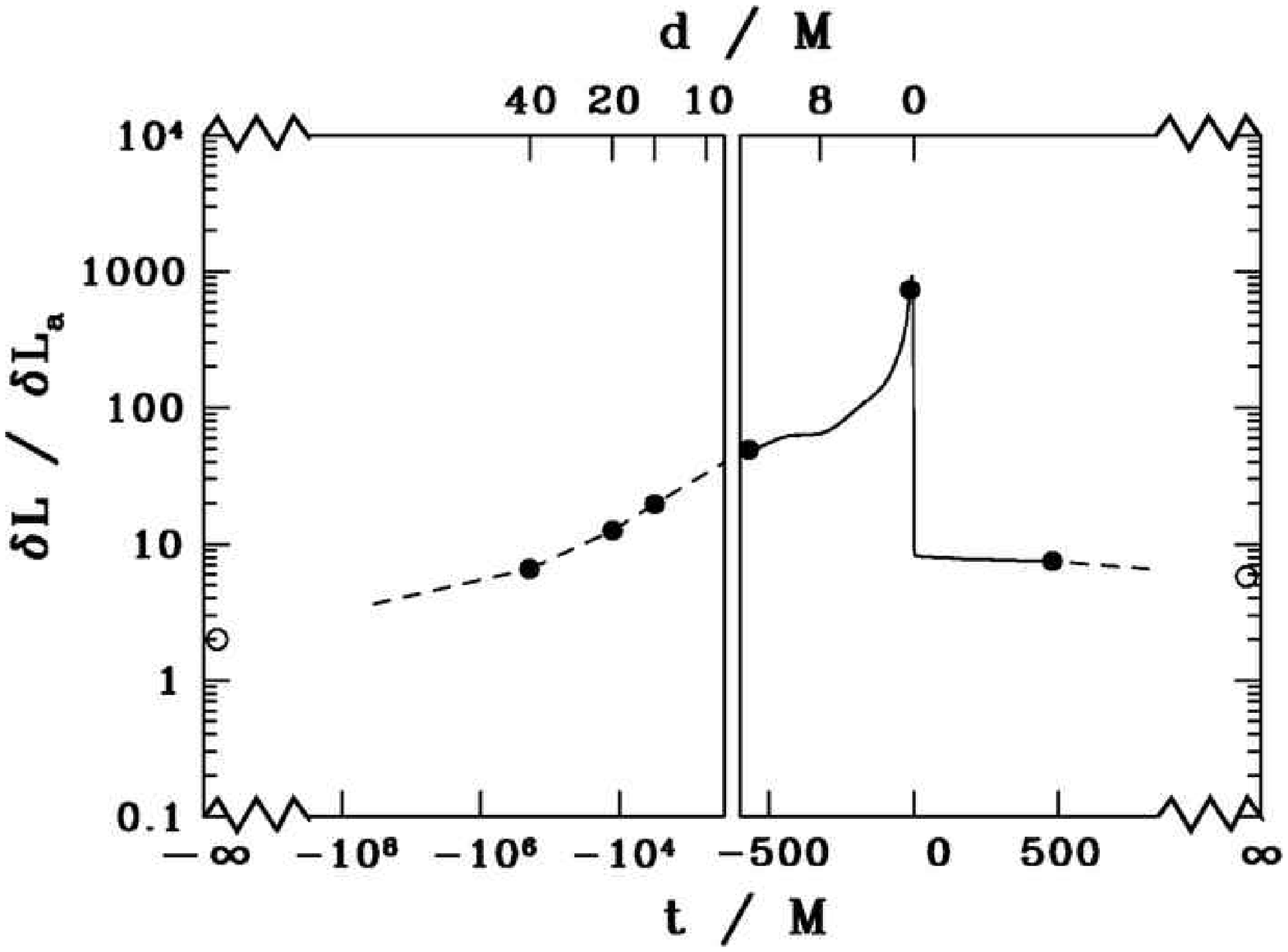}
\end{center}
\caption{Time evolution of $\dot{M}$ and $\delta L$ for binary Bondi accretion for the prototype case with $\Gamma = 13/9$.  Time $t / M$ is measured relative to the time at which the merger
  occurs.  $\dot{M}_a$ and  $\delta L_a$  are the accretion rate and luminosity enhancement over the background for a single isolated
  black hole with mass equal to the initial ADM mass of the binary.  The left-hand box shows values from ``snapshots'' in regimes 1 and 2.   The right-hand box shows results for the final inspiral and merger from an initial separation of $d=10M$ (regime 3).  Solid lines denote numerical data, dashed lines denote extrapolated data.  Solid dots correspond to profile plots highlighted in Fig.~\ref{fig:snap_bondi_13o9}, while open circles show expected values at $t=\pm \infty$.  The asymptotic sound speed is set at  $\sscloud=0.148$, for which $\Racc /M = 22.7$.}
\label{fig:tplot_proto_13o9}
\end{figure}
\begin{figure}
\begin{center}
\epsfxsize=3.2in
\leavevmode
\vspace{6mm}
\epsffile{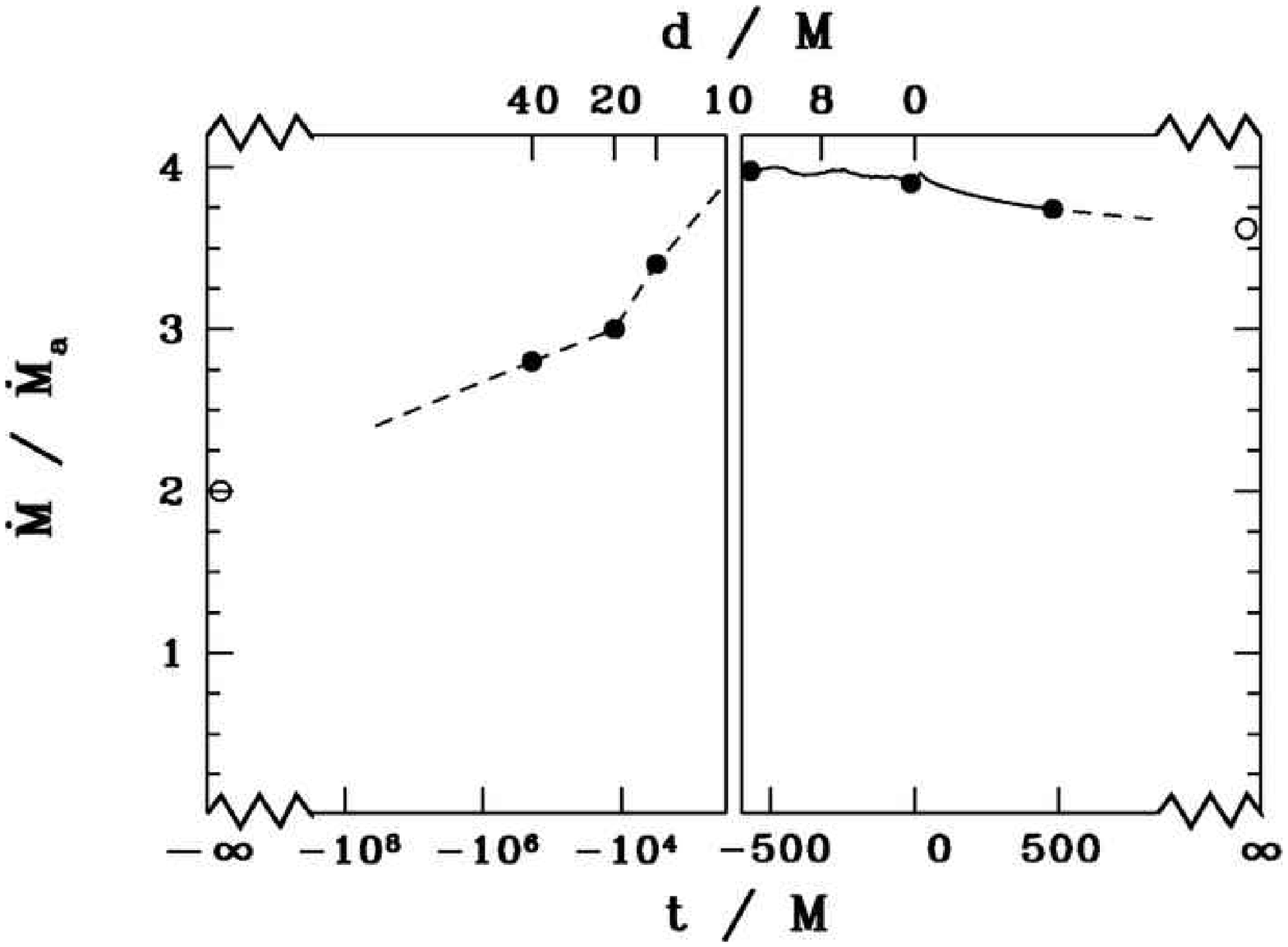}
\epsfxsize=3.2in
\leavevmode
\epsffile{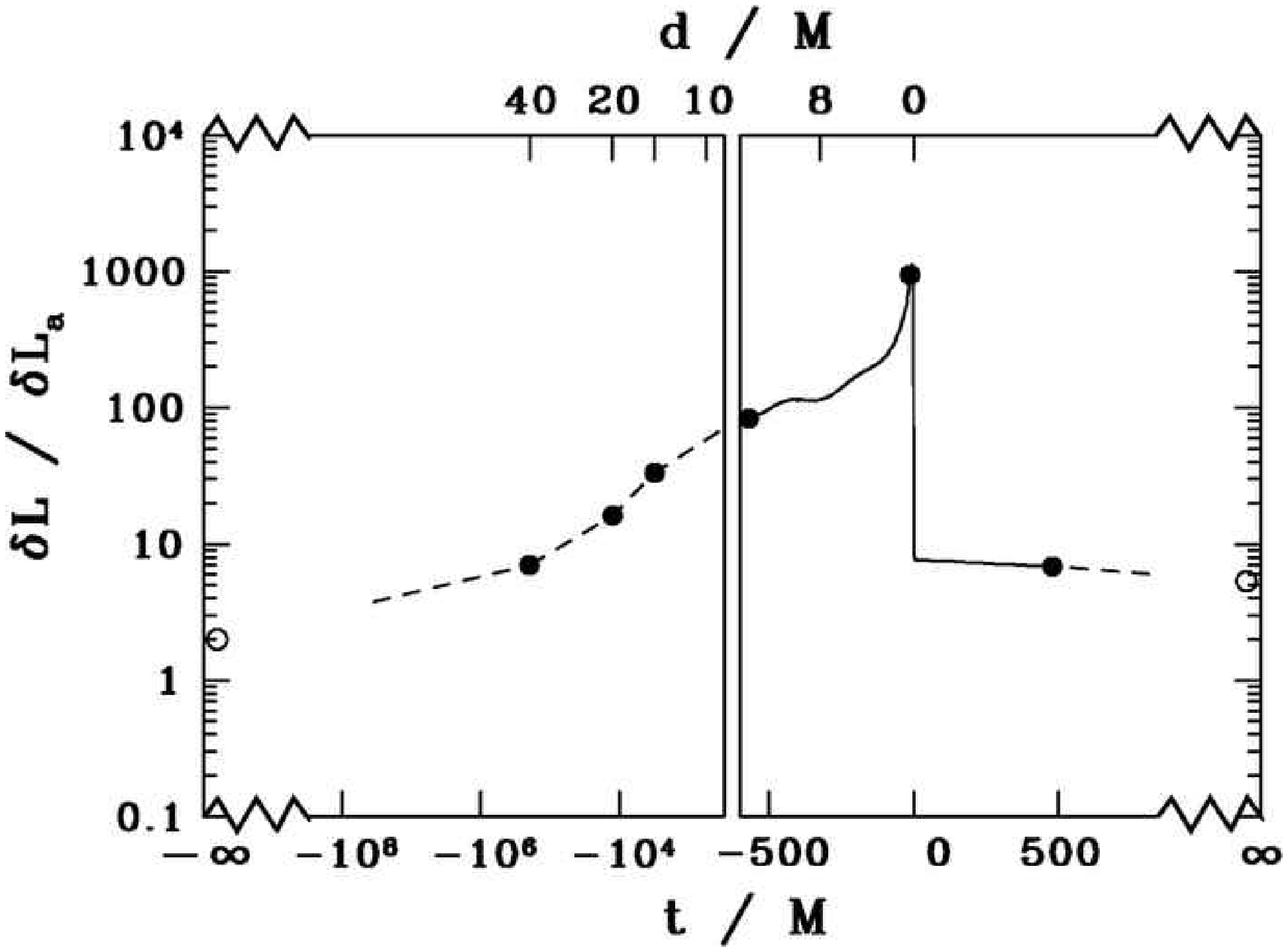}
\end{center}
\caption{Same as Fig.~\ref{fig:tplot_proto_13o9}, but for $\Gamma=4/3$.}
\label{fig:tplot_proto_4o3}
\end{figure}
\begin{figure}
\begin{center}
\epsfxsize=3.2in
\leavevmode
\vspace{6mm}
\epsffile{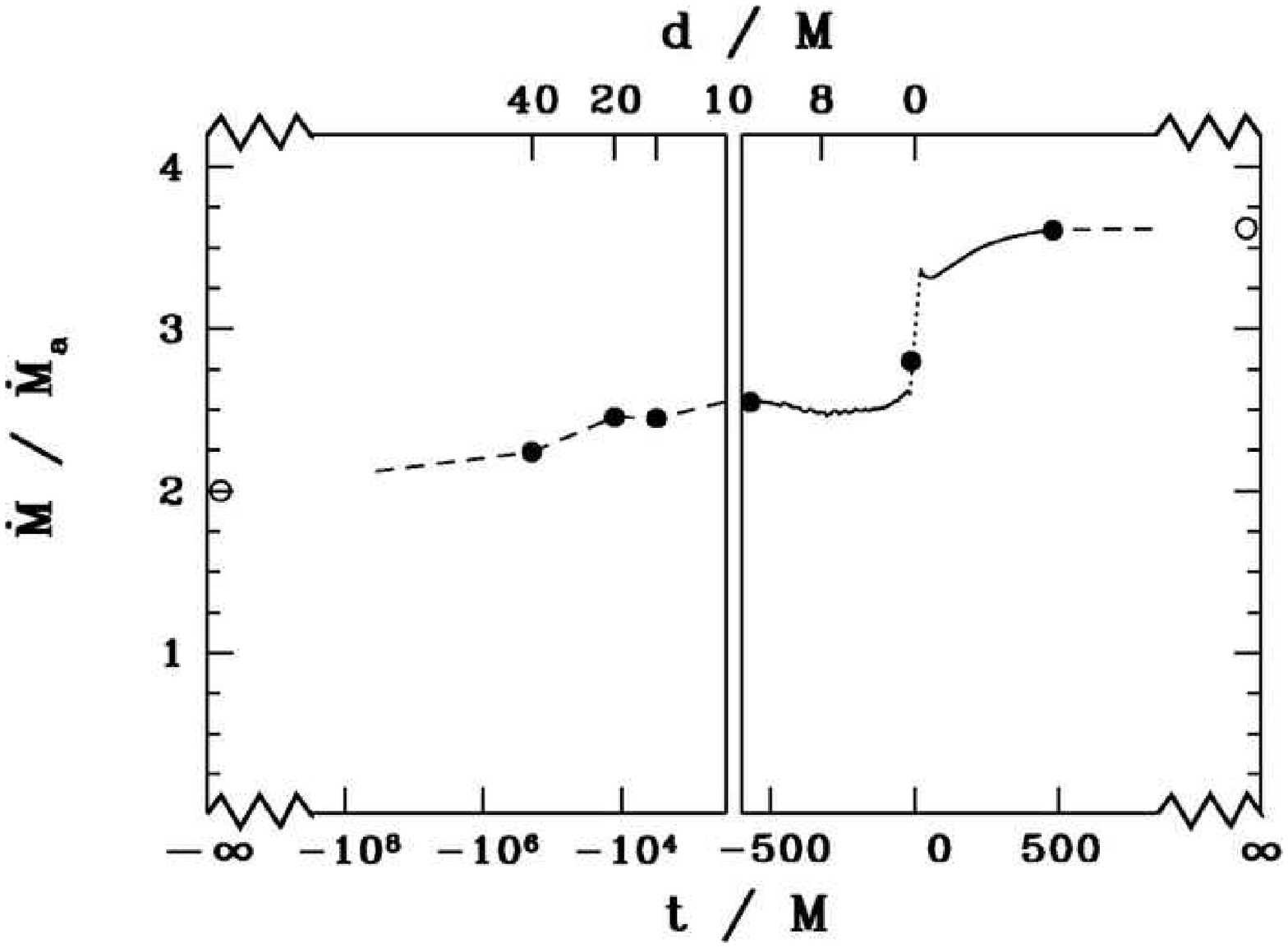}
\epsfxsize=3.2in
\leavevmode
\epsffile{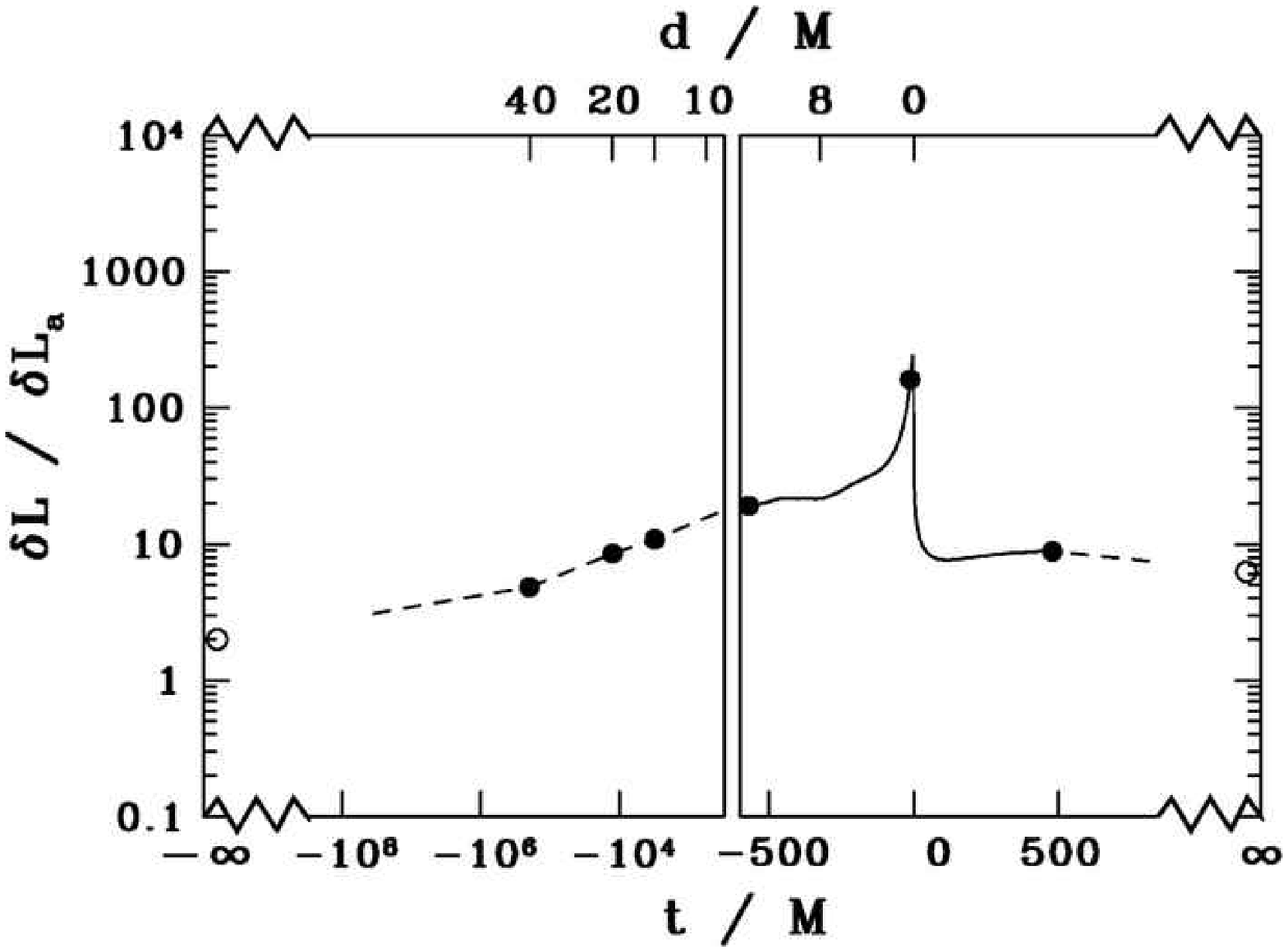}
\end{center}
\caption{Same as Fig.~\ref{fig:tplot_proto_13o9}, but for $\Gamma=5/3$.}
\label{fig:tplot_proto_5o3}
\end{figure}

\begin{figure*}
\begin{center}
  \epsfxsize=6.0in
  \leavevmode
  \epsffile{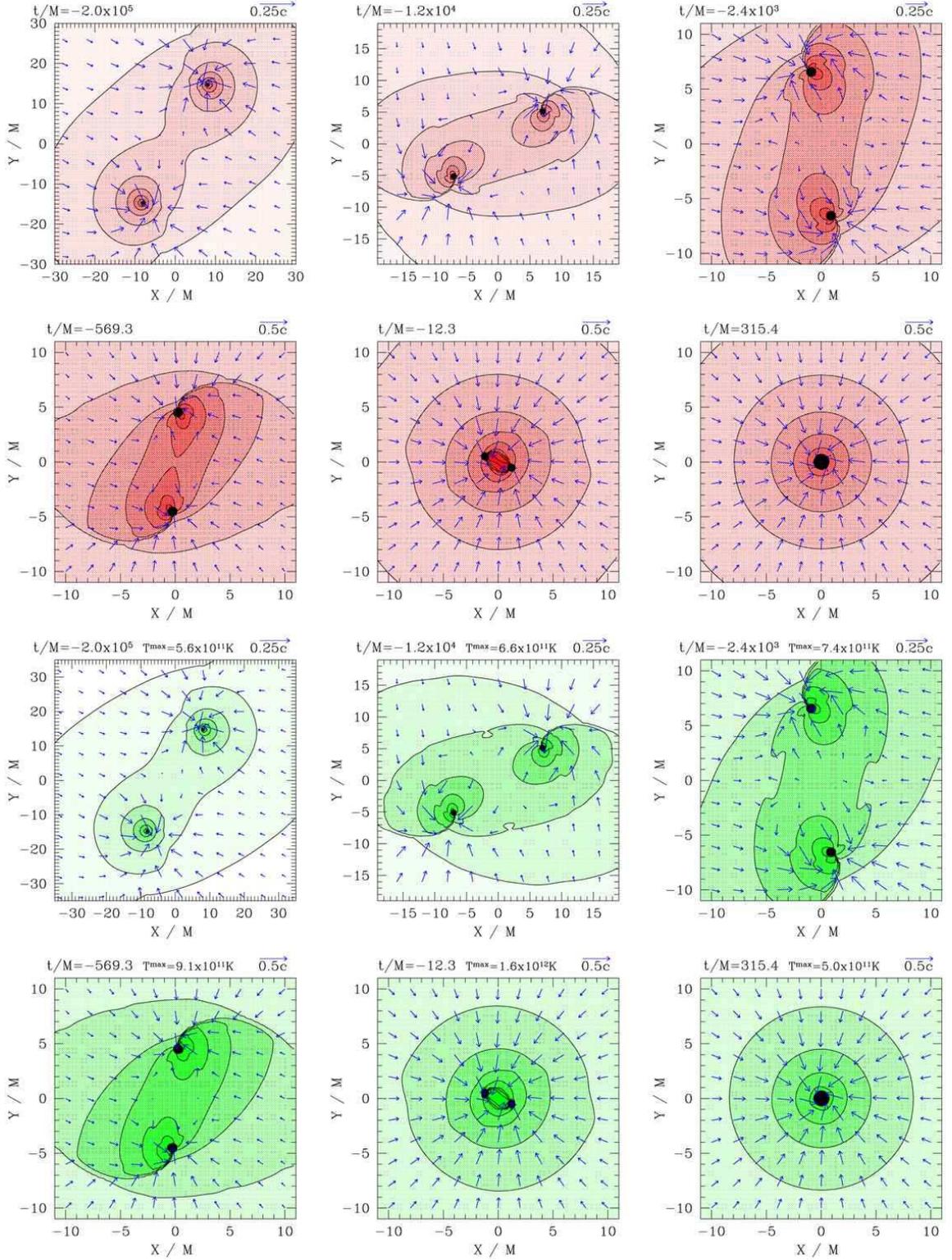}
\end{center}
\caption{Snapshots of rest-mass density $\rho_0$ and temperature $T$ contours in the orbital plane for
the binary Bondi ``prototype'' case with $\Gamma=13/9$.  First and second
rows show density contours and velocity profiles, third and fourth rows show snapshots of temperature
 $T$.  Density contours are plotted at $\rho_0=\rho_{0,\infty}10^{0.25 j}\ \ (j=1,2,..
 ..,12)$.  Temperature contours are plotted at $T=10^{11+0.125 j} K \ \  (j=1,2,....,12)$. 
 Contours of highest density and temperature are near the BHs.  Arrows denote velocity vectors.  The AH
 interior is marked by a filled black circle. }
\label{fig:snap_bondi_13o9}
\end{figure*}

\begin{figure*}
\begin{center}
  \epsfxsize=6.0in
  \leavevmode
  \epsffile{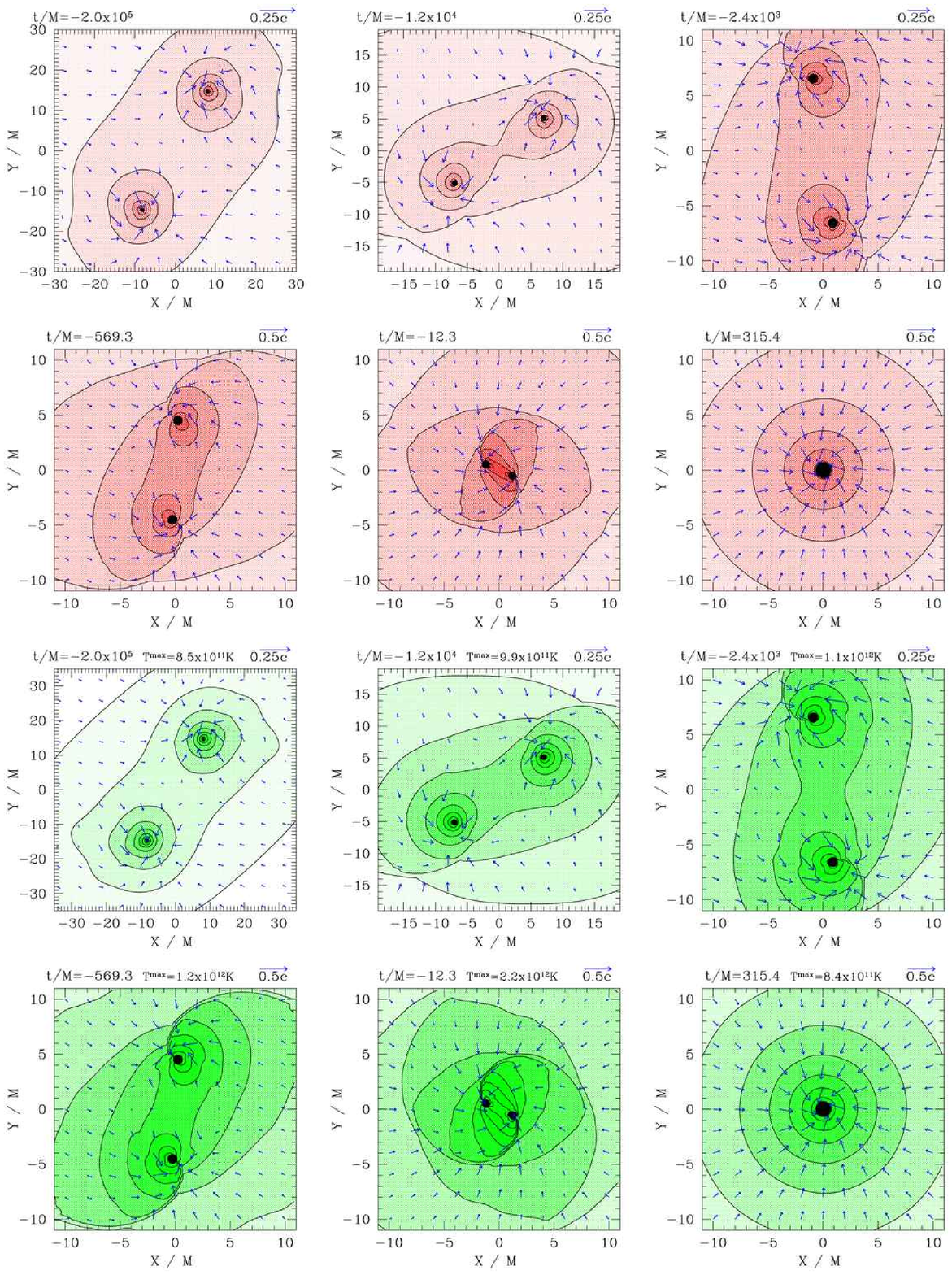}
\end{center}
\caption{Same as Fig.~\ref{fig:snap_bondi_13o9}, but for $\Gamma=5/3$.}
\label{fig:snap_bondi_5o3}
\end{figure*}

\begin{figure*}
\begin{center}
  \epsfxsize=6.0in
  \leavevmode
  \epsffile{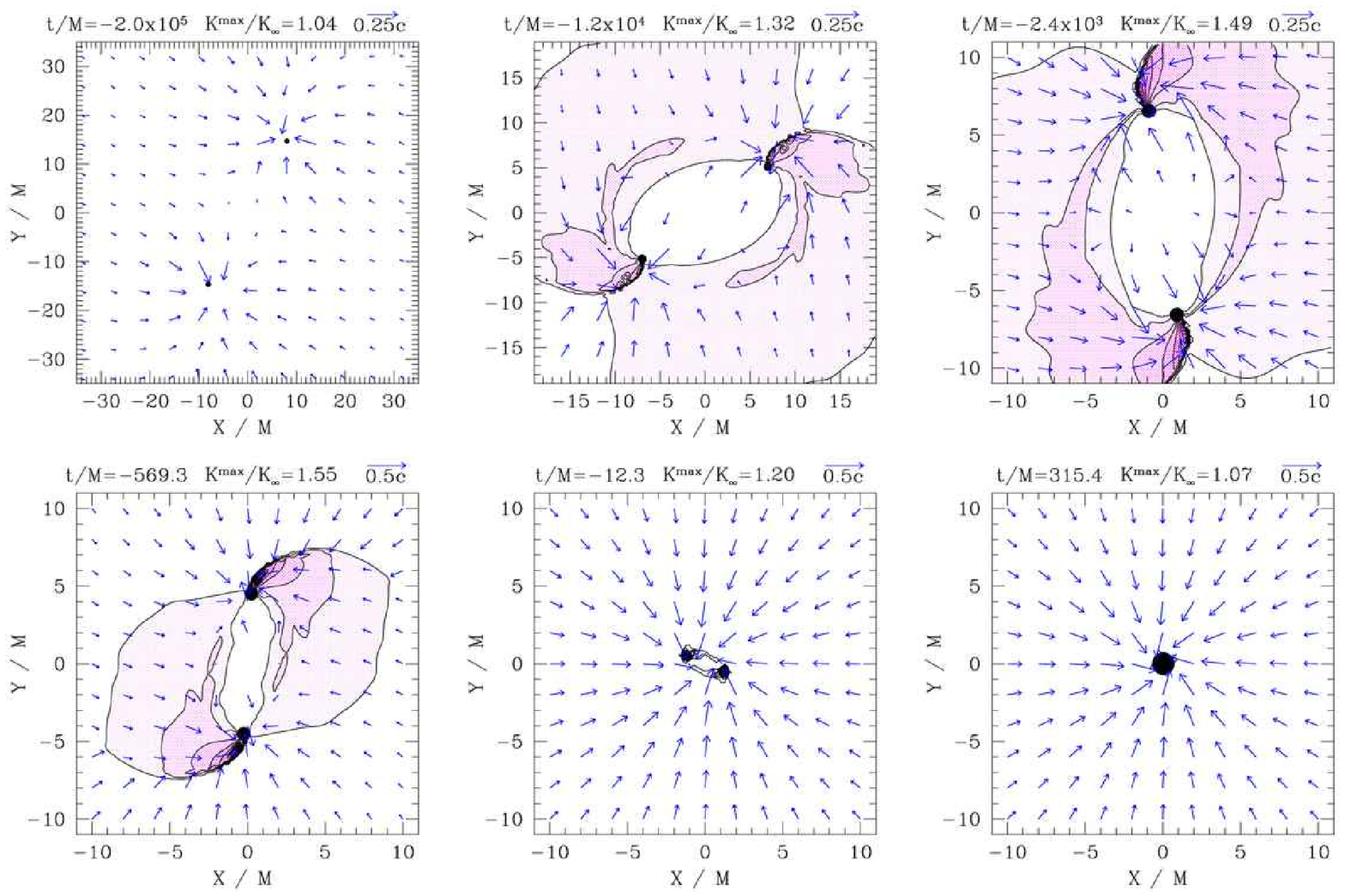}
\end{center}
\caption{Snapshots of  $K / K_{\infty}$ contours in the orbital plane for 
the binary Bondi ``prototype'' case with $\Gamma=13/9$.  Contours are drawn for $K/K_{\infty}=1+0.05 j \ \  (j=1,2,..
..,12)$.  Arrows denote velocity vectors.  The AH
 interior is marked by a filled black circle. }
\label{fig:K_contour}
\end{figure*}

The ``prototype'' calculations provide valuable qualitative insight into the evolution of the accretion flow as the binary separation decreases, passing through the three regimes identified above.  For the case of ``binary Bondi accretion'', in which the gas at infinity is at rest relative to the binary center of mass, we obtain a sequence of ``snapshots'' for different binary separations.  Each snapshot is generated by evolving the binary long enough to allow the gas to relax to a quasistationary flow, but not long enough for the separation $d$ to change significantly.  By sandwiching together snapshots we can trace the evolution in accretion rate $\dot{M}$ and luminosity $L$ as the binary transitions from regime 1 to regime 2.  The final transition from regime 2 to regime 3 is captured by a single simulation which follows the binary through inspiral and merger.  We perform these simulations for cases PA1 ($\Gamma=13/9$), PA2 ($\Gamma=4/3$), and PA3 ($\Gamma=5/3$) in order to study the effect of the EOS on the evolution (see Table~\ref{table:combined}).  Numerical results for all cases are given in Table~\ref{table:table2}.  The time variations in the accretion rate and total luminosity for these cases are shown in Figs.~\ref{fig:tplot_proto_13o9}--\ref{fig:tplot_proto_5o3}.

The evolution in $\dot{M}$ can be understood as follows.  Define $\dot{M}_a$ to be the accretion rate onto a single, isolated BH of mass $M/2$.  Then it follows that the total accretion rate onto two infinitely separated BHs of mass $M/2$ will be given by $\dot{M} = 2 \dot{M}_a$. However, late in the inspiral, when the separation satisfies $d \ll \Racc$, the binary can be treated as a single gravitating object.  From Eq.~(\ref{Mdot_formula}) we see the accretion rate is proportional to $M^2$, so we expect that the accretion rate will approach $\dot{M} = 4 \dot{M}_a$.  During the final stage of the inspiral, mass-energy is radiated away in the form of gravitational waves.  Thus we expect that the final post-merger accretion rate will be approximately given by $\dot{M} = 4 \dot{M}_a (1 - \delta M / M)^2$.  In our simulations, we find $\delta M / M \approx 0.05$, consistent with the values reported in \cite{tichy08}.  We observe the expected behavior for both our PA1 (see Fig.~\ref{fig:tplot_proto_13o9}) and PA2 (see Fig.~\ref{fig:tplot_proto_4o3}) runs.  For PA3, the accretion rate never reaches $4 \dot{M}_a$ before the merger.  We attribute this to the fact that for $\Gamma=5/3$, gas is more efficiently heated, allowing $P \sim \rho_0 v^2$ and $a \sim v$, even in the absence of shocks.  When shocks do form, the flow is much more easily disrupted as the kinetic energy of the flow does not dominate the thermal energy, contrary to cases PA1 and PA2.  Thus, some matter is swept away from the vicinity of the binary, causing a lowering of the accretion rate (see Fig.~\ref{fig:tplot_proto_5o3}).  We note, however, that after the merger when the shocks have a chance to dissipate, the accretion rate does settle to its expected value, taking mass loss into account.  

We also plot the luminosity enhancement due to bremsstrahlung and synchrotron emission in Figs.~\ref{fig:tplot_proto_13o9}--\ref{fig:tplot_proto_5o3}.  
Because the high-temperature homogeneous background gas in our prototype simulations has an intrinsic, nonnegligible emissivity, we subtract it from the total luminosity measured.  We define $\delta L \equiv L - L_{bg}$, where $L_{bg}$ is the background luminosity which would be present in our computational domain for a homogeneous gas cloud of density $\rho_{0,\infty}$ and temperature $T_{\infty}$, with no BH present.  We normalize $\delta L$ by $\delta L_a$, which we define as the luminosity above background which would be present for a single, isolated BH of mass $M/2$.  If we take the limit in which the binary separation $d \rightarrow \infty$, we expect that $\delta L / \delta L_a \rightarrow 2$.  This limiting value is indicated by the open circle at $t/M=-\infty$  in Figs.~\ref{fig:tplot_proto_13o9}--\ref{fig:tplot_proto_5o3}.  We also calculate the expected value of $\delta L / \delta L_a$ for a single BH of mass $M-\delta M$ and plot it for reference with an open circle at $t/M=+\infty$.  We see that for each $\Gamma$ the luminosity enhancement increases by several orders of magnitude over the course of the inspiral.  While the numerical value of this variation is not physically meaningful due to the unrealistic temperatures used in these ``prototype'' calculations, this behavior provides strong qualitative evidence of a significant enhancement in luminosity that can be expected to accompany such an inspiral.  Mergers in realistic clouds, yielding realistic luminosities whose values are physically meaningful, will be treated in Sec.~\ref{sec:real_bondi}.

Figures~\ref{fig:snap_bondi_13o9} and \ref{fig:snap_bondi_5o3} show snapshots of density and temperature contours for cases PA1 and PA3. We do not show snapshots for the PA2 case because they look very similar to the PA1 case.  We can see that in the early phases of the inspiral, the accretion flow resembles two independent spherical Bondi flows.  As the separation decreases and becomes comparable to the transonic radius, the orbital velocity of each BH becomes comparable to the sound speed.  Within $\Racc$ shocks begin to form, which grow in strength until the merger.  It is the heating from these shocks which contributes to the dramatic increase in the luminosity observed.  Note that the final accretion flow near the BH is not spherically symmetric due to the spin of the merged BH.  The BH spin does not significantly change the final accretion rate predicted by Eq.~(\ref{Mdot_formula}), as this quantity is determined by gas parameters at $r \sim \Racc \gg M$, where the effect of the BH spin is negligible.  This result conforms with the findings of \cite{shapiro74}.

In order to highlight the role that shock heating plays during phases 2 and 3 of the merger, we also present contours of $K / K_{\infty}$ for our PA1 case (see Fig.~\ref{fig:K_contour}).  
Here $K \equiv P / \rho_0^{\Gamma}$ and $K_{\infty}$ is the value of $K$ at infinity.  Because $K / K_{\infty}=1$ everywhere for adiabatic flow in the absence of shocks, this quantity serves as a useful tracer for the amount of shock heating which is taking place.  The quantity $K=K(s)$, where $s$ is the gas entropy, and is constant in the absence of shocks; shock heating yields $K/K_{\infty} > 1$ (see Appendix B of \cite{etienne09}).  As expected, we see that $K / K_{\infty}$ increases steeply near the shock front.  For each snapshot, we compute the maximum value of  $K / K_{\infty}$ outside the horizon.  We find that  $K_{max} / K_{\infty}$ initially increases as the separation decreases and shocks become stronger, as expected.  This trend terminates in the very late stage of the merger, when $d \sim 5 M \sim d_{ISCO}$ \cite{baumgarte00}.  This is likely due to the fact that kinetic energy dissipated as heat is confined to a small region and is being quickly consumed by the BHs at this stage.  After the merger, the gas relaxes to laminar spherical Bondi flow and $K / K_{\infty}$ returns to unity everywhere.

\subsubsection{Binary Bondi-Hoyle-Lyttleton Accretion}

\begin{figure*}
\vspace{-4mm}
\begin{center}
\epsfxsize=6.0in
\leavevmode
\epsffile{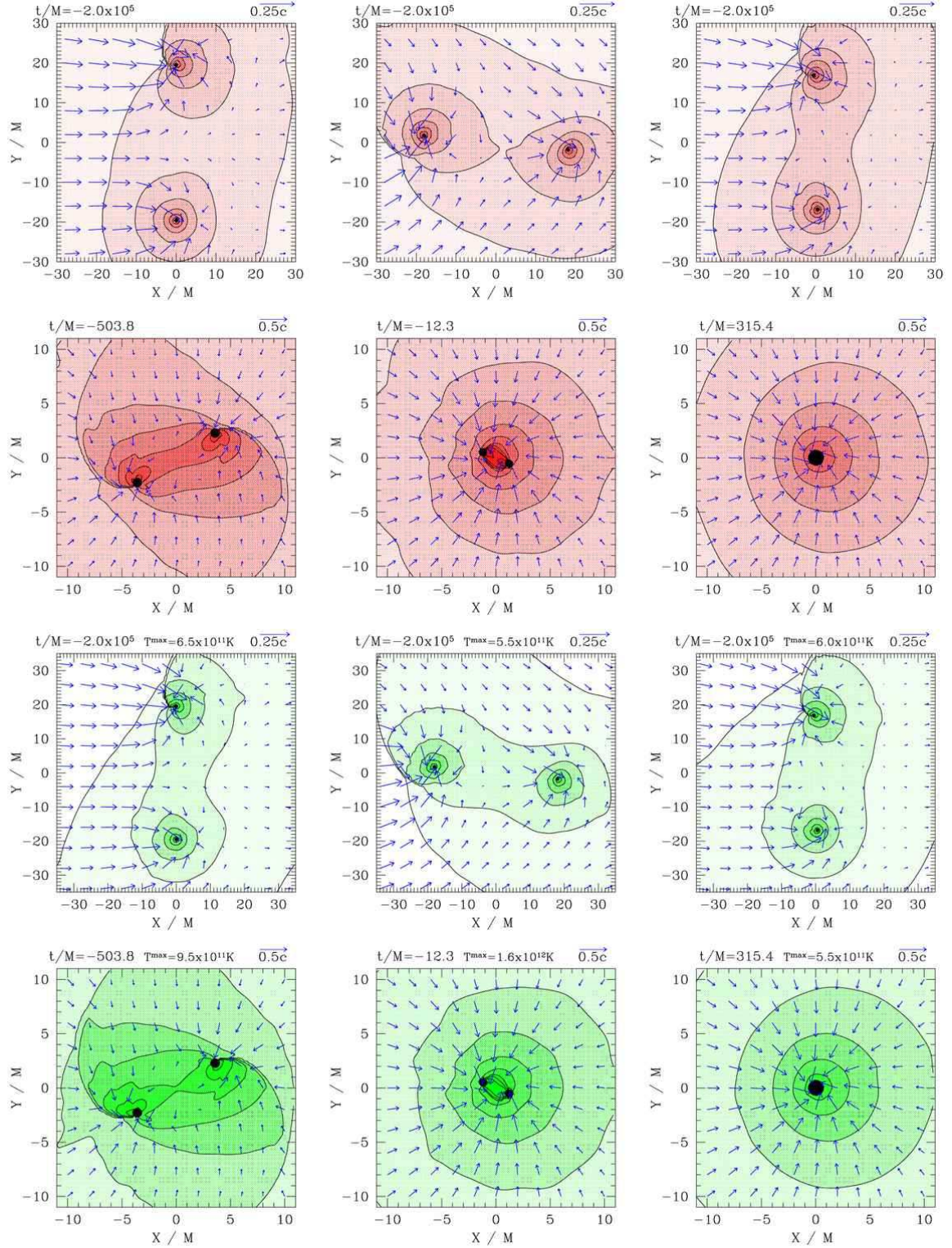}
\end{center}
\caption{Same as Fig.~\ref{fig:snap_bondi_13o9}, but for subsonic BHL accretion with $\Gamma=13/9$ and $\vcloud=0.1$.  The asymptotic velocity $\vcloud$ is in the $+\hat{x}$ direction.}
\label{fig:snap_sub_13o9}
\end{figure*}

\begin{figure*}
\vspace{-4mm}
\begin{center}
\epsfxsize=6.0in
\leavevmode
\epsffile{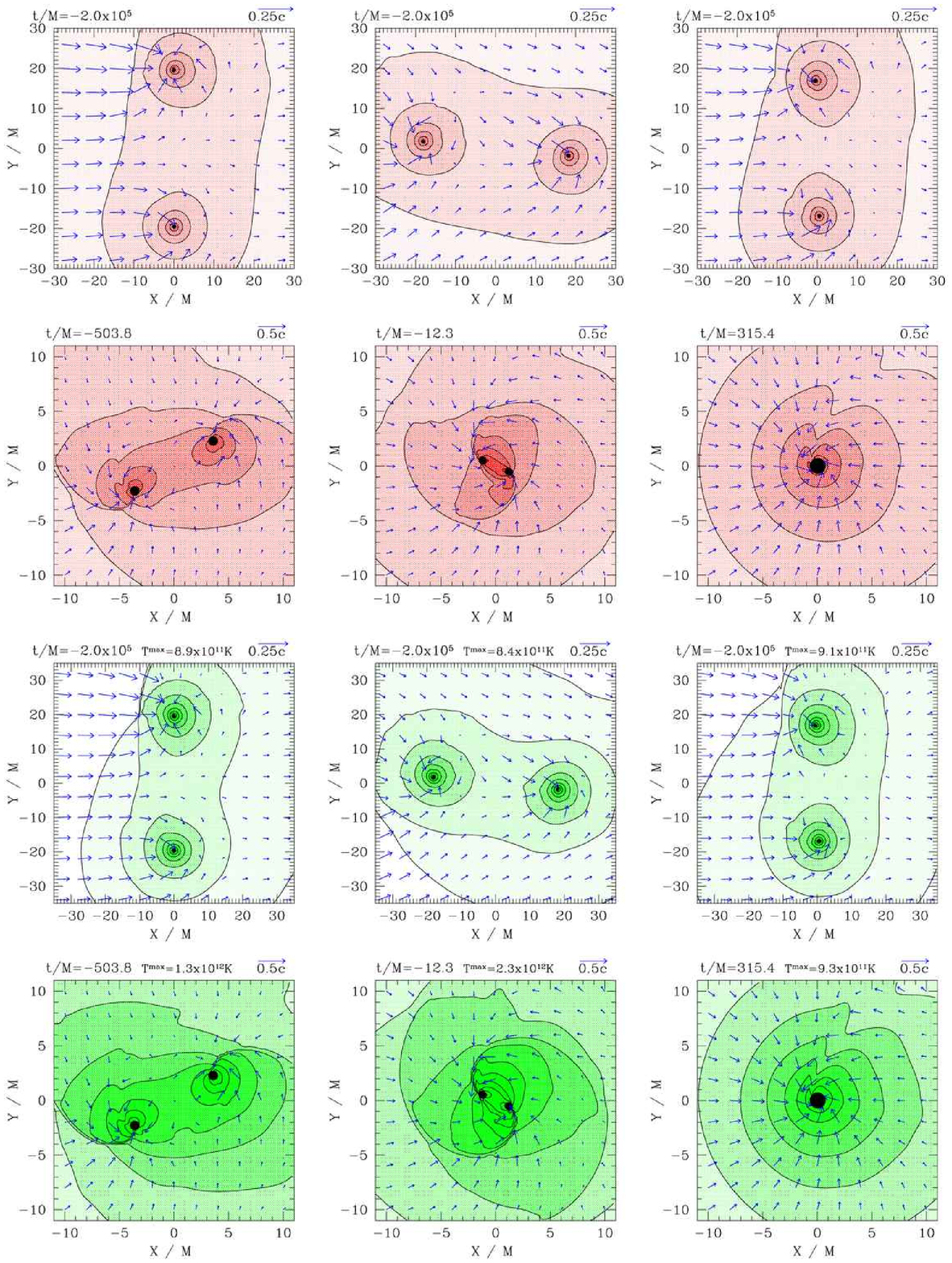}
\end{center}
\caption{Same as Fig.~\ref{fig:snap_bondi_13o9}, but for subsonic BHL accretion with $\Gamma=5/3$ and $\vcloud=0.1$.  The asymptotic velocity $\vcloud$ is in the $+\hat{x}$ direction.}
\label{fig:snap_sub_5o3}
\end{figure*}

\begin{figure}
\vspace{-4mm}
\begin{center}
  \epsfxsize=3.2in
  \leavevmode
  \epsffile{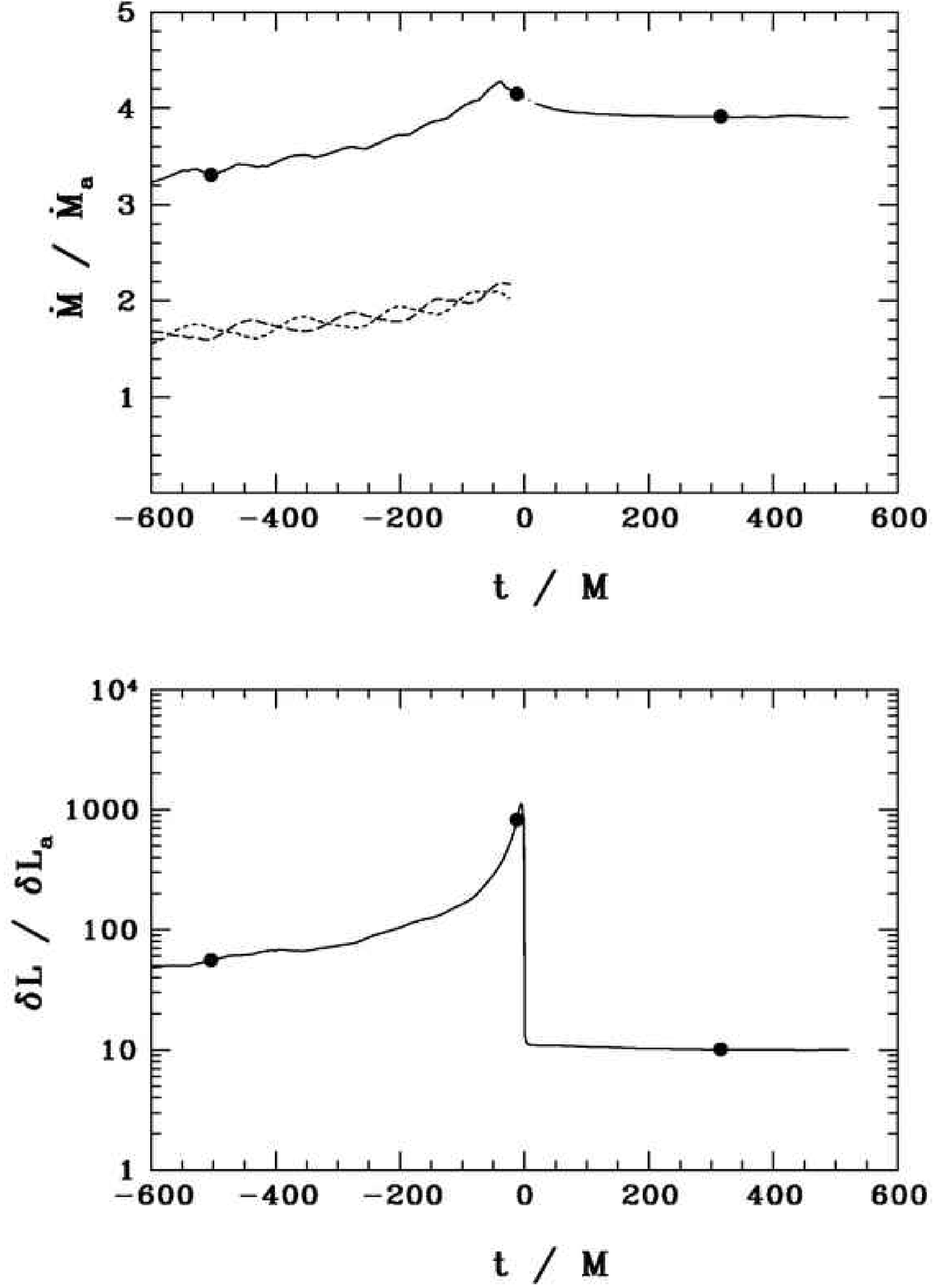} 
\end{center}
\caption{
  Time evolution of $\dot{M}$ and
  $\delta L$ for binary BHL inspirals with $\Gamma=13/9$, $\sscloud=0.148$, and $\vcloud=0.1$.  The initial binary
  separation is $d=10M$ and the BHs evolve to merger. 
  $\dot{M}_a$ and  $\delta L_a$  are the accretion rate and luminosity
  enhancement over the background for a single isolated
  black hole with mass equal to the initial ADM mass of the binary.
  Dashed lines in $\dot{M}$ plot represent accretion rates onto
  individual BHs, the solid line represents the total accretion rate. Dots represent the times highlighted in the last three snapshots in Fig.~\ref{fig:snap_sub_13o9}.
}
\label{fig:tplot_sub_13o9}
\end{figure}

\begin{figure}
\vspace{-4mm}
\begin{center}
 \epsfxsize=3.2in
  \leavevmode
  \epsffile{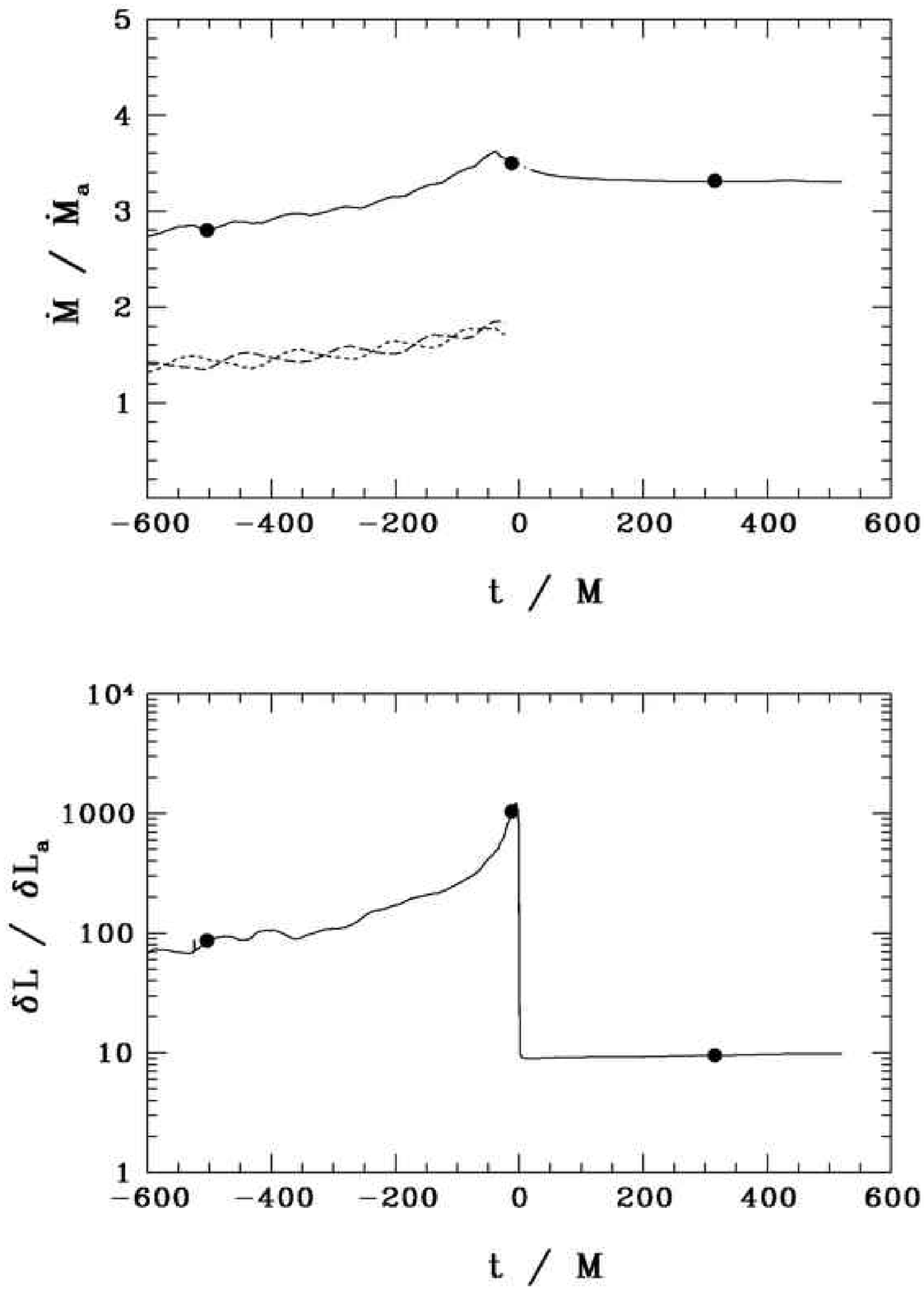} 
\end{center}
\caption{Same as Fig.~\ref{fig:tplot_sub_13o9}, but for $\Gamma=4/3$.}

\label{fig:tplot_sub_4o3}
\end{figure}
\begin{figure}
\vspace{-4mm}
\begin{center}
 \epsfxsize=3.2in
  \leavevmode
  \epsffile{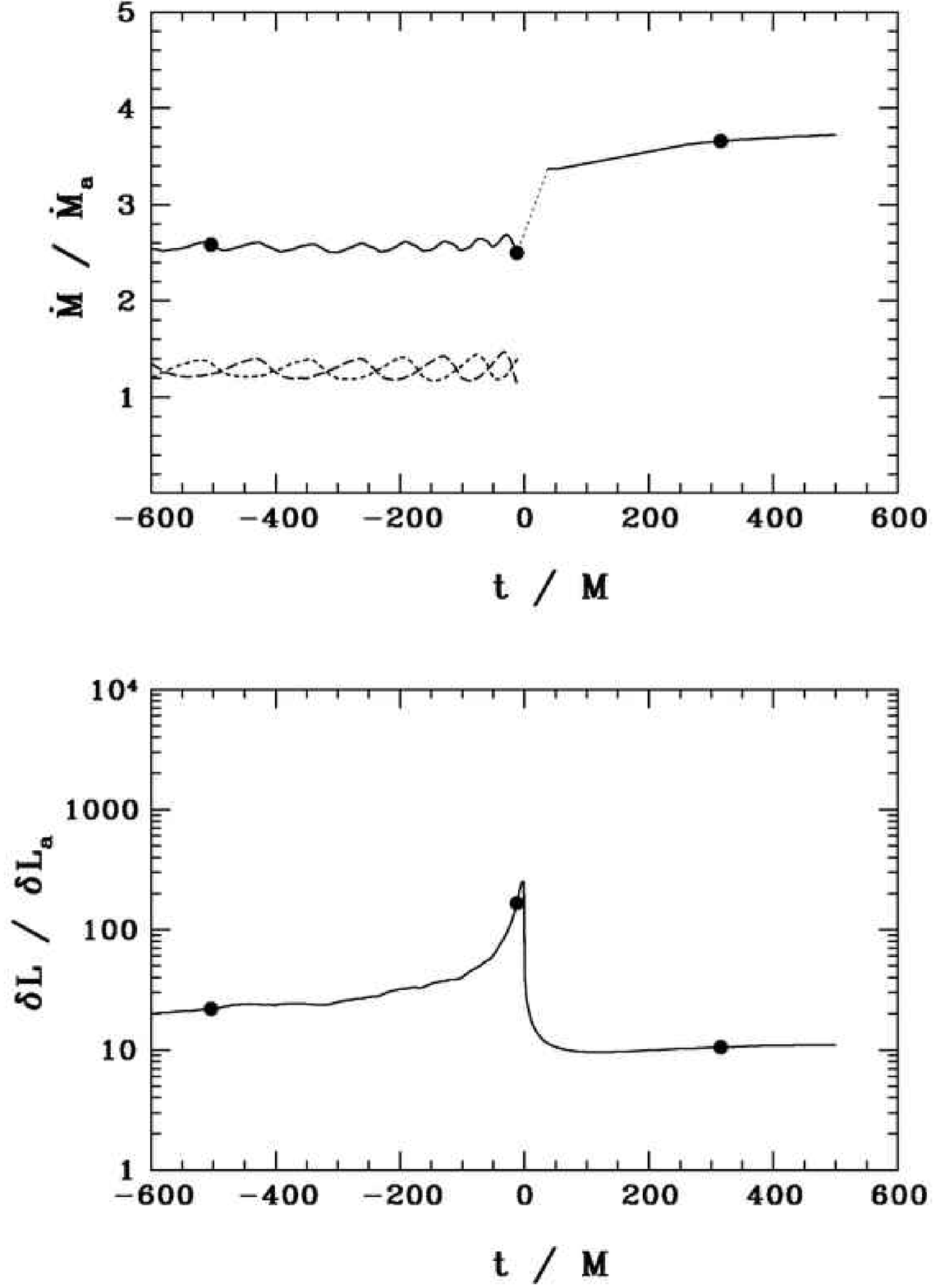} 
\end{center}
\caption{Same as Fig.~\ref{fig:tplot_sub_13o9}, but for $\Gamma=5/3$.}
\label{fig:tplot_sub_5o3}
\end{figure}

\begin{figure*}
\vspace{-4mm}
\begin{center}
\epsfxsize=6.0in
\leavevmode
\epsffile{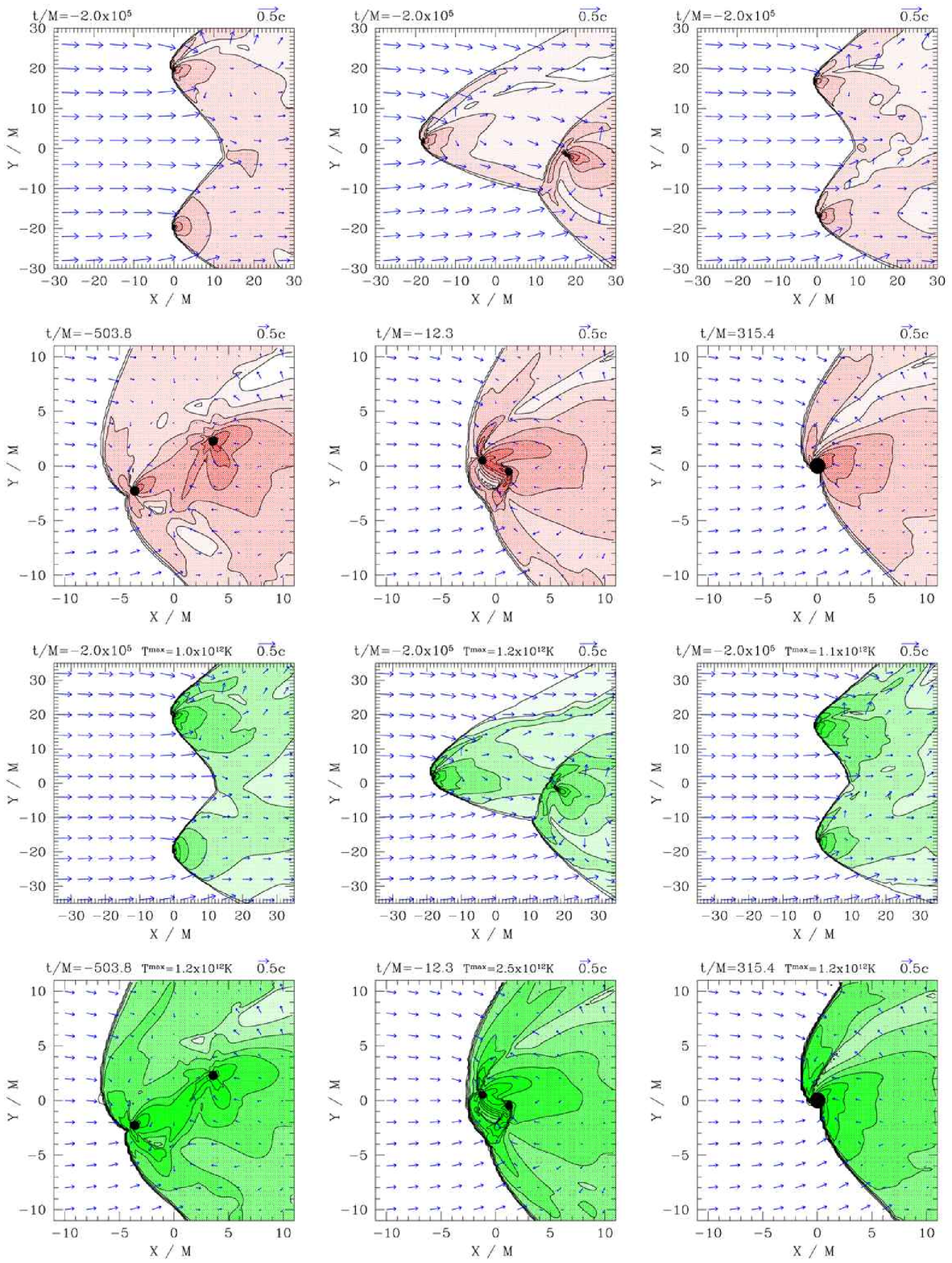}
\end{center}
\caption{Same as Fig.~\ref{fig:snap_bondi_13o9}, but for subsonic BHL accretion with $\Gamma=13/9$ and $\vcloud=0.4$.  The asymptotic velocity $\vcloud$ is in the $+\hat{x}$ direction.}
\label{fig:snap_sup_13o9}
\end{figure*}

\begin{figure*}
\vspace{-4mm}
\begin{center}
\epsfxsize=6.0in
\leavevmode
\epsffile{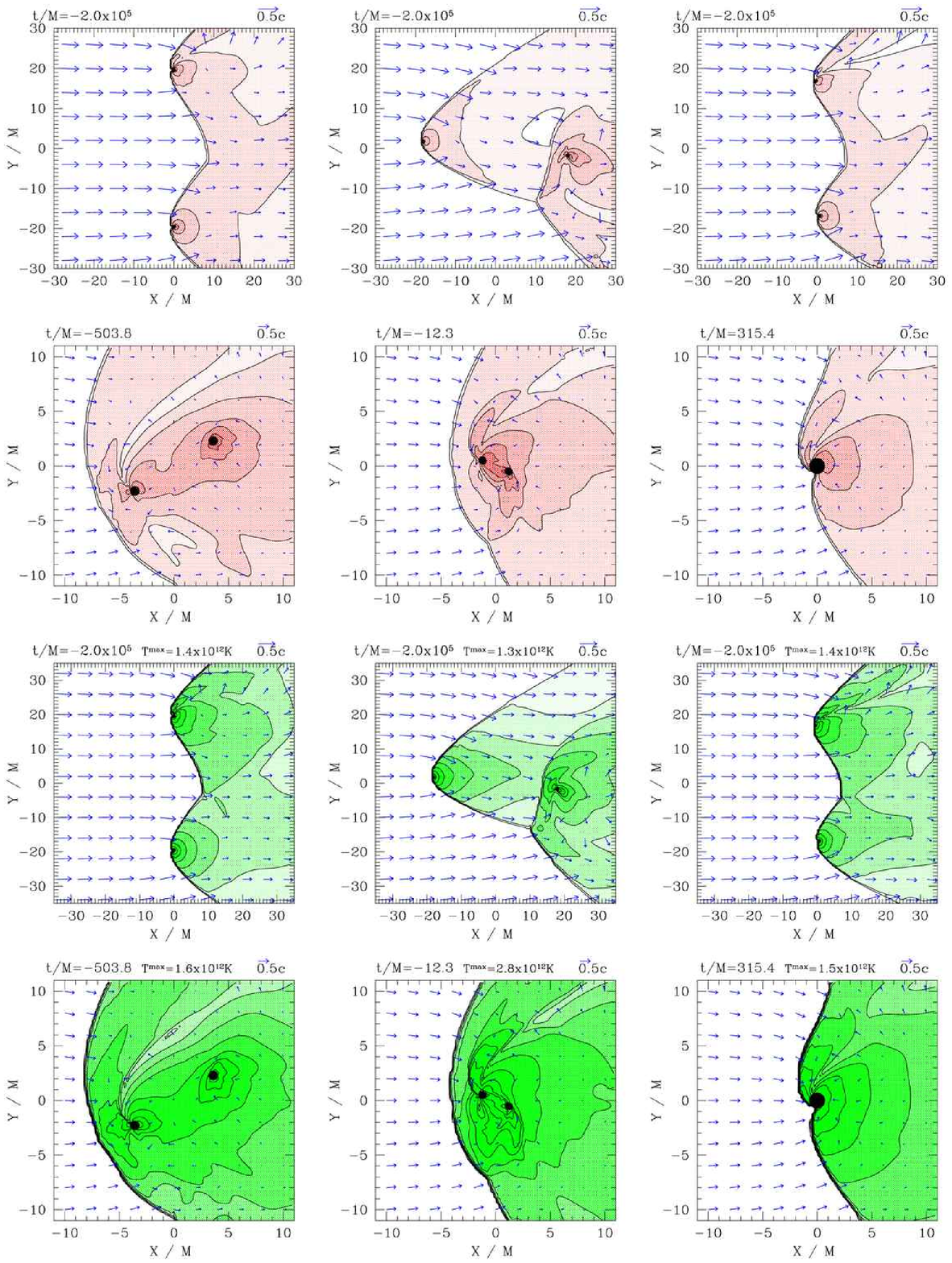}
\end{center}
\caption{Same as Fig.~\ref{fig:snap_bondi_13o9}, but for subsonic BHL accretion with $\Gamma=5/3$ and $\vcloud=0.4$.  The asymptotic velocity $\vcloud$ is in the $+\hat{x}$ direction.}
\label{fig:snap_sup_5o3}
\end{figure*}

\begin{figure}
\vspace{-4mm}
\begin{center}
  \epsfxsize=3.2in
  \leavevmode
  \epsffile{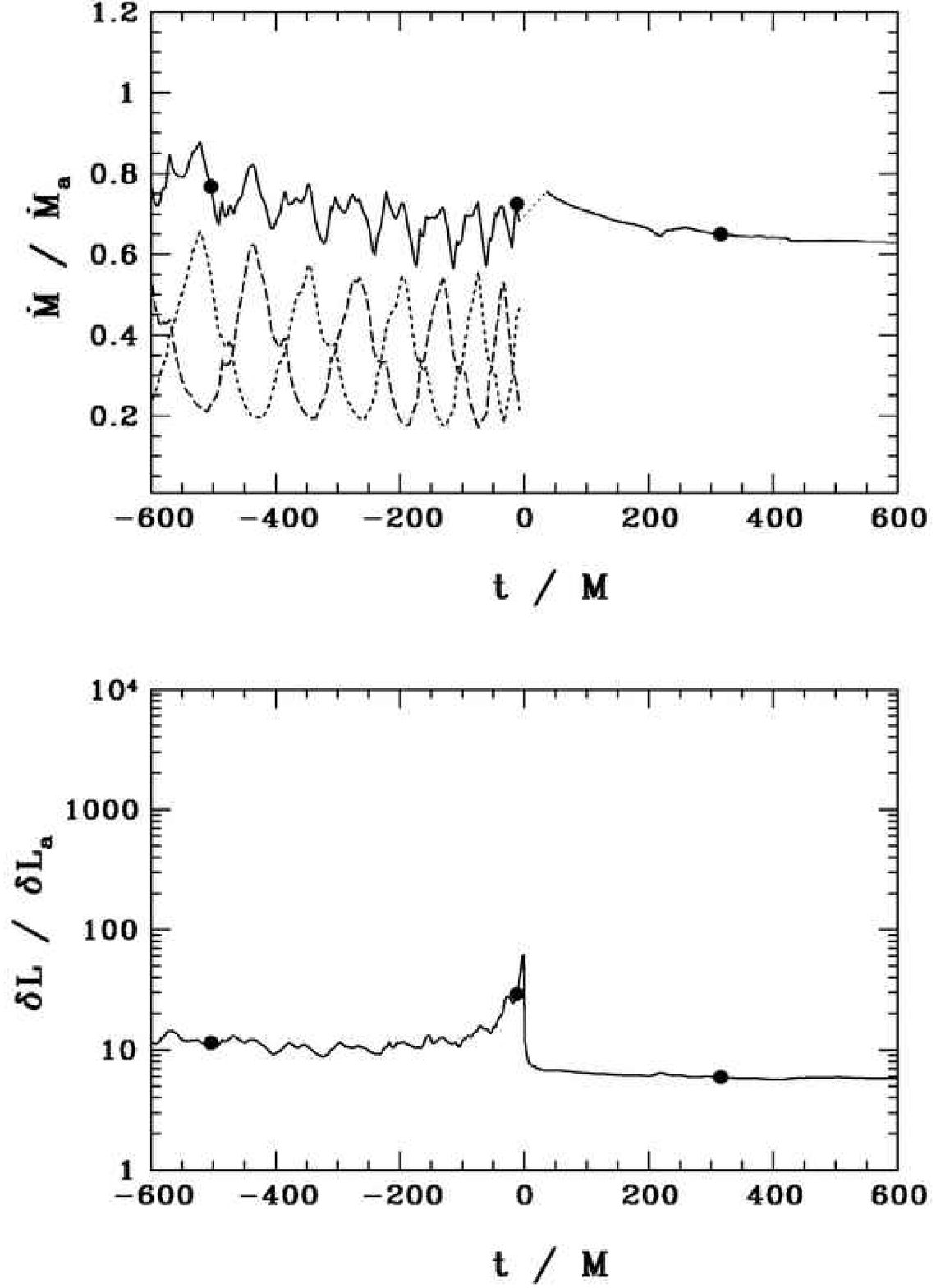} 
\end{center}
\caption{Same as Fig.~\ref{fig:tplot_sub_13o9}, but for $\Gamma=13/9$, $\vcloud=0.4$.}
\label{fig:tplot_sup_13o9}
\end{figure}
\begin{figure}
\vspace{-4mm}
\begin{center}
  \epsfxsize=3.2in
  \leavevmode
  \epsffile{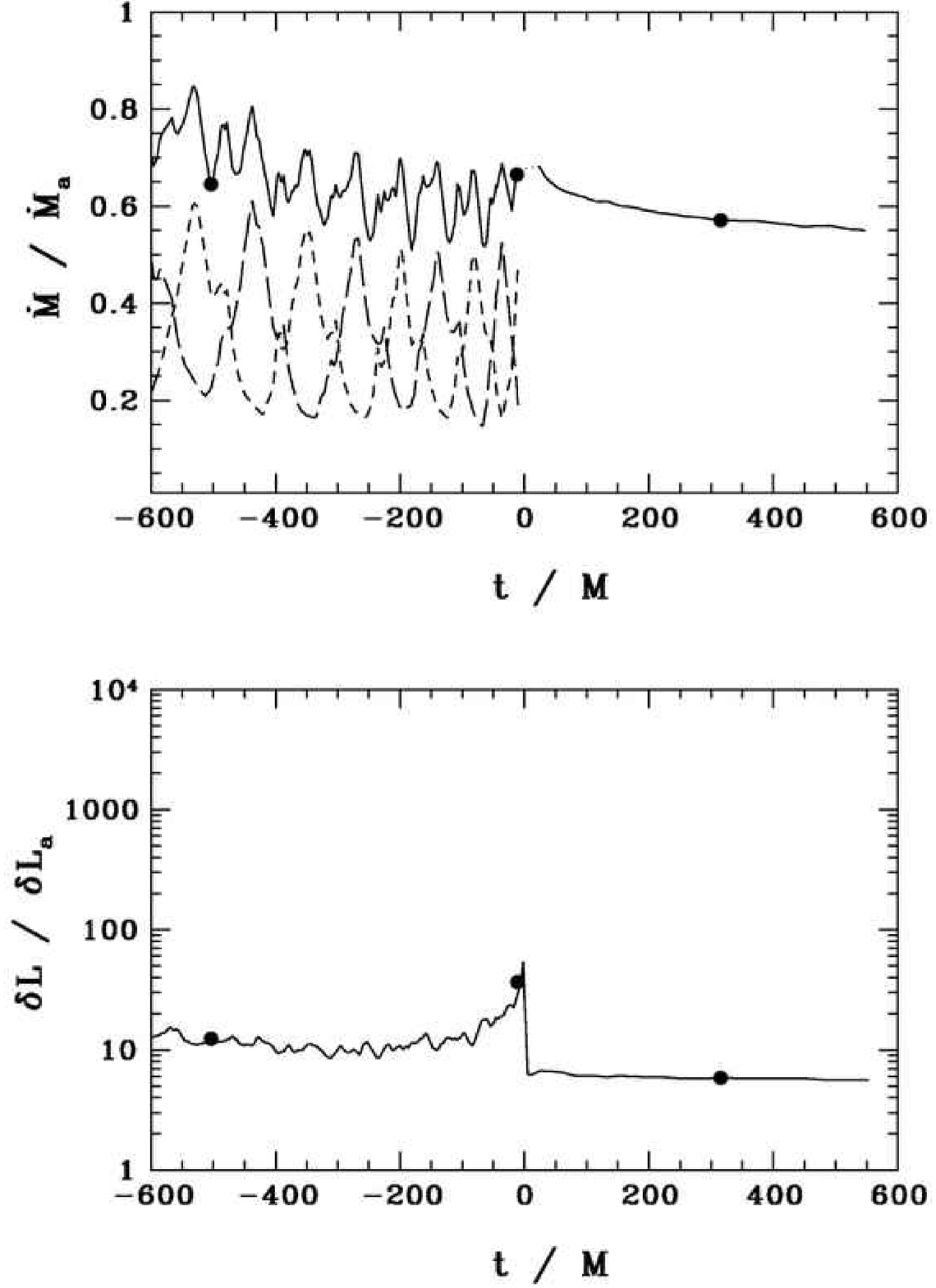} 
\end{center}
\caption{Same as Fig.~\ref{fig:tplot_sub_13o9}, but for $\Gamma=4/3$, $\vcloud=0.4$.}
\label{fig:tplot_sup_4o3}
\end{figure}
\begin{figure}
\vspace{-4mm}
\begin{center}
  \epsfxsize=3.2in
  \leavevmode
  \epsffile{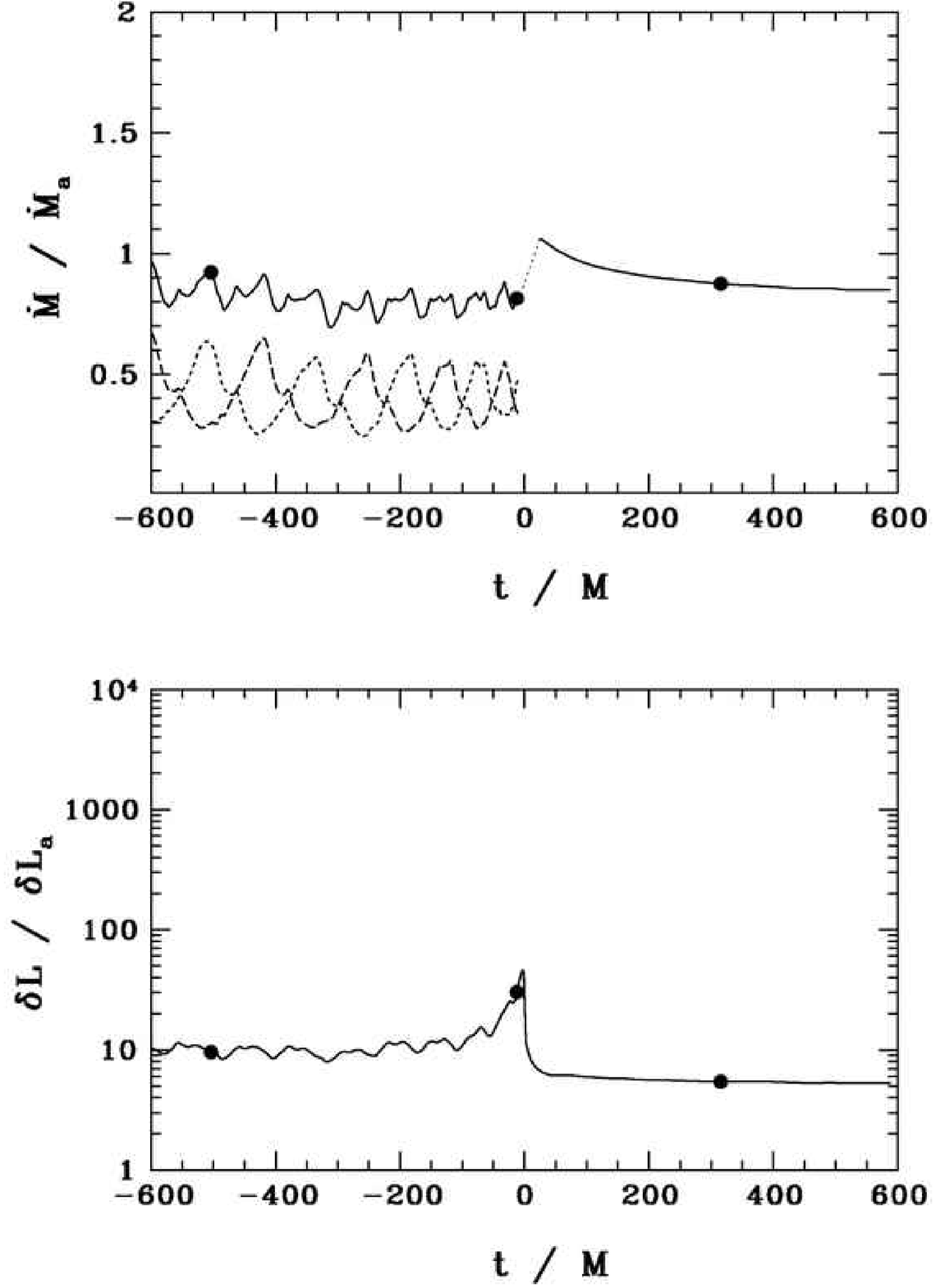} 
\end{center}
\caption{Same as Fig.~\ref{fig:tplot_sub_13o9}, but for $\Gamma=5/3$, $\vcloud=0.4$.}
\label{fig:tplot_sup_5o3}
\end{figure}

In order to investigate additional electromagnetic signatures which may be present due to the motion of the binary relative to the cloud, we have performed a series of ``prototype'' simulations of BHL accretion onto merging binaries.  We consider  both subsonic cases ($\vcloud / \sscloud = 0.7$) and supersonic cases ($\vcloud / \sscloud = 2.7$).  We again consider $\Gamma=13/9$ (PB1 and PC1), $\Gamma=4/3$ (PB2 and PC2), and $\Gamma=5/3$ (PB3 and PC3) in order to assess the influence of the EOS on the flow. We perform both wide separation runs ($d=40 M$) to produce snapshots that are quasistationary in the corotating frame of the binary, as well as close separation runs ($d=10 M$) in which we evolve the binary from inspiral to merger.

For our (asymptotically) subsonic cases, the departures from the binary Bondi case are subtle. Figures~\ref{fig:snap_sub_13o9} and \ref{fig:snap_sub_5o3} show snapshots of density and velocity field for cases PB1 and PB3. Snapshots for case PB2 are similar to PB1 and hence are now shown here. At wide separation $d=40 M$, we see an asymmetry in the accretion flow for cases PB1 and PB2, as shocks develop around one BH as it moves against the flow of the gas, but not around the other as it moves in the same direction as the flow. This phenomenon continues up to the merger.  At this separation the orbital Keplerian velocity is $\vkep \approx 0.08$. Thus, these shocks are formed when one BH moves supersonically against the flow of the gas and $(\vkep+\vcloud)/a \gtrsim 1$.  The reason that this behavior is not seen in case PB3 is that the gas is adiabatically heated more efficiently for $\Gamma=5/3$ as it flows toward the BHs, causing the sound speed of the gas to be greater near each BH.  We find that for $\Gamma=5/3$, $(\vkep+\vcloud)/a < 1$ near the black hole, preventing any shocks from forming.  The post-merger accretion flow exhibits some departure from spherical symmetry due to BH spin and $\vcloud \neq 0$, but all shocks dissipate.  In Figs.~\ref{fig:tplot_sub_13o9}--\ref{fig:tplot_sub_5o3}, we once again observe an increase in accretion rate and luminosity over the course of the merger.  We have also plotted in these figures the individual accretion rates onto each BH in order to demonstrate the effect of the binary motion relative to the wind. 

For our supersonic cases (PC1-PC3), the departure from the binary Bondi case is more dramatic.  When the BHs are widely separated ($d>\Racc$), a bow shock forms around each individual BH, as seen in Figs.~\ref{fig:snap_sup_13o9} and \ref{fig:snap_sup_5o3}.  Late in the inspiral, when $d<\Racc$, these shocks merge and form a single bow shock surrounding the binary, which persists after the merger.  The final flow resembles the solution found by \cite{petrich89} for steady accretion onto a moving BH with $\vcloud / \sscloud = 2.5$, although in our case the remnant is spinning.  During the inspiral, the modulation in $\dot{M}$ and $ \delta L$ is more pronounced than in the subsonic case (see Figs.~\ref{fig:tplot_sup_13o9}--\ref{fig:tplot_sup_5o3}).

\subsection{Realistic Binary Bondi accretion}
\label{sec:real_bondi}

\begin{figure}
\vspace{-4mm}
\begin{center}
  \epsfxsize=3.2in
  \leavevmode
  \epsffile{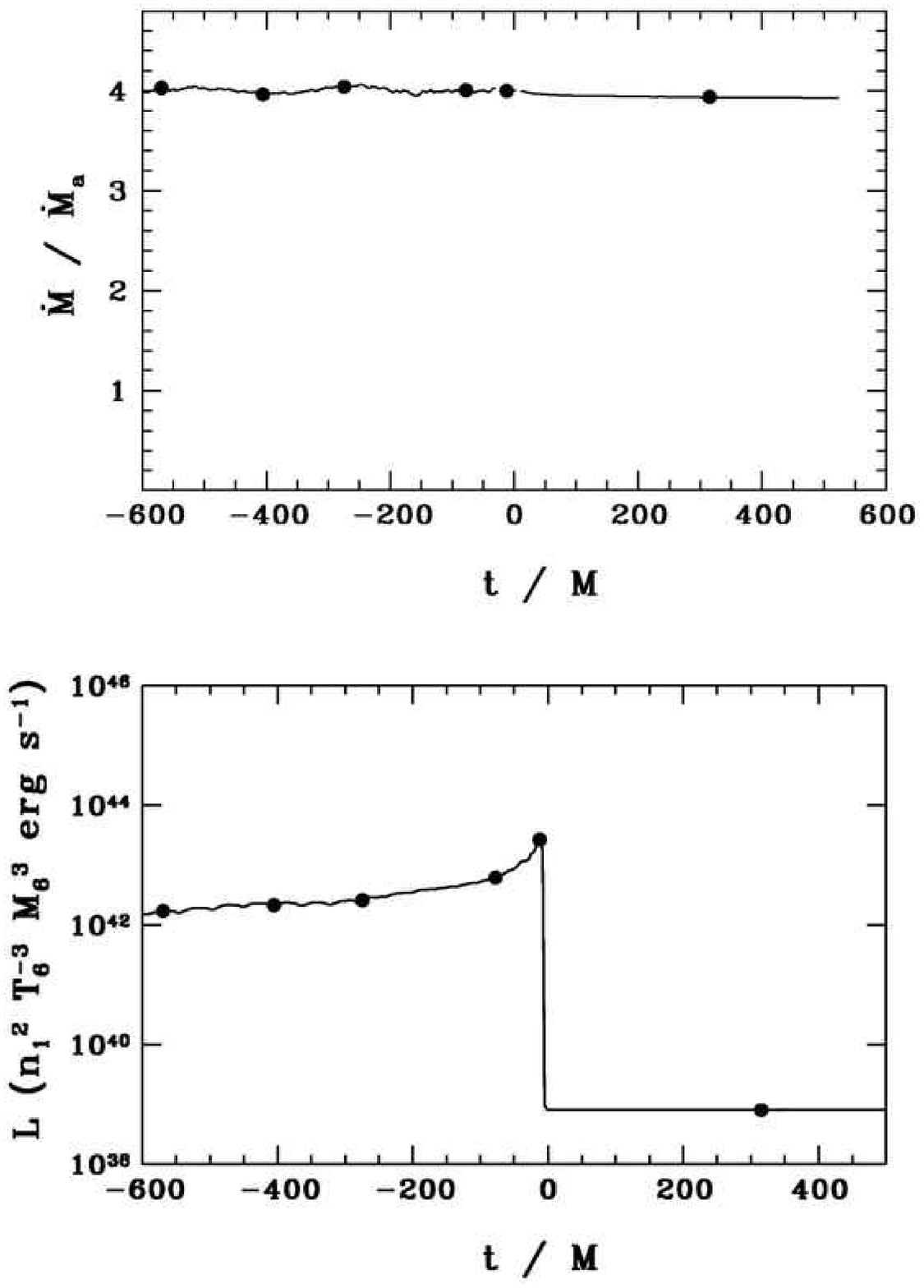}
\end{center}
\caption{
  Plots showing time evolution of $\dot{M}$ and $\dot{L}$.  Here time is measured relative to the time at which the merger
  occurs.  
  Asymptotic temperature is $T=10^6K$.
  Adiabatic index given by $\Gamma=\Gamma^*$.  $n_1 \equiv n_{\infty} / 10 \mbox{cm}^{-3}$, $T_6 \equiv T_{\infty}/10^6K$, $M_6 \equiv M / 10^6 M_{\odot}$.
}
\label{fig:tplot_real_13o9}
\end{figure}

\begin{figure}
\vspace{-4mm}
\begin{center}
  \epsfxsize=3.2in
  \leavevmode
  \epsffile{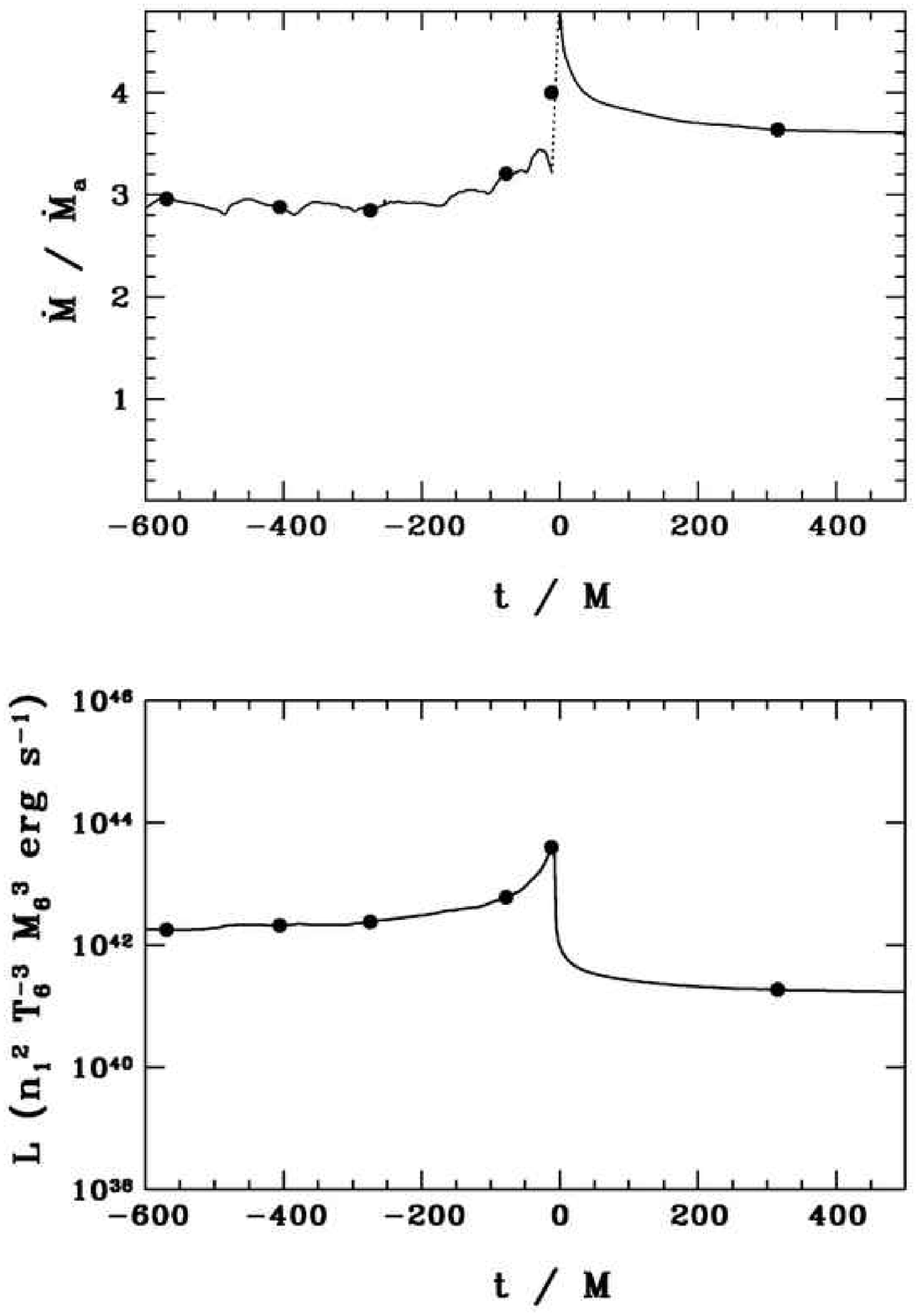}
\end{center}
\caption{Same as Fig.~\ref{fig:tplot_real_13o9}, but for $\Gamma=5/3$.}
\label{fig:tplot_real_5o3}
\end{figure}

\begin{figure*}
\vspace{-4mm}
\begin{center}
\epsfxsize=6.0in
\leavevmode
\epsffile{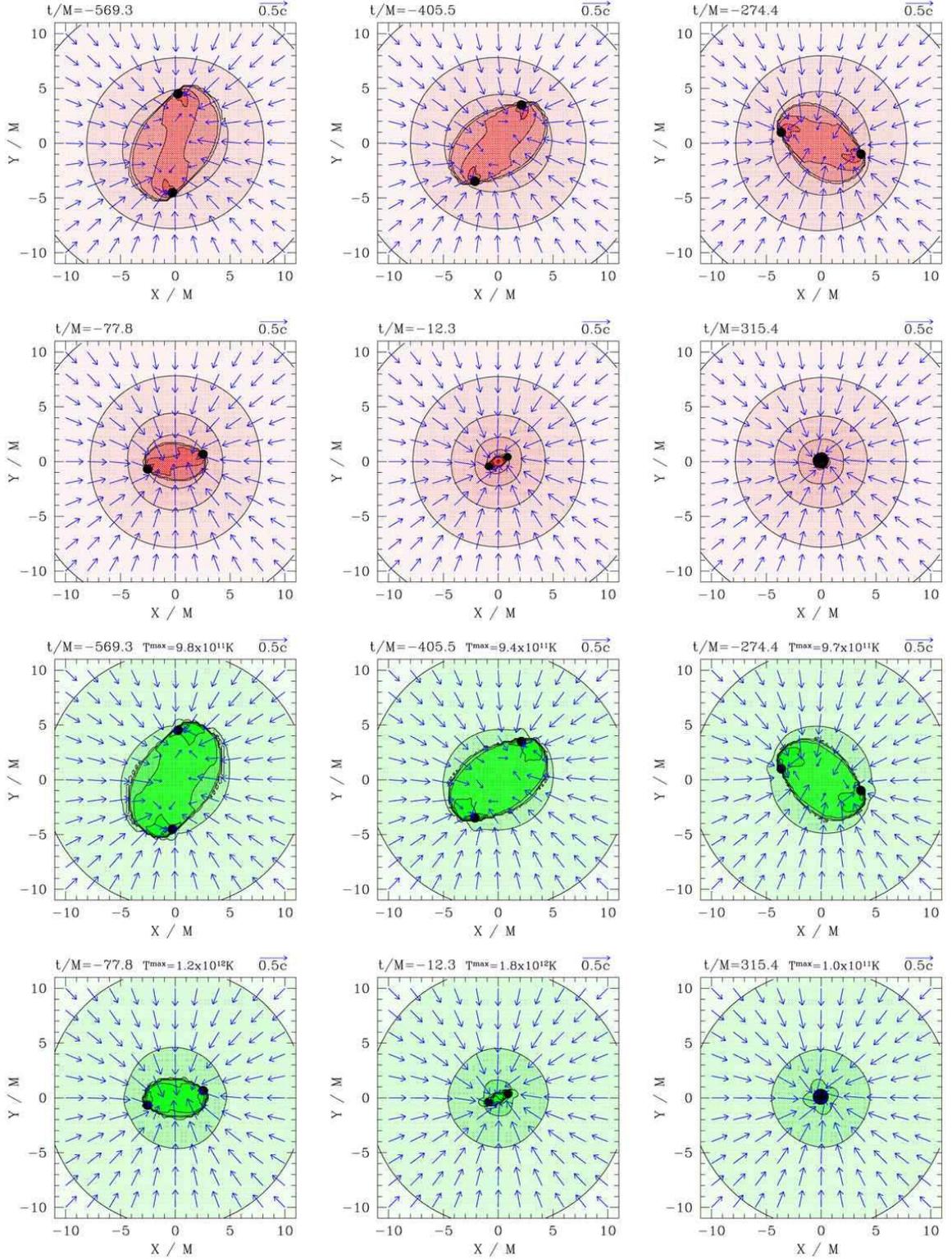}
\end{center}
\caption{
  Snapshots of rest-mass density $\rho_0$ and temperature $T$ contours for
  the $\Gamma=\Gamma^*$ case.  First and second
  rows show snapshots of density contours and velocity profiles in the
  orbital plane.  Third and fourth rows show snapshots of
  $T$.  Density is plotted according to $\rho=\rho_{0,\infty}10^{7+0.33 j}\ \ (j=1,2,.
...,12)$.  Temperature contours are plotted according to $T=10^{10+0.25j}K \ \ (j=1,2,....,12)$. 
  Arrows represent the velocity field in the given plane.  
}
\label{fig:snap_real_13o9}
\end{figure*}

\begin{figure*}
\vspace{-4mm}
\begin{center}
\epsfxsize=6.0in
\leavevmode
\epsffile{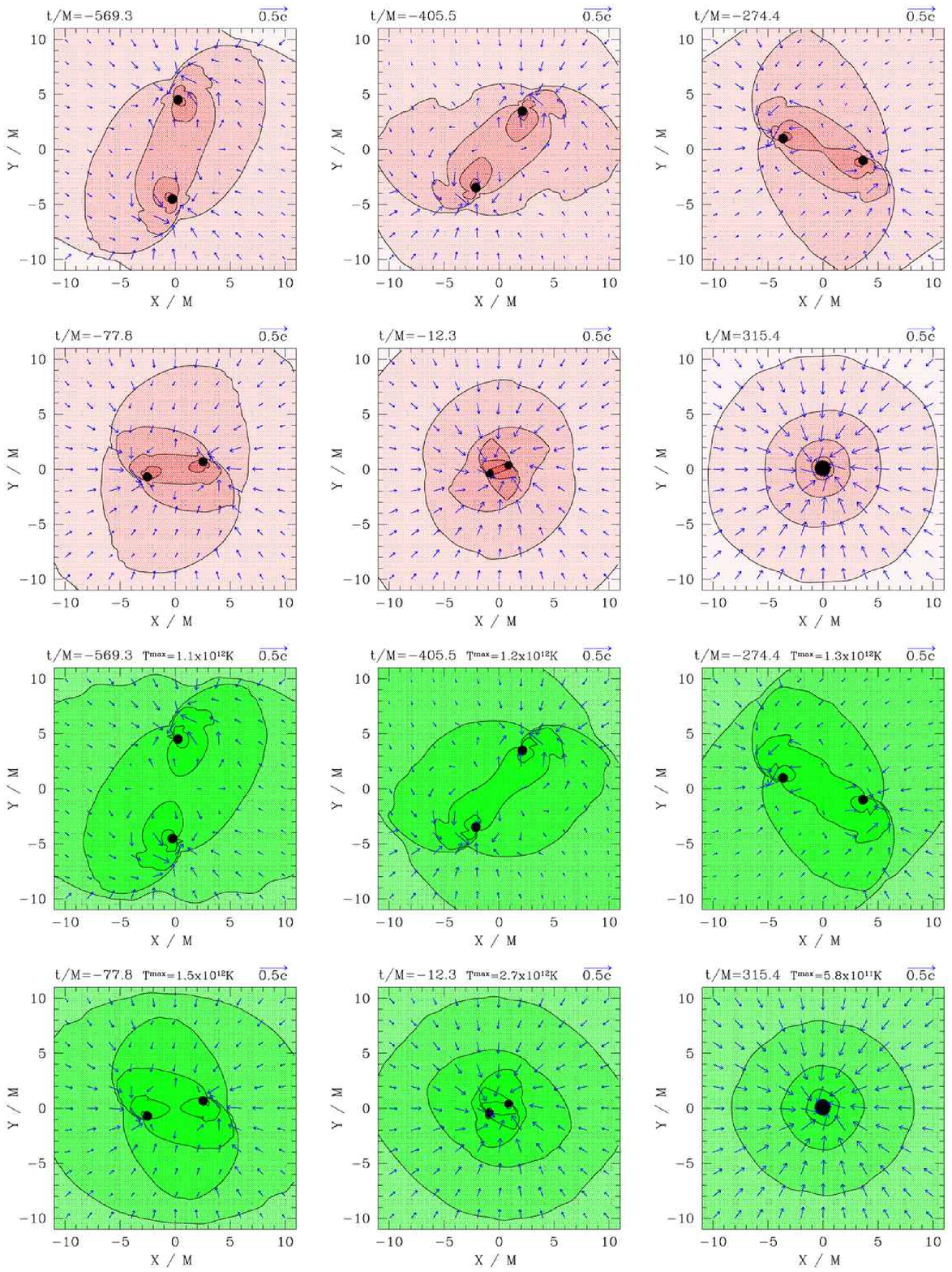}
\end{center}
\caption{Same as Fig.~\ref{fig:snap_real_13o9} but for $\Gamma=5/3$.}
\label{fig:snap_real_5o3}
\end{figure*}

While the high temperature prototype runs give us valuable insight into the general nature and different phases of the accretion flows onto inspiralling BHBH binary systems in gas clouds, we require simulations with more realistic gas temperatures in order to identify observational signatures.  Accordingly, we have performed simulations of the binary Bondi problem for a gas cloud with asymptotic density $n_{\infty}=10~\mbox{cm}^{-3}$ and temperature $T=10^6 K$.  This choice is consistent with the proposed ``cooling flow model of quasar fueling'' describing the hot interstellar gas found in galaxies \cite{merritt05,nulsen00}.  As mentioned previously, computational limitations demand that our outer boundary be placed inside the transonic radius for these simulations.  As a result, we focus only on the final phase of the inspiral and merger, in which $d \ll \Racc$.  We use both a $\Gamma=\Gamma^*$ (case RA1) and $\Gamma=5/3$ (case RA2) EOS. As explained in Sec.~\ref{effective_index}, we still set $\Gamma=13/9$ in the computational domain for the $\Gamma=\Gamma^*$ case since the $\Gamma=5/3$ and transition region is outside our computational domain. However, the initial hydrodynamic profile of the flow is very different from a pure $\Gamma=13/9$ EOS with the same asymptotic temperature. We focus here on the BHBH Bondi problem only ($\vcloud=0$), and postpone a study of binary BHL accretion in a realistic gas cloud for a later analysis.  As in the prototype calculations, we find evidence for a strong enhancement in the luminosity due to shock heating of the gas (see Fig.~\ref{fig:tplot_real_13o9} and Fig.~\ref{fig:tplot_real_5o3}).  For these runs, we plot the luminosities in cgs units.

We note that for case RA1, strong shocks form, but are confined to the immediate vicinity of the binary.  This is because the inward gas flow onto the binary is strongly supersonic, making it difficult for shocks to propagate outward (see Fig.~\ref{fig:snap_real_13o9}).  For case RA2, on the other hand, the radial component of the 4-velocity $u \sim a$ everywhere, making it much easier for shocks to spread outward, as seen in Fig.~\ref{fig:snap_real_5o3}.

\subsection{Scaling and Detectability}
All of our quoted results for the accretion rate $\dot{M}$ are normalized by the value for a single black hole of mass $M/2$ undergoing stationary, spherical Bondi accretion.  Using Eq.~(\ref{Mdot_formula}) we see that this quantity scales with asymptotic temperature and density according to 
\begin{equation}
\dot{M}_a c^2 = 4.6 \times 10^{40} \ \lambda_{5/3} \ n_1 \ T_6^{-3/2} \ M_6^2 \ \mbox{erg s}^{-1} \ ,
\end{equation}
where we define $\lambda_{5/3} \equiv \lambda(\Gamma,a_{\infty})/\lambda(5/3,a_{\infty})$, $n_1 \equiv n_{\infty} / 10 \mbox{cm}^{-3}$ , $T_6 \equiv T_{\infty} / 10^6 K$, and $M_6 \equiv M / 10^6 M_{\odot}$.  Recall that in the Newtonian limit, when $\Racc \gg M$, $\lambda$ depends only on $\Gamma$.

It is also possible to derive simple scaling relations for the luminosities.  In each of our simulations, most of the electromagnetic luminosity is generated near the horizons of the BHs, as the temperature and density both rise sharply when approaching the horizon, and achieve their maximum values there. By examining Eqs.~(\ref{brememissivity_q})--(\ref{brememissivity_F}) (ignoring the logarithmic terms), and Eq.~(\ref{synch_high_T}), we see that in the high temperature limit ($kT > m_e c^2$), which applies near the horizon in all of our simulations, the bremsstrahlung and synchrotron emissivities depend on the density and temperature according to,
\beqn
q_{ff} \propto n_h^2 T_h \label{eq:qbr_highT} \\
q_{syn} \propto n_h^2 T_h^3 \beta^{-1} \label{eq:qs_highT}
\eeqn
Here $n_h$ and $T_h$ refer to the density and temperature at the horizon, respectively.  Integrating Eqs.~(\ref{eq:qbr_highT}) and (\ref{eq:qs_highT}), we find
\beqn
L_{ff} &\approx& \int dV q_{ff} \propto   n_h^2 T_h M^3\\
L_{syn} &\approx& \int dV q_{syn}  \propto n_h^2 T_h^3 \beta^{-1} M^3 \ ,
\eeqn
where we have assumed that the radiation is generated near the horizon and hence the radiative volume scales as $M^3$. To estimate $n_h$, we use the fact that $r_h \ll \Racc$, and so the fluid 4-velocity is approximately given by its free-fall value
\beq
\hat{u} \approx \left(\frac{2 M}{r_h}\right)^{1/2} \ .
\eeq
Substituting this into Eq.~(\ref{continuity}) and Eq.~(\ref{Mdot_formula}), we find that 
\beq
\frac{n_h}{n_{\infty}} \sim \left(\frac{\Racc}{r_h}\right)^{3/2} \propto \left(\frac{kT_{\infty}}{m_B c^2}\right)^{-3/2} \label{eq:nh}
\eeq
Here we have used $a^2 \approx \Gamma P/\rho_0 = 2 \Gamma kT/m_B c^2$.    This scaling should remain reasonably accurate, even in the presence of strong shocks, as the shocks cause density enhancements of $\lesssim (\Gamma+1)/(\Gamma-1)$ (see Sec. 89 of \cite{landau59}), which is of order unity.  

The temperature at the horizon $T_h$, on the other hand, will be strongly affected by the presence of shocks.  The gas in any shocked region will be heated to $kT\sim m_B v^2$. Since $v \lesssim c$ near the horizon, shock heating guarantees that $k T_h \lesssim m_B c^2 \sim 10^{13}~\mbox{K}$ {\it independant} of the gas temperature at infinity $T_{\infty}$.  This result has important consequences for the scaling of the maximum luminosities.  Once the shock-heated gas is accreted following the merger, $T_h$ drops below this value for all $\Gamma < 5/3$ (see e.g. Fig~\ref{fig:semi_analytic_plot}) and the luminosity settles down to a value below the maximum.

Even in the absence of shocks, $kT_h \lesssim m_B c^2$ for $\Gamma=5/3$.  This is because for this particular EOS, an appreciable fraction of the gravitational potential energy is converted into thermal energy (both scale as $r^{-1}$ inside $\Racc$: $kT \sim M m_B / r$) \cite{shapiro_book_83}. For $\Gamma=\Gamma^*$,  it can be shown that the temperature at the horizon for spherical Bondi flow is approximately given by \cite{shapiro_book_83}
\beq
k T_h  \approx \frac{2}{3} \left(\frac{9}{80}\right)^{2/3}\left(\frac{m_e}{m_B}\right)^{1/3} {m_B c^2}= 0.013 \ m_B c^2
\label{kTh_Gamma_star}
\eeq
Thus, we find that the temperature at the horizon, $T_h$ is independent of gas parameters at infinity for both $\Gamma=5/3$ and $\Gamma = \Gamma^*$.  However, for $\Gamma=\mbox{const} < 5/3$, $T_h$ does exhibit scaling with $T_{\infty}$ in the absence of shocks.  In this case, $P=K \rho_0^{\Gamma}$, so we find
\beq
\frac{T_h}{T_{\infty}} = \left(\frac{n_h}{n_{\infty}}\right)^{\Gamma-1}  \propto \left(\frac{kT_{\infty}}{m_B c^2}\right)^{-3(\Gamma-1)/2}
\eeq

We can now apply these results to see how the luminosity scales in different phases of the inspiral.  During both the very early pre-merger phase of the inspiral, when $d > \Racc$, and during the post-merger phase after the fluid has settled to a quasiequilibrium state, there are no shocks present.  We can therefore use the above results to see that in these phases,

\begin{eqnarray}
  L_{ff}&\propto& \left\{\begin{array}{l l}
      \displaystyle  n_1^2 \ T_6^{-3} M_6^3 \ ,& \ \ \ \Gamma = 5/3 \mbox{ or } \Gamma=\Gamma^* \ ,  \\
      \displaystyle  n_1^2 \ T_6^{-(3\Gamma+1)/2} M_6^3 \ , & \ \ \ \Gamma = \mbox{const} < 5/3 \ , 
    \end{array}\right. \label{Lbr_scaling}\\
  L_{syn}&\propto& \left\{\begin{array}{l l}
 \displaystyle  n_1^2 \ T_6^{-3} \beta_1^{-1} M_6^3 \ , & \ \ \ \Gamma = 5/3 \mbox{ or } \Gamma=\Gamma^* \ , \\
      \displaystyle  n_1^2 \ T_6^{-9(\Gamma-1)/2} \beta_1^{-1} M_6^3 \ , & \ \ \ \Gamma = \mbox{const} < 5/3 \ .
\end{array}\right. \label{Ls_scaling}
\end{eqnarray}

Here $\beta_1 \equiv \beta / 10$.  During the late premerger phase when $d \sim M \ll \Racc$ shocks will be present and $T_h$ will no longer depend on $T_{\infty}$ for any EOS.  In this case, we find
\begin{equation}
  L_{ff}\propto  n_1^2 \ T_6^{-3} M_6^3  \ \ \ , \ \ \ 
  L_{syn}\propto  n_1^2 \ T_6^{-3} \beta_1^{-1} M_6^3 \ .
\end{equation}

Using the above scaling relations along with the results of our realistic temperature simulations, we find that the peak luminosity shortly before the merger of an equal mass BHBH binary in a gas cloud for case RA1 ($\Gamma=\Gamma^*$) is given by
\begin{eqnarray}
L_{ff}^{max} &\approx&  3 \times 10^{37} \ n_1^2 \ T_6^{-3} M_6^3 \ \mbox{erg s}^{-1}  \ ,\\
L_{syn}^{max} &\approx& 3 \times 10^{43} \ n_1^2 \ T_6^{-3} \beta_1^{-1} M_6^3 \ \mbox{erg s}^{-1}  \ .
\end{eqnarray}
Similarly, the peak luminosity for case RA2 ($\Gamma=5/3$) is given by
\begin{eqnarray}
L_{ff}^{max} &\approx&  4 \times 10^{37} \ n_1^2 \ T_6^{-3} M_6^3 \ \mbox{erg s}^{-1} \ , \\
L_{syn}^{max} &\approx& 4 \times 10^{43} \ n_1^2 \ T_6^{-3} \beta_1^{-1} M_6^3 \ \mbox{erg s}^{-1} \ .
\end{eqnarray}

At the late post-merger phase, the fluid relaxes to a stationary flow. The scaling relations~(\ref{Lbr_scaling}) and (\ref{Ls_scaling}) hold. Combining these scaling relations and our simulation results, we find that during the post-merger phase
\begin{eqnarray}
  L_{ff}&\approx&  3 \times 10^{35} \ n_1^2 \ T_6^{-3} M_6^3 \ \mbox{erg s}^{-1} \ , \label{Lbr_RA1} \\
L_{syn} &\approx& 8 \times 10^{38} \ n_1^2 \ T_6^{-3} \beta_1^{-1} M_6^3 \ \mbox{erg s}^{-1}
\label{Ls_RA1}
\end{eqnarray}
for case RA1, and
\begin{eqnarray}
L_{ff} &\approx&  3 \times 10^{36} \ n_1^2 \ T_6^{-3} M_6^3 \ \mbox{erg s}^{-1} \ ,  \label{Lbr_RA2}\\
L_{syn} &\approx& 2 \times 10^{41} \ n_1^2 \ T_6^{-3} \beta_1^{-1} M_6^3 \ \mbox{erg s}^{-1}
\label{Ls_RA2}
\end{eqnarray}
for case RA2.  We note that in each case, the total luminosity is dominated by the synchrotron emission.

In each of our calculations, we have ignored the effects of radiative cooling on the gas dynamics.  We can estimate the error induced by this by comparing the rate of thermal energy transport, $\dot{E}_{th}$, to the luminosity.  Here we define 
\begin{equation}
  \dot{E}_{th} = \dot{M} \epsilon 
  = \dot{M} \frac{P}{\rho_0(\Gamma-1)} \ .
\end{equation} 

Since the luminosity is dominated by emission near the horizon, we are primarily concerned about the region near the horizon.  Using Eqs.~(\ref{Mdot_formula}), (\ref{eq:nh}), (\ref{kTh_Gamma_star}), (\ref{Lbr_RA1}) and (\ref{Ls_RA1}), we find that for case RA1, at late times after the merger when the flow has reached equilibrium, 
\begin{equation}
\frac{L_{syn}+L_{ff}}{\dot{E}_{th}}\sim 0.1 \ n_1 \ T_6^{-3/2} \ \beta_1^{-1} M_6 \ .
\end{equation}
Thus, we see that in these regimes, it is a good approximation to neglect the effects of radiative cooling for our canonical model parameters.  At the moment of maximum luminosity, shortly before merger, we find that
\begin{equation}
\frac{L_{syn}+L_{ff}}{\dot{E}_{th}}\sim 40  \ n_1 \ T_6^{-3/2} \beta_1^{-1}\ M_6 \ .
\end{equation}
Thus during the final stages of the merger, the validity of our assumption of adiabatic flow begins to break down for canonical parameters.  In future work, we will address this by including cooling terms in our gas evolution to account for energy losses due to radiation.  

We also note that in order for us to be able to neglect radiation pressure in the momentum equation, we require that the luminosity be small compared to the Eddington luminosity.  We find that for case RA1, 
\begin{equation}
  \frac{L^{max}_{syn}+L^{max}_{ff}}{L_{Edd}} \sim 0.2 \ n_1^2 \ T_6^{-3} \beta_1^{-1} M_6^2 \ ,
\end{equation}
which suggests that radiation pressure may begin to play a role for parameters close to our canonical choices.

In calculating the luminosity, we have assumed that the gas is optically thin.  We can verify this assumption by estimating the optical depth.  For the gas parameters chosen for this study ($n_{\infty} = 10 \mbox{ cm}^{-3}$ and $T_{\infty} = 10^6 \mbox{K}$), we find that the dominant source of opacity is electron scattering.  We estimate that the optical depth for electron scattering of synchrotron photons is 
\beq
\tau_{es} \approx n_h \sigma_T R \sim \ 10^{-3} n_1 T_6^{-3/2}M_6  ,
\eeq
where  $n_h \sim 10^{11} n_1 T_6^{-3/2}$ is the density at the horizon, $\sigma_T = 0.67 \times 10^{-24} \mbox{ cm}^2$ is the Thomson scattering cross-section, and $R \sim 10^{11} M_6 \mbox{ cm}$ is the characteristic size of the emission region.  Thus, our assumption of an optically thin gas is valid.

To estimate the characteristic frequencies at which the emission occurs we again note that the maximum emission comes from near the horizons and compute the characteristic frequency produced in this region.  For bremsstrahlung emission, the characteristic observed frequency of the emission is given by $h \nu \sim kT_h/(1+z)$ for a source at redshift $z$. We measure the maximum temperature near the horizon for our case RA1, and find that at the moment of maximum luminosity in the late pre-merger phase,
\begin{equation}
  h \nu_{ff}^{max} \approx \frac{150 \ {\rm MeV}}{1+z} \ (\mbox{RA1}) \ ,
\end{equation}
After the merger, in the quasistationary phase, we find that the characteristic frequency drops to
\begin{equation}
 h \nu_{ff} \approx \frac{10 \ {\rm MeV}}{1+z} \ (\mbox{RA1}) \ .
\end{equation}
Following the same procedure for case RA2, we find that at the moment of maximum luminosity in the late pre-merger phase,
\begin{equation}
  h \nu_{ff}^{max} \approx \frac{230 \ {\rm MeV}}{1+z} \ (\mbox{RA2}) \ ,
\end{equation}
and in the post-merger phase, the frequency drops to,
\begin{equation}
 h \nu_{ff} \approx \frac{50 \ {\rm MeV}}{1+z} \ (\mbox{RA2}) \ .
\end{equation}
Thus, we see that the bremsstrahlung emission will be predominantly in $\gamma$-rays, in agreement with \cite{shapiro73}. Given the bremsstrahlung luminosity calculated above, we estimate that the flux from this emission will be $\sim 10^{-21} \mbox{erg cm}^{-2} \mbox{ s}^{-1}$ for a source at $z=1$.  Unfortunately, it is unlikely that this emission is strong enough to be detectable.

We can estimate the characteristic frequency of the synchrotron emission by noting that Eq.~(\ref{angaveraged}) is maximized when  $x_M \approx 1.09$.  For case RA1, this corresponds to an observed frequency 
\begin{eqnarray}
  h \nu_{syn}^{max} &=& \frac{1.09}{1+z}\frac{3 e h B}{4 \pi m_e c} \left( \frac{kT}{m_e c^2}\right)^2\\
  &=&  \frac{80}{1+z} \ n_1^{1/2} \ T_6^{-3/4} \beta_1^{-1/2}\ \mbox{eV} \ (\mbox{RA1})
\end{eqnarray}
during the late pre-merger phase at the moment of maximum luminosity.  In the post-merger phase, the frequency drops to,
\begin{equation}
  h \nu_{syn} = \frac{0.008}{1+z} \ n_1^{1/2} \ T_6^{-3/4} \beta_1^{-1/2}\ \mbox{eV} \ (\mbox{RA1}) \ ,
\end{equation}
For case RA2, we find the characteristic synchrotron frequency to be
\begin{equation}
  h \nu_{syn}^{max} =\frac{100}{1+z} \ n_1^{1/2} \ T_6^{-3/4} \beta_1^{-1/2}\ \mbox{eV} \ (\mbox{RA2}) 
\end{equation}
during the late pre-merger phase at the moment of maximum luminosity, and 
\begin{equation}
  h \nu_{syn} =\frac{0.75}{1+z} \ n_1^{1/2} \ T_6^{-3/4} \beta_1^{-1/2} \ \mbox{eV} \ (\mbox{RA2}) 
\end{equation}
during the post-merger phase.

This corresponds to infrared and visible radiation.  For a binary at $z=1$ with the luminosity calculated above, this source has an apparent magnitude of $m=24$ and should be observable by the proposed LSST instrument \cite{LSST}.  Our simulations follow the late stage of the inspiral in which the binary separation decreases from $d=10M$ to merger.  For a $10^6 M_{\odot}$ binary, this corresponds to a timescale of $\Delta t \sim 1.3 \ \mbox{hrs}$ during which the radiation should achieve peak values.

We note that all the scalings derived above will not apply when $\vcloud \gg \sscloud$, but can be derived in a similar fashion.  For a realistic gas with $T \gtrsim 10^6 K$ ($\sscloud \approx 100 \mbox{km/s}$), this regime will might never be realized, so we neglect here.

\section{Summary and Discussion}
\label{sec:discussion}
In this paper we have performed a set of fully general relativistic simulations of BHBH binary mergers in gaseous environments, using our relativistic hydrodynamics code with AMR capability.  Our focus has been on demonstrating a mechanism for an observable electromagnetic signal which may accompany the gravitational wave signal from a black hole merger.  We have restricted our attention to gas clouds which are asymptotically uniform and either at rest (BHBH Bondi accretion), or moving  (BHBH BHL accretion) with respect to the binary center of mass.  We have performed ``prototype'' high-temperature ($T_{\infty} \sim 10^{11} \mbox{ K}$) simulations in order to gain a qualitative understanding of the different flow regimes characterizing such systems.  In these simulations, the accretion radius $\Racc$ is placed within the computational domain, which allows us to sample all the various regimes, studying how the luminosity and accretion rate change when the binary separation $d$ goes from $d > \Racc$ to $d < \Racc$.  We have also performed ``realistic'' simulations with astrophysically plausible asymptotic temperatures ($T_{\infty}=10^6 K$) to identify observable electromagnetic signals.  For each simulation, we have calculated the time-varying rest-mass accretion rate, as well as the electromagnetic luminosity due to bremsstrahlung and synchrotron emission.  We also derive scaling relations for the luminosity for the realistic cases, enabling our results to be used for different asymptotic gas parameters and BH masses.
 
In each case, we find evidence for a time-varying electromagnetic signature accompanying the BHBH binary merger.  For our ``realistic'' temperature simulations, we find that during the final inspiral from separation $d=10 M$ to merger, there is an enhancement in the total luminosity of $\sim1-2$ orders of magnitude.  This is followed by a sharp drop in the luminosity by a factor of $\sim 2-5$ orders of magnitude immediately following the merger.  In each case, we find that the luminosity is dominated by the synchrotron component, and that this signature should be detectable by the proposed LSST instrument.

We note that the nature of the electromagnetic emission may be altered
significantly when one considers different accretion geometries, such
as disk accretion, for example.  We intend to study such systems in future work.

By solving the BHBH binary analogs of the classic
Bondi and BHL accretion problems we have sought to
lay a natural and rigorous foundation for our future simulations
involving BHBH mergers in gaseous environments. Our focus
here was to perform ``prototype'' simulations to
identify the different regimes characterizing
the gas flow in these classic scenarios and then to perform a
``realistic'' simulation for a merger occuring in an astrophysically
plausible gaseous setting. For the latter case we also determined the scaling
behavior of the gas dynamical parameters, emitted luminosities
and characteristic frequencies so that they could be used to apply our results
to other environments. We noted that for asymptotic gas parameters
not too different from the ones that characterize
the interstellar medium in recent models of
merging galaxies ($n_{\infty} \sim 10~\mbox{cm}^{-3}$ and $T_{\infty} \sim 10^6 \mbox{K}$)
some of the assumptions that go into our calculation may be breaking
down, like the assumptions of adiabaticity, sub-Eddington luminosities
and an optically thin medium. These are complications we intend to address
in future work. We also hope to compare our findings to other recent
investigations performed simultaneously with ours (e.g. \cite{bode09}),
although these treat a different scenario and adopt very different
initial data, making a direct comparison difficult.
Finally, we hope to use the simulations reported here as point of
comparison with future simulations involving BHBH mergers in ambient gaseous
disks.

\acknowledgments
We would like to thank Z. Etienne and C. Gammie for useful discussions.  We would also like to thank M. Ansorg
for providing the {\tt TWOPUNCTURES} code for generating BHBH metric initial
data. This paper was supported in part by NSF Grants PHY02-05155 and
PHY06-50377 as well as NASA Grants NNG04GK54G and NNX07AG96G. Simulations were
performed under TeraGrid Grant TG-MCA99S008.

\appendix
\section{Derivation of $\dot{M}$ Expression}
\label{sec:Mdot_appendix}
\begin{figure}
  \includegraphics[width=6cm]{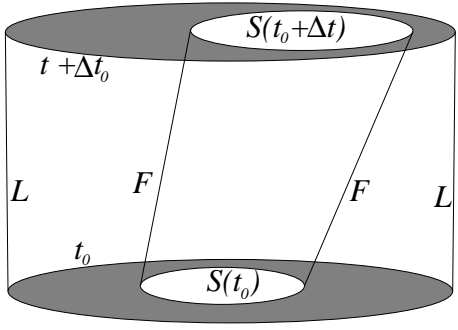}
  \caption{Spacetime diagram depicting the hypersurfaces relevant to 
calculating the mass accretion rate. The 2D hypersurface $S(t_0)$ and 
$S(t_0+\Delta t)$ (white regions) represent BH horizons on 
neighboring time slices. They are enclosed by spacelike 3D hypersurfaces 
$\Sigma_{t_0}$ and $\Sigma_{t_0+\Delta t}$ (shaded region) on 
these slices.}
  \label{fig:4dMdot_fig}
\end{figure}

Consider a 3D hypersurface $F$ given by $f(t,x,y,z)= constant$ in a spacetime diagram (see 
Fig.~\ref{fig:4dMdot_fig}).  
The intersection of $F$ with a $t=\mbox{constant}$ time slice is a 2-surface $S(t)$. 
Below we will identify $S(t)$ to be the apparent horizon of a black hole at time $t$. 
Let $L$ be a 3D hypersurface which is a worldtube enclosing $F$. 
Let us further define $\Sigma_t$ to be the 3D region on the time slice $t$ between the surface $S(t)$ 
and $\Omega(t)$, the 2D cross section of $L$ on the time slice $t$. 
We can imagine a 4-volume $\mathcal{V}$ to be the region bounded by 
the hypersurfaces $\Sigma_{t_0}, \Sigma_{t_0+\delta t}, L,$ and $F$.  
Consider a 4-vector flux $j^{\mu}$ with a vanishing divergence, 
$\nabla_{\mu}j^{\mu}=0$. At a given time $t$, define the function 
$q(t)$ according to
\begin{equation}
  q(t) = \int_{\Sigma_t}  j^{\mu} d^3 \Pi_{\mu}
\end{equation}
From Gauss's law, 
\begin{eqnarray}
  0&=& \int_{\mathcal{V}} \nabla_{\mu} j^{\mu} d \mathcal{V} = \int_{\partial \mathcal{V}} j^{\mu} d^3 \Pi_{\mu} \nonumber \\
  &=&q(t_0+\delta t)-q(t_0) \nonumber \\
  &+&\int_{F} j^{\mu} d^3 \Pi_{\mu}-\int_{L} j^{\mu} d^3 \Pi_{\mu} \ .
\end{eqnarray}

To evaluate the integral on $F$, it is convenient to introduce a coordinate 
system $(t,f,a,b)$, where $a=a(t,x,y,z)$ and $b=b(t,x,y,z)$ are two 
other coordinates. We may write
\begin{eqnarray}
  \int_F j^\mu  d^3 \Pi_{\mu} &=& 
  \frac{1}{3!}\int j^{\mu} \epsilon_{\mu \nu \rho \sigma} dx^\nu dx^\rho 
  d x^\sigma \cr 
  &=& \frac{1}{3!}\int j^f \epsilon_{ftab} \ dt \wedge da \wedge db \cr
  &=& -\int \sqrt{-g'}j^f \ dt da db \nonumber \\
  &=& -\int \sqrt{-g} j^\mu \partial_{\mu}f J \ dt da db \ ,
\end{eqnarray}
where $g'$ is the determinant of the metric in the $(t,f,a,b)$ 
coordinate system, $g$ is the determinant in the $(t,x,y,z)$ 
coordinate system, $J$ is the Jacobian
\begin{equation}
  J = \left|
    \frac{\partial(t,f,a,b)}{\partial(t,x,y,z)} \right|^{-1} 
    = \left|\frac{\partial(f,a,b)}{\partial(x,y,z)} \right|^{-1}  \ ,
\end{equation}
and $j^f=j^\mu \partial_{\mu}f$ follows from the usual transformation 
formula for a vector field between the $(t,f,a,b)$ and $(t,x,y,z)$ 
coordinate systems. 
Similarly, if $L$ is given by $l(t,x,y,z)=0$, then 
\begin{equation}
  \int_{L} j^{\mu} d^3 \Pi_{\mu} = -\int \sqrt{-g}j^\mu \partial_{\mu}l 
J_l \ dt da db \ ,
\end{equation}
where $J_l=|\partial(l,a,b)/\partial(x,y,z)|^{-1}$.
Taking the limit $\delta t \rightarrow 0$, we obtain
\begin{equation}
   \frac{dq}{dt} = -{\cal F}_F + {\cal F}_L \ ,
\end{equation}
where 
\begin{eqnarray}
  {\cal F}_F &=& -\int_F \sqrt{-g} j^\mu \partial_{\mu}f J \ da db \ , \\
  {\cal F}_L &=& -\int_L \sqrt{-g} j^\mu \partial_{\mu}l J_l \ da db \ .
\end{eqnarray}

Consider a fluid accreting onto a BH. Let $F$ be the BH horizon world tube. 
At any given time $t$, consider 
the fluid in the region $\Sigma_t$ between the BH (apparent) horizon 
$S(t)$ and a distant 2-surface $\Omega(t)$ surrounding the BH. Let $L$ 
be the 3D hypersurface formed by stacking $\Omega$ with time. The 
continuity equation gives $\nabla_\mu (\rho_0 u^\mu)=0$. 
Setting $j^\mu = \rho_0 u^\mu$, we have 
\begin{equation}
  q(t) = \int_{\Sigma_t} \rho_0 u^\mu d^3 \Pi_{\mu} = \int_{\Sigma_t} \rho_* d^3 x = M_0(t)
\end{equation}
is the rest mass bounded by the surface $\Omega$ and the horizon, 
where $\rho_* = \sqrt{-g}\, \rho_0 u^0$. Hence we have 
\begin{equation}
  \frac{dM_0}{dt} = {\cal F}_L  - {\cal F}_F \ .
\label{eq:dM0dt}
\end{equation}
This equation states that the rate of change of the rest mass is equal 
to the amount of rest mass flowing into $\Omega$ per unit time 
(${\cal F}_L$) minus the amount of rest mass flowing into the 
horizon per unit time (${\cal F}_F$). Hence we define the rest-mass accretion rate 
onto the BH according to 
\begin{equation}
\dot{M} \equiv {\cal F}_F = -\int_F \alpha \sqrt{\gamma} 
\rho_0 u^{\mu} \partial_{\mu} f J d\theta d\phi \ ,
\end{equation}
which is the expression given in Eq.~(\ref{Mdot_expression}). 
Here we have used the identity $\sqrt{-g}=\alpha \sqrt{\gamma}$, 
and choose $a$ and $b$ to be the spherical angular coordinates 
$\theta$ and $\phi$ with the origin at the BH center.

It is apparent from the definition that in general $\dot{M}$ depends on 
how the spacetime is sliced near the horizon. However, in some cases 
$\dot{M}$ may be time independent. Consider the cases 
where the fluid's mass is negligible compared to the BH's mass and 
there exists a timelike Killing vector $\ve{\xi}=\partial/\partial \lambda$ 
in the vicinity of a BH. This Killing vector could be the time Killing 
vector describing a stationary BH, or a helical Killing vector which 
approximates the BHBH spacetime in the inspiral phase.
One might choose to measure $\dot{M}$ using a coordinate system in which 
$t=\lambda$ (at least locally). In this case, 
\beq
  \dot{M}\equiv \dot{M}_\lambda = -\int_F \sqrt{-g_\lambda} \rho_0 u^f 
d\theta d\phi \ ,
\eeq
where $g_\lambda$ is the determinant of the spacetime metric in the 
$(\lambda,f,\theta,\phi)$ coordinate system. Suppose the flow of the 
fluid also achieves a stationary state near the horizon in which 
$\partial_\lambda (\rho_0 u^f)=0$ everywhere on $F$, i.e.\ 
$\rho_0 u^f=\rho_0 u^f(\theta,\phi)$ on $F$.
Since $\partial_\lambda g_\lambda=0$, $\dot{M}_\lambda$ is a constant 
independent of the coordinate time $\lambda$. On the other hand, 
in a coordinate system in which $\lambda$ is not the time coordinate, we have 
\beq
  \dot{M} = -\int_F \sqrt{-g'} \rho_0 u^f d\theta d\phi = -\int_F (\xi^t)^{-1} \sqrt{-g_\lambda} \rho_0 u^f d\theta d\phi \ ,
\eeq
where $g'=g_\lambda/(\xi^t)^2$ is the determinant of the spacetime metric 
in the $(t,f,\theta,\phi)$ coordinate system, and $\xi^t$ is the time 
component of the Killing vector $\ve{\xi}$. Note that both $g_\lambda$ and 
$\rho_0 u^f$ are still time independent on $F$, but $\dot{M}$ is 
time dependent if $\xi^t$ is time dependent. The accretion rate 
$\dot{M}$ is time independent only if a gauge is chosen 
so that $\partial_t \xi^t=0$ on $F$. Furthermore, if $\xi^t$ is constant 
everywhere on $F$, then $\dot{M}=\dot{M}_\lambda/\xi^t_h$, 
where $\xi^t_h$ is the value of $\xi^t$ on $F$.
As an example, consider a stationary accretion flow onto a Kerr BH. 
In the boosted Kerr-Schild coordinates, we have 
$\xi^t=\gamma_b=1/\sqrt{1-v_b^2}$ everywhere in the spacetime, 
where $v_b$ is the boost velocity. Hence we have 
$\dot{M}=\dot{M}_\lambda/\gamma_b$ in the boosted 
Kerr-Schild coordinates.

We point out that in a numerical simulation, even if a Killing 
vector exists, the adopted gauge (i.e.\ time slicing) may not 
correspond to the gauge in which $\partial/\partial t$ is the 
Killing vector. However, in some situations there exists a gauge in which a (quasi)stationary 
flow is expected. Such situations include the Bondi accretion 
onto a BHBH binary in the inspiring phase and the Bondi accretion 
onto a single BH following the binary merger. In these situations, 
the mass accretion rate onto a distant, fixed surface $\Omega$ 
(i.e.\ ${\cal F}_L$) 
is time independent for any gauge choices that give rise to a spacetime 
that is asymptotically Minkowsky. 
A ``well-behaved'' gauge should give a ${\cal F}_F$ (or the sum of 
two ${\cal F}_F$'s in the BHBH case) equal to ${\cal F}_L$ 
(when averaged over time); otherwise, Eq.~(\ref{eq:dM0dt})
implies that there will be an accumulation (if ${\cal F}_F<{\cal F}_L$) 
or depletion (if ${\cal F}_F>{\cal F}_L$) of 
rest mass in the interior of $\Omega$ as a result of a pure gauge 
effect. In our numerical simulations, we do not see such a gauge 
effect. We compute ${\cal F}_L$ on 
spherical surfaces of various radii. We find that after the flow 
reaches a (quasi)stationary state, ${\cal F}_L$ is 
slowly changing with time in the binary inspirling
phase and is approximately time independent after merger. 
The computed fluxes at various radii are also the same. Moreover, 
the sum of the computed fluxes at the BH horizons agree with the value 
${\cal F}_L$, indicating that our adopted puncture gauge conditions 
are well-behaved.
\section{Emissivities}
\label{sec:emissivity_appendix}
\subsubsection{Bremsstrahlung emissivity}
In order to estimate the electromagnetic emission due to bremsstrahlung, we use the following expressions for electron-ion, and electron-electron cooling rates given in \cite{narayan95}
\begin{eqnarray}
  \label{brememissivity_q}
q_{ff} &=& q_{ei} + q_{ee}\\
q_{ei} &=& n^2\frac{8 \pi}{3} (\alpha_f r_e^2 c) \
 (m_e c^2)
F_{ei}(\theta) 
\mbox{ ergs cm}^{-3} \ \mbox{s}^{-1}\ \ \ \ \ \\
q_{ee} &=&n^2(\alpha_f r_e^2 c) (m_e c^2) F_{ee}(\theta)
\mbox{ ergs cm}^{-3} \ \mbox{s}^{-1}
\end{eqnarray}
Here, $\alpha_f=e^2/\hbar c$ is the fine structure constant, $r_e=e^2/m_e c^2$ is the classical electron radius, $\theta \equiv k T / m_e c^2$,  $n=\rho_0 / m_B$ is the baryon number density, and
\begin{eqnarray}
  F_{ei}(\theta) &=& \left\{\begin{array}{l l}
      \displaystyle   4\left(\frac{2\theta}{\pi^3}\right)^{1/2}(1+1.781 \ \theta^{1.34}) & \ \ \ \theta < 1\\
      \displaystyle  \frac{9 \theta}{2 \pi} [ \mbox{ln}(1.123 \theta + 0.48) + 1.5] & \ \ \ \theta >  1
    \end{array}\right.\\
  F_{ee}(\theta) &=& \left\{\begin{array}{l l}
      \displaystyle  \frac{20}{9 \pi^{1/2}}(44-3 \pi^2)\theta^{3/2} &\\
      \displaystyle \times (1+1.1 \theta + \theta^2 - 1.25 \ \theta^{5/2}) & \ \ \ \theta < 1\\
      \displaystyle      24 \theta(\mbox{ln } 1.123\theta + 1.28) & \ \ \ \theta > 1 \ .
   \label{brememissivity_F}
 \end{array}\right.
\end{eqnarray}

\subsubsection{Synchrotron emissivity}
We use the estimates for synchrotron cooling rates given by \cite{pacholczyk70}:

\begin{equation}
q_{\nu,\mbox{s}} =  \frac{4 \pi n e^2 \nu}{\sqrt{3} c K_2(1/\theta)} I\left( \frac{x_M}{\mbox{sin} \theta}\right) \mbox{ergs cm}^{-3} \ \mbox{ s}^{-1} \ \mbox{ Hz}^{-1} \ ,
\end{equation}
where
\begin{eqnarray}
\nu_0 &=& \frac{e B}{2 \pi m_e c} = \mbox{cyclotron frequency}\\
x_M &=&\frac{2 \nu}{3 \nu_0 \theta^2} \ .
\end{eqnarray}
From \cite{mahadevan96}, we get the following approximation for $I(x_M)$.
\begin{equation}
I(x_M) = 2.561\left(1 + \frac{1.92}{x_M^{1/3}} + \frac{0.9977}{x_M^{2/3}}\right)\mbox{exp}(-1.8899x_M^{1/3}) \ ,
\end{equation}
and the angle averaged version,
\begin{equation}
\label{angaveraged}
I'(x_M) = \frac{4.0505}{x_M^{1/6}}\left( 1 + \frac{0.40}{x_M^{1/4}} + \frac{0.5316}{x_M^{1/2}}\right)\mbox{exp}(-1.8899x_M^{1/3}) \ .
\end{equation}

We integrate over frequency to get the total cooling rate. 

Let us denote,
\begin{eqnarray}
  A&=&4.0505 \frac{4 \pi n e^2}{\sqrt{3}c K_2(1/(\theta))}\\
  a_1&=&\frac{2}{3 \nu_0 \theta^2}\\
  a_2&=&1.8899 \ .
\end{eqnarray}
Thus, we see,
\begin{eqnarray}
  \int_0^{\infty} q_{\nu,\mbox{s}} d \nu &=&
  \frac{A}{a_1^2}\int_0^{\infty} x_M^{5/6} \left(1 
    + \frac{0.4}{x_M^{1/4}}
    + \frac{0.5316}{x_M^{1/2}}
  \right) \nonumber\\
&& \ \ \ \ \ \ \times \ \mbox{exp}\left(-a_2 x_M^{1/3}\right)
  d x_M \nonumber \\
  &=&
  \frac{3A C}{a_1^{2}}
\end{eqnarray}
Where
\begin{equation}
  C \equiv \frac{\Gamma(11/2)}{a_2^{11/2}} 
    + \frac{0.4 \ \Gamma(19/4)}{a_2^{19/4}}
    + \frac{0.5316 \ \Gamma(4)}{a_2^{4}}
    = 2.151889 
\end{equation}
Thus, we find
\begin{equation}
q_{\mbox{s}} = \int_0^{\infty} q_{\nu,\mbox{s}} d \nu = 
4.0505 \ C \frac{108 n^2 e^4 c}{\sqrt{3} K_2(1/\theta) }\frac{\theta^5 }{\beta m_e c^2} \ .
\end{equation}
Here we have used
\begin{equation}
  P=\beta P_M \equiv \beta \frac{B^2}{8 \pi} \ .
\end{equation}
As discussed in Sec.~\ref{sec:numerical}, we have chosen $\beta=10$ for our simulations based on simulations of magnetized accretion flows which have demonstrated that the magnetic fields do not typically reach their full equipartition value \cite{mckinney04}.  We further assume that the hydrodynamic flow is turbulent due to the magneto-rotational instability (MRI), causing the frozen-in field lines to be tangled, and randomly oriented \cite{shapiro73b}.  This justifies our use of the angle averaged function in Eq.~(\ref{angaveraged}).
Note that for $x \ll 1$, $K_2(x) \approx 2/x^2$, so that for sufficiently large temperatures we may approximate:
\begin{equation}
\label{synch_high_T}
q_{\mbox{s}} \approx \int_0^{\infty} q_{\nu,\mbox{s}} d \nu = 
4.0505 \ C \frac{54 n^2 e^4 c}{\sqrt{3} }\frac{\theta^3 }{\beta m_e c^2} \ .
\end{equation}

\bibliography{bbh_bondi_paper}
\end{document}